\documentclass[12pt]{iopart}
\usepackage{graphicx}
\usepackage{caption}

\begin{document}


\title{Dibaryons and where to find them}
\author{M. Bashkanov}
\ead{mikhail.bashkanov@york.ac.uk}
 \address{Department of Physics, University of York, Heslington, York, Y010 5DD, UK}
 
\author{D.P. Watts}%
\address{Department of Physics, University of York, Heslington, York, Y010 5DD, UK}

\author{G. Clash}%
\address{Department of Physics, University of York, Heslington, York, Y010 5DD, UK}
\author{M. Mocanu}%
\address{Department of Physics, University of York, Heslington, York, Y010 5DD, UK}
\author{M. Nicol}%
\address{Department of Physics, University of York, Heslington, York, Y010 5DD, UK}


\date{\today}

\begin{abstract}
In recent years there has been tremendous progress in the investigation of bound  systems of quarks with multiplicities beyond the more usual two- and three-quark systems.  Experimental and theoretical progress has been made in the four-, five- and even six-quark sectors. In this paper, we review the possible lightest six-quark states using a simple ansatz based on SU(3) symmetry and evaluate the most promising decay branches. The work will be useful to help focus future experimental searches in this six-quark sector. 
\end{abstract}


\submitto{\jpg}
\maketitle


\section{\label{sec:Intro} Introduction}
Theories of the strong interaction indicate there could be more ways to combine multiple quarks into a single object than established experimentally. The recent experimental discoveries of four- and five-quark systems has led to renewed interest in the field of multiquark states - where "multiquark" refers to objects with "non-standard" quark configurations beyond that of baryons and mesons.  Unfortunately, for many of these multiquark systems, the internal structure is not currently established. Due to this, CERN has adopted a new scheme in referring to these states. Tetraquarks refer to all objects with four quarks inside regardless if it is a genuine "tetraquark" or a meson-meson molecule, pentaquarks for five-quark objects (genuine pentaquarks and meson-baryon molecules) and hexaquarks (genuine hexaquarks and baryon-baryon molecules). We adopt this system in the current paper - but note that some of the earlier literature adopted different systems with hexaquark referring only to the non-molecular states.  

The hexaquark family can be further decomposed by quark content since the six-quark system can be arranged in two possible ways with three quarks and three antiquarks ({\it baryonium}) and six quarks ({\it dibaryon}). In this paper we consider the latter configuration and develop a simple theoretical ansatz so that we can infer the likely decay properties of such "dibaryon" six quark states.

It is interesting to note that hexaquarks were the first multiquark states to be established. The deuteron, a trivial hexaquark composed (predominantly) of a molecular state of a proton and neutron, was discovered in 1931. The earliest paper on non-trivial hexaquarks is attributed to Dyson and Xuong~\cite{DYS}, who predicted the existence of six non-strange hexaquarks based on the SU(6) model, just half a year after the discovery of quarks~\cite{Gell-Mann}. All of the states predicted by Dyson and Xuong have been the subject of experimental searches in the subsequent years. In Table.~\ref{d&x} below we summarise the experimental signals for the states of different isospin $I$ and spin $J$, along with their associated SU(3) multiplet.

\begin{table}[ht]
\centering \protect\caption{Six non-strange dibaryons}
\vspace{2mm}
\scalebox{0.75}{%
\resizebox{\textwidth}{!}{
\begin{tabular}{|l|c|c|c|r|}
\hline
$I,J$  &  D. and X. prediction [MeV] & Experimental results [MeV] & SU(3) multiplet &  Ref.\\
\hline
\hline
$0,1$    & 1876 & 1876 & $10^*$ & \cite{Urey} \\
$1,0$    & 1876 & 1878 & $27$  & \cite{FSI1,FSI2} \\
\hline
$1,2$    & 2160 & 2160 & $27$  & \cite{Arndt1,Arndt2} \\
$2,1$    & 2160 & 2160 & $35$  & \cite{isoT} \\
\hline
$0,3$    & 2350 & 2380 & $10^*$  & \cite{mb,MB,MBE1}\\
$3,0$    & 2350 & 2464 & $28$  & \cite{iso3}\\
\hline
\end{tabular}} \label{d&x}
}
\end{table}
As shown in the table, all states in this u,d quark sector have given experimental signatures with properties broadly consistent with the predicted properties. Signatures for some of the members have only been obtained very recently. 

Dibaryons, like any other particles, appear in SU(3) multiplets. Since SU(3) symmetry works reasonably well, one can use this symmetry to provide expectations of the properties of the family of particles in the multiplet (spin, isospin, mass). 

In their pioneering paper, Dyson and Xuong~\cite{DYS} focused on the u,d quark sector and did not consider the properties of strange quark containing members of dibaryon multiplets or their decay branches. However, modern experimental facilities now offer the prospect of experimental study of dibaryons in the strange quark sector. In this paper, we will review what is known about these mysterious strange quark containing particles, point to the most promising experimental channels and, where feasible, estimate their decay branches.

The paper is structured as follows. In Section~\ref{sec:Deut} we review the current experimental evidence and theoretical interpretation of the deuteron anti-decuplet. In Section~\ref{sec:FSI} we cover the complimentary NN-FSI 27-plet. Section~\ref{sec:dStar} addresses the anti-decuplet based on the $d^*(2380)$ hexaquark. Following a review we present our new theoretical estimate based on SU(3) symmetry considerations for the decay branches of the strange quark containing partners in this anti-decuplet under assumptions of pure hexaquark and pure molecular natures of the $d^*$ multiplet members. We than use the same theoretical approach to investigate decay properties of the $\Delta\Delta$ 28-plet (Section~\ref{sec:DD}), $N\Delta$ 27-plet (Section ~\ref{sec:ND27}) and $N\Delta$ 35-plet (Section~\ref{sec:ND35} ).

\section{Deuteron antidecuplet}\label{sec:Deut}

\begin{figure}[!h]
\begin{center}
        \includegraphics[width=0.5\textwidth,angle=0]{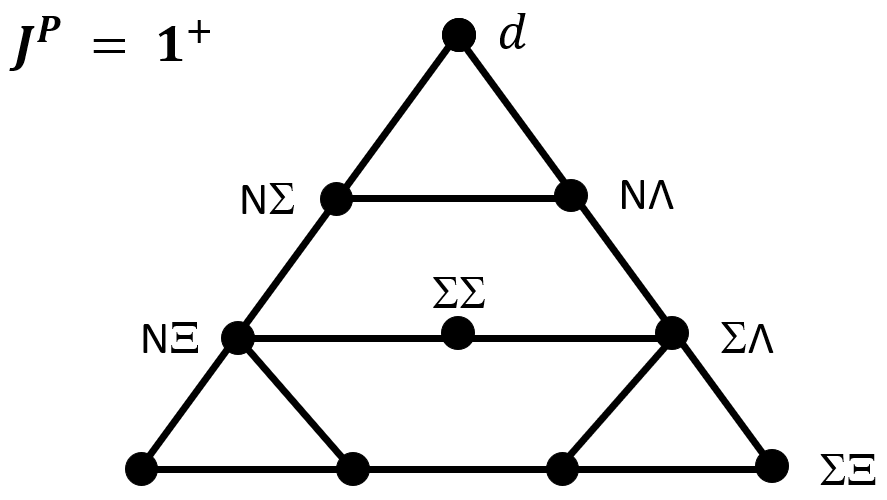}
\end{center}
\caption{Deuteron multiplet}
\label{DeuMult}
\end{figure}

The deuteron SU(3) multiplet is probably the most well-studied hexaquark SU(3) family. We know that the deuteron, and hence all the other objects in this multiplet, are predominantly baryon-baryon molecules. The estimated genuine hexaquark $|6q>$ component of the deuteron wavefunction is predicted to be only around $0.15\%$~\cite{Miller}.
Fig.~\ref{DeuMult} shows the members of the deuteron multiplet in standard convention (x-axis - 3rd projection of isospin, y-axis - strangeness). For each member of this antidecuplet we also show which $8\oplus8$ states can contribute.  
The SU(3) decomposition of this multiplet is also summarised in Table ~\ref{DeuTab}.
\begin{table}[ht]
\centering \protect\caption{Deuteron SU(3) multiplet}
\vspace{2mm}
{%
\scalebox{0.75}{
\begin{tabular}{|l|c|c|c|}
\hline
Strangeness  &  Decomposition & Masses of components[MeV] & Binding [MeV]\\
\hline
\ 0    & $\textrm{N}\textrm{N}$  & $\textrm{N}\textrm{N}$(1878) & -2.2\\
-1    & $\frac{1}{\sqrt{2}}(\textrm{N}\Lambda-\textrm{N}\Sigma)$ & $\textrm{N}\Lambda$(2054) $\textrm{N}\Sigma$(2129) & +0.166 \\
-2    & $\frac{1}{\sqrt{6}}(\sqrt{2}\textrm{N}\Xi+\Sigma\Sigma+\sqrt{3}\Sigma\Lambda)$  & $\textrm{N}\Xi$(2253) $\Sigma\Sigma$(2382) $\Sigma\Lambda$(2305)&  +0.5 \\
-3    & $\Sigma\Xi$  & $\Sigma\Xi$(2504) & +1.0 \\
\hline
\end{tabular}} \label{DeuTab}
}
\end{table}

Out of all multiplet members, only the deuteron appears to be bound. The $\Lambda-\textrm{N}$ interaction is not attractive enough to make a bound state e.g. the hyperdeuteron appears to be unbound. \footnote{With additional particles the nuclear binding would be expected to increase -  as a result hypertritium does becomes bound but by only $150 keV$.} The $\Lambda\textrm{N}$ state therefore presents as a virtual state or a Final-State Interaction(FSI)~\cite{Haid}. The $\Sigma-\textrm{N}$ interaction is suggested to be less attractive than $\Lambda-\textrm{N}$~\cite{Haid}. Due to the large SU(3) mass splitting in this decuplet ($M_\Sigma-M_\Lambda\sim 77$~MeV), compared to the MeV-keV range of bound/virtual states, it is difficult to treat these states in an SU(3) limit.

Aside from the deuteron, none of the states from the multiplet form a bound state, so all of them can be observed only in elastic/quasi-elastic reactions or via the correlation observations in heavy ion collisions. There were several recent publications from STAR and ALICE about extraction of the $\Lambda \textrm{N}$~\cite{NLambda}, $\Sigma \textrm{N}$~\cite{NLambda} and $\Xi \textrm{N}$~\cite{NXi} scattering lengths. Due to very unfavourable decay branches in the $\Sigma$ family, it is extremely difficult to access any correlation distributions where the $\Sigma$ baryon is involved. However, one may get some information in $\Xi \textrm{N}\to \Sigma\Sigma$ or $\Xi \textrm{N}\to \Lambda\Sigma$ from KLF or JPARC facilities from tertiary $\Xi$ beams.

\section{NN-FSI 27-plet}\label{sec:FSI}
\begin{figure}[!h]
\begin{center}
\includegraphics[angle=0,width=0.3\textwidth]{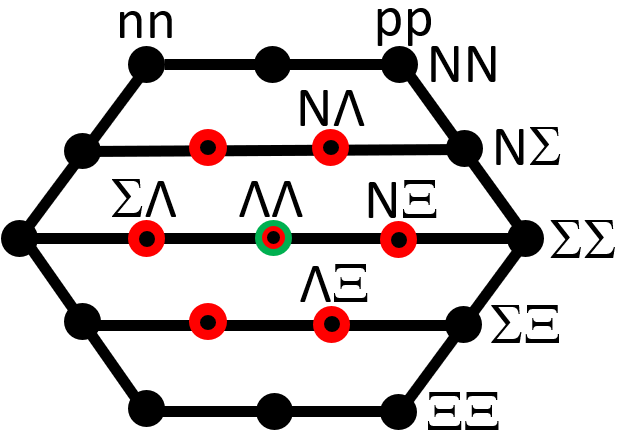}  
\end{center}
\caption{27-plet $8\oplus8$}
\label{27_oo_multiplet}
\end{figure}

The 27-plet $8\oplus 8$ of spin zero states is shown schematiclaly in Fig.~\ref{27_oo_multiplet} and Table \ref{Tab2} details the quark compositions. A strong attraction in the spin-zero pp-system was probably first realised by Migdal~\cite{FSI2} and Watson~\cite{FSI1} in their description of $\textrm{N}\textrm{N}$-FSI (Final State Interactions). Indeed it appears the absence of the tensor force in the $\textrm{N}\textrm{N}$ $^1S_0$ channel implies the "demon deuteron" would not form, as this system is unbound by 66 keV. Thus this multiplet is expected to be comprised of unbound molecular-like virtual states since the strangeness zero members are established to be unbound. As is the case for the deuteron multiplet(Section~\ref{sec:Deut}), this SU(3) family suffers from large SU(3) mass splitting. All the members can be observed only in elastic/quasi-elastic reactions and/or correlations  in heavy-ion collisions.

\begin{table}[ht]
\centering \protect\caption{Expected decay branches of the $8\oplus8$ SU(3) multiplet}
\vspace{2mm}
{%
\resizebox{\textwidth}{!}{
\begin{tabular}{|l|c|c|c|c|}
\hline
S  & Max Isospin & Med Isospin & Min Isospin & Mass [MeV] \\
\hline
\ 0     & $\textrm{NN}$ & & & $\textrm{NN}(1876)$\\
-1    & $\textrm{N}\Sigma$ & $\frac{1}{\sqrt{10}}(3\textrm{N}\Lambda - \textrm{N}\Sigma)$ & & $\textrm{N}\Sigma$(2127) $\textrm{N}\Lambda(2054)$ \\
-2    & $\Sigma\Sigma$ & $\frac{1}{\sqrt{10}} (\sqrt{6}\Sigma\Lambda+2\textrm{N}\Xi)$ & $\frac{1}{\sqrt{10}}(\frac{3\sqrt{3}}{2}\Lambda\Lambda - \sqrt{3}\textrm{N}\Xi - \frac{1}{2}\Sigma\Sigma$) & $\Sigma\Sigma(2378) \Sigma\Lambda(2305) \textrm{N}\Xi(2253) \Lambda\Lambda(2232)$ \\
-3    & $\Sigma\Xi $ & $\frac{1}{\sqrt{10}}(3\Lambda\Xi - \Sigma\Xi)$ & & $\Sigma\Xi(2504) \Lambda\Xi(2431)$\\
-4    & $\Xi\Xi$ & & & $\Xi\Xi$(2630)\\

\hline
\end{tabular}} \label{Tab2}
}
\end{table}

 Although this multiplet in isolation would not be expected to produce bound states the presence of overlapping members with the same quark content but different isospin raises the question about mixing. Effects established in the meson sector, such as $\rho-\omega-\phi$ mixing in the case of meson nonet, could in principle have corollaries here. Several members of this 27-plet can be mixed, but the most famous case proposed is the so-called H-dibaryon, a potential $\Lambda-\Lambda$ deeply bound state~\cite{Hdib}. Experimentally this is now shown to be unbound~\cite{LL1,LL2,LL3,LL4}. Recent advances in Lattice-QCD calculations for this sector indicate it to be located in the vicinity of the $\textrm{N}\Xi$-threshold,~\cite{HdibLQCD}. 
 
 In a simplified picture of this multiplet, we have 3 states with isospins 2,1,0 (plus a SU(3) singlet isosingle H-dibaryon state). These states can mix, but only in an isospin violating way, analagous to $\Lambda-\Sigma$ mixing in the baryon octet or $\rho-\omega$ mixing in meson nonet. The redistribution of wave function components, similar to the $\omega-\phi$ vs $\phi_0-\phi_8$ in the meson nonet, is not allowed here since all states have different isospin(besides $^{27}d_{ss}(I=0)-H$-dibaryon mixing). 

All three states have dissimilar properties: the $I=2$ is a pure $\Sigma\Sigma$ state, the $I=1$ is ($\sqrt{\frac{2}{5}}|\textrm{N}\Xi>+\sqrt{\frac{3}{5}}|\Sigma\Lambda>$) and $I=0$ ($\sqrt{\frac{12}{40}}|\textrm{N}\Xi>-\sqrt{\frac{1}{40}}|\Sigma\Sigma>+\sqrt{\frac{27}{40}}|\Lambda\Lambda>$). The $|I,I_z>=|1,0>$ state does not couple to $\Sigma\Sigma$ and $|I,I_z>=|0,0>$ state does not couple to $\Sigma\Lambda$. The $\Sigma\Sigma$ component of an $I=0$ state is tiny - such that no mixing with $I=2$ state is expected. The $I=0$ and the $I=1$ states can mix only via the $N\Xi$ component since it is the only common part in their wave functions. The H-dibaryon($|H>=\sqrt{\frac{1}{2}}|\textrm{N}\Xi>-\sqrt{\frac{3}{8}}|\Sigma\Sigma>-\sqrt{\frac{1}{8}}|\Lambda\Lambda>$) belongs to a singlet state, but it has exactly the same quantum numbers, including isospin, as the central state of the 27-plet. One could therefore expect some mixing between these states. However, since no H-dibaryon was found in the vicinity of the $\Lambda\Lambda$ threshold, the $I=0$ state from the 27-plet should be even less bound and should have much weaker experimental signatures.  

Since, for an experimental observation of this multiplet we are limited to elastic scattering and heavy-ion correlations, there are very few states which we can currently access. An elastic scattering with either direct beams or rescattering of secondary beams within target material~\cite{NZ} can give us access to $\textrm{NN},\Lambda \textrm{N},\Sigma \textrm{N}, \Xi \textrm{N}$  scattering data. The latter reaction can be performed at the recently proposed KLF facility~\cite{KLF} with $K_L \textrm{N}\to K^+ \Xi$ as a first step and $\Xi \textrm{N}\to X$ as a second step reaction within a large target volume (or with similar methodology at JPARC). 

For the heavy-ion correlation searches, all events with neutral final states need to be excluded from realisable experimental study, which essentially removes all channels with $\Sigma$ baryons. The $\textrm{N}\textrm{N}$, $\Lambda \textrm{N}$, $\Xi \textrm{N}$ and $\Lambda\Lambda$ channels were already investigated, see Refs~\cite{NLambda,NXi,LL3,LL4}. The only other channels pending analysis for this multiplet is  $\Xi \Xi$ (the $\Xi \Lambda$ ALICE analysis was recently published~\cite{LambdaXi}, but it suffers from low statistics). We do not expect any breakthrough here - with expectation of loosely unbound states. However, these data would help to fix several ChPT coupling  constants in the baryon-baryon sector and also improve our understanding of the nuclear equation of state for a system with strangeness, essential for neutron star physics.


\section{$d^*(2380)$ antidecuplet}\label{sec:dStar}

The antidecuplet of states built on the six-quark containing $d^*(2380)$ (hexaquark) are shown schematically in Fig. \ref{mult}. Signatures of the $d^*(2380)$ have been established quite rigorously in recent years following its initial observation in proton-neutron scattering and pionic fusion reactions~\cite{mb,MB,MBC,TS1,TS2,MBA,MBE1,MBE2,BCS}. It has a mass of $M_{d^*}=2380$~MeV, vacuum width $\Gamma = 70$~MeV and quantum numbers $I(J^P)=0(3^+)$. All of the strong decay branches have been identified and measured in experiment~\cite{BCS}. The electromagnetic properties of the $d^*(2380)$ were also investigated recently from measurements of its photoexcitation from deuteron targets~\cite{mbMainz, MBPy, MBCx, BWP20}, with further programmes planned~\cite{MAMI_proposal}. This particle has also recently been suggested  to have a potential impact in astrophysics~\cite{nstars,nstars1,Cond}.

The search for strange SU(3) partners of the $d^*(2380)$ is a natural next step for hexaquark studies. Characterisation of additional members of the antidecuplet would provide valuable insights into the underlying physics of hexaquark systems. All previous multiplets mentioned above have at best a very low binding suggesting its molecular nature. In contrast the $d^*(2380)$ has a very large binding energy, which unavoidably means it should have a large $|6q>$ component even if one considers it from a molecular picture. Moreover, the majority of theoretical papers in different ansatz suggest that the  $d^*(2380)$ might indeed be a genuine hexaquark-dominated object~\cite{BBC,Kuk2,Dong}. Its evidence in photoproduction (e.g. relative contribution of multipoles) also favours this possibility~\cite{BWP20,mbMainz}.   

However, as there is an abundance of states and associated decay channels from the many states in the $d^*(2380)$ antidecuplet, some prioritisation of initial searches is beneficial. In this paper, we develop a simple model to identify the most appropriate final states for each member of the antidecuplet, including estimates of the partial decay widths. Due to conservation laws, the production of any of the multiplet members is expected to proceed via associated many-body reactions at rather high energies. Deriving accurate predictions of the production cross-sections are therefore challenging. In their absence, an experimental strategy to search for such states is to look for structure in the cross sections consistent with the predicted locations, widths and branching ratios derived from theoretical models. In the following sections we present the first such calculations to guide future measurement strategies. The properties of the decouplet are inferred under assumption of a hexaquark (Sections~\ref{sec:dStar.hex}) and a molecular state (Sections ~\ref{sec:dStar.mol}). 

\subsection{$d^*(2380)$ multiplet masses}\label{sec:dStar.mass}

\begin{figure}[!h]
\begin{center}
        \includegraphics[width=0.5\textwidth,angle=0]{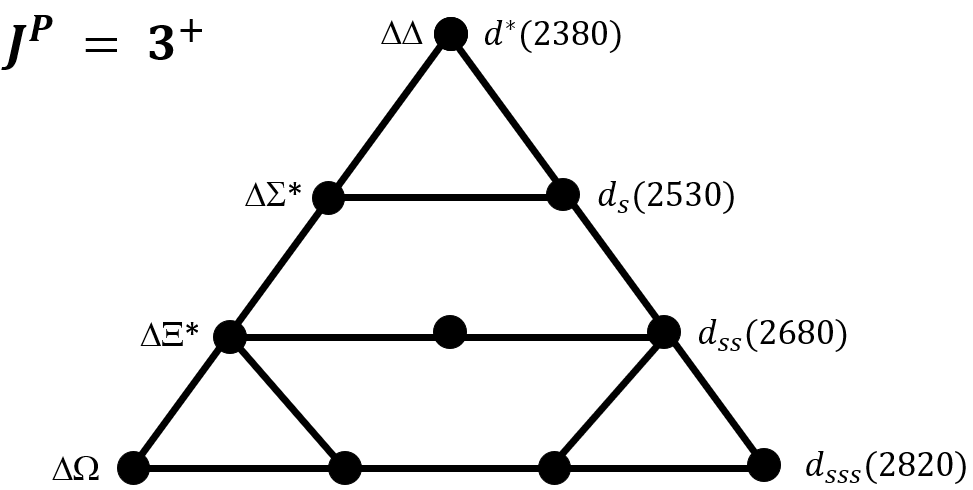}
\end{center}
\caption{$d^*(2380)$ multiplet}
\label{mult}
\end{figure}

From the unitary group theory of the strong interaction for the light quark (u,d,s) sector any strongly interacting particle, such as the  $d^*(2380)$, should be part of a SU(3) multiplet. For the case of the $d^*(2380)$ it would be expected to be a member of an antidecuplet, Fig.~\ref{mult}. The spectroscopic study of the other multiplet members would provide important new constraints on the $d^*$ internal structure, complementary to that potentially achievable in Form-Factor studies. In a simple molecular picture, the $d^*$ SU(3) multiplet would derive from the coupling of two baryon decuplet members bound by long-range pion exchange: corresponding to $\Delta\Delta$ for the $d^*(2380)$ to $\Delta\Omega$ for the $d_{sss}$. However, since the pion does not have a coupling to strange quarks, there is an expectation for such molecular systems that the binding energy should decrease with increasing strangeness content of the state. Conversely, in a genuine hexaquark (non-molecular) picture the binding energy should {\em increase} with increasing strangeness content
as the presence of heavier s-quarks would imply stronger binding.
In these two cases, a rough evaluation can be made for the masses of the $d^*(2380)$ multiplet.
\subsubsection{$d^*(2380)$ multiplet as genuine hexaquarks}\label{sec:dStar.hex}

The work of Gell-Mann presented a simplified way to calculate masses within an SU(3) multiplet~\cite{Gell-Mann}. In this approach the masses originate from two factors - the masses of the constituent quarks ($M_q$) and a contribution from hyperfine splitting ($\frac{K}{M_{q_1}M_{q_2}}$). In this approach the mass difference between the nucleon and the $\Delta$ (which have the same quark content) can be reproduced: $M_N=3M_q-\frac{3K}{M_{q}^2}$, $M_{\Delta}=3M_q+\frac{3K}{M_{q}^2}$. Here $M_q=363 MeV$ is the masses of light, unflavored constituent quarks and the $K$ is the parameter responsible for hyperfine splitting $\frac{K}{M_{q}^2}=50 MeV$. One can immediately see that if we substitute the light quark mass in the hyperfine term with the heavier strange quark mass, the splitting gets smaller and the baryon decuplet members get lighter. A hexaquark system has a lot more permutations between quarks, which is why the splitting term is essential. Using this ansatz we have combined  the evaluation of the $d^*(2380)$ multiplet masses in Table.~\ref{DMass1Tab}.
\begin{table}[ht]
\centering
\captionsetup{justification=centering,margin=0.5cm}
\protect\caption{Expected masses of the $d^*(2380)$ SU(3) multiplet in pure genuine hexaquark picture}
\vspace{2mm}
{%
\scalebox{0.8}{%
\begin{tabular}{|l|c|c|r|}
\hline
Particle  &  Mass structure & Mass value [MeV] & Binding\footnote{relative to Decuplet-Decuplet pole} [MeV]\\
\hline
$d^*$    & $6M_n+\frac{15K}{M_{n}^2}-B_h$  &  2380 & 84\\
$d_s$    & $5M_n+M_s+\frac{10K}{M_{n}^2}+\frac{5K}{M_{n}M_{s}}-B_h$  &  2474 & 141 \\
$d_{ss}$    & $4M_n+2M_s+\frac{6K}{M_{n}^2}+\frac{8K}{M_{n}M_{s}}+\frac{K}{M_{s}^2}-B_h$  & 2573 & 193\\
$d_{sss}$    &  $3M_n+3M_s+\frac{3K}{M_{n}^2}+\frac{9K}{M_{n}M_{s}}+\frac{3K}{M_{s}^2}-B_h$ & 2677 &  234 \\
\hline
\end{tabular}} \label{DMass1Tab}
}
\end{table}
For the entries in the Table.~\ref{DMass1Tab}, note that $M_n=363 MeV$ is the mass of a light constituent quark, $M_s=538 MeV$ is the mass of a strange constituent quark, $\frac{K}{M_{n}^2}=50 MeV$ is the splitting parameter and $B_h\sim-550 MeV$ is the hexaquark binding which reproduces the observed $d^*(2380)$ mass. This latter parameter is taken to be the same for all members of the multiplet (similar to a bag constant of Ref~\cite{Bag}).

Note that under these assumptions the mass of the $d_{sss}$ is 236 MeV lower than a $\Delta\Omega$ threshold and only 160 MeV higher than the $\Sigma\Xi$ threshold. This mass difference with the $\Sigma\Xi$ threshold is not large, especially considering the $d_{sss}\to\Sigma\Xi$ decay should proceed via a D-wave in the $\Sigma\Xi$-system. 

Although this is a simplified picture of the antidecuplet, its flexibility does enable lower limits for multiplet masses to be evaluated. We note there are a large variety of estimates with different theoretical ansatz presented in Refs.~\cite{Goldman,Mulders,Maltman}.

\subsubsection{$d^*(2380)$ multiplet as molecules}\label{sec:dStar.mol}

For the case of a molecular state, the dependence of the mass of the states with strangeness content should be very different than for a genuine hexaquark. Due to the larger spatial extent of molecular states, one can assume that the binding is mainly driven by long-range pion exchange. In the SU(6) approximation, the pion couples only to light quarks, so the strength of the interaction $\Delta:\Sigma^*:\Xi^*:\Omega$ should scale as $3:2:1:0$, see Ref.~\cite{nstars1}. Since the interaction potential ($V$) is proportional to the product of coupling constants and the binding energy $B\sim M_{Red}\cdot V^2$, where $M_{Red}$ is the reduced mass~\cite{Krane}. One can therefore estimate binding energies for various states under these assumptions. The $\Delta\Omega$ state is predicted have a binding energy $B=0~MeV$. However, since $d_{sss}$ has not only $\Delta\Omega$ but also a $\Xi^*\Sigma^*$ component, we expect some binding also in this case. The results of these simplified calculations are summarised in Table ~\ref{DMass2Tab}. Note that $M_{Red}$ is the reduced mass and $f$ is an effective meson baryon coupling constant fixed to reproduce the $d^*(2380)$ mass.

\begin{table}[ht]

\centering \protect
\caption{Expected masses of the $d^*(2380)$ SU(3) multiplet in pure molecular picture}
\vspace{0.5mm}
{%
\scalebox{0.8}{%
\begin{tabular}{|l|c|c|r|}
\hline
Particle  &  Binding energy structure & Mass value [MeV] & Binding\footnote{relative to the lightest member of the Decuplet-Decuplet state} [MeV]\\
\hline
$d^*$    & $M_{Red}(\Delta\Delta)(3f\cdot3f)^2$  &  2380 & 84\\
$d_s$    & $M_{Red}(\Delta\Sigma^*)(3f\cdot2f)^2$  & 2578   &  39\\
$d_{ss}$    & $2/3M_{Red}(\Delta\Xi^*)(3f\cdot1f)^2+1/3M_{Red}(\Sigma\Sigma^*)(2f\cdot2f)^2$  & 2753 & 13\\
$d_{sss}$    &  $0+1/2M_{Red}(\Sigma^*\Xi^*)(2f\cdot1f)^2$ & 2909 & 2  \\
\hline
\end{tabular}} \label{DMass2Tab}
}
\end{table}

\begin{figure}[!h]
\begin{center}
        \includegraphics[width=0.6\textwidth,angle=0]{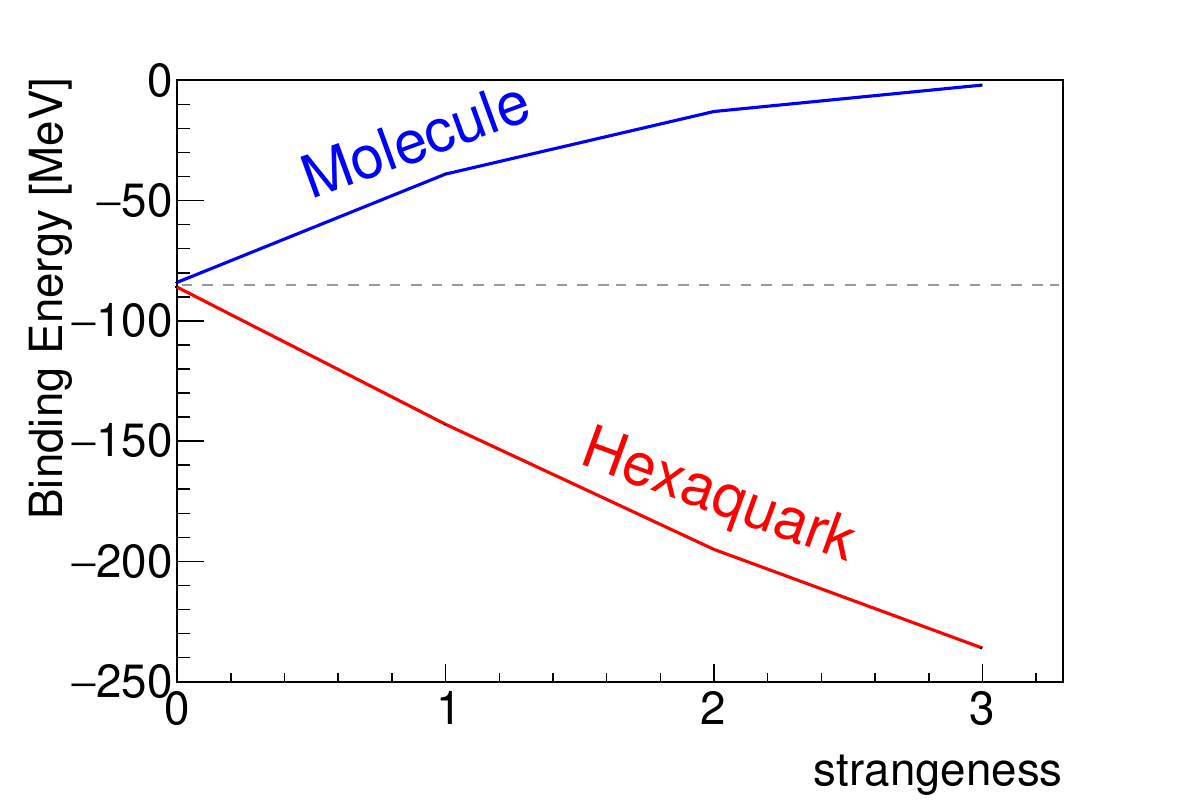}
\end{center}
\captionsetup{justification=centering,margin=2cm}
\caption{$d^*(2380)$ multiplet binding under assumption of pure genuine hexaquark (red) and pure molecule (blue).}
\label{mult_bind}
\end{figure}

The observed mass of each member of the antidecuplet would be expected to lie between our estimations for a pure "genuine hexaquark" and a pure "molecule", Fig.~\ref{mult_bind}. We note that group theory actually prohibits a 100\% genuine hexaquark state~\cite{BBC}, with the maximum purity state expected to contain 20\% of molecular admixture. More elaborate recent calculations based on hidden colour quark models predict a 30\% molecular contribution to the $d^*$~\cite{YB1}. 

In the following sections, we evaluate the widths and branching ratios as a function of binding energy. As a benchmark case, we take the same binding for all hexaquark members which, in this simplified ansatz, would correspond to nearly equal molecular/hexaquark contributions.

\subsection{$d^*(2380)$ multiplet decays}

Regardless of the inferred structure, the $10^*$ antidecuplet decay properties are largely driven by the SU(3) symmetry and available phase space for the decay products. Being a $10^*$  SU(3) antidecuplet, the decay couplings are limited to either octet+octet, $8\oplus8$ or decuplet+decuplet, $10\oplus10$ baryons. The possible decay branches, deriving from the established $8\oplus8$ and $10\oplus10$ baryonic members are summarised in Table.~\ref{tab:10dec}.
\begin{table}[ht]
\footnotesize
\centering \protect\caption{Expected decay branches of the $d^*(2380)$ SU(3) multiplet}
\vspace{2mm}
{%
\scalebox{0.85}{%
\begin{tabular}{|l|c|c|}
\hline
Particle  &  $8\oplus8$ & $10\oplus10$\\
\hline
$d^*$    & $pn$  & $\Delta\Delta$ \\
$d_s$    & $\frac{1}{\sqrt{2}}(\textrm{N}\Lambda-\textrm{N}\Sigma)$  & $\Delta\Sigma^*$ \\
$d_{ss}$    & $\frac{1}{\sqrt{6}}(\sqrt{2}\textrm{N}\Xi+\Sigma\Sigma+\sqrt{3}\Sigma\Lambda)$  & $\frac{-1}{\sqrt{3}}(\sqrt{2}\Delta\Xi^*+\Sigma^*\Sigma^*)$ \\
$d_{sss}$    & $\Sigma\Xi$  & $\frac{1}{\sqrt{2}}(\Delta\Omega+\Sigma^*\Xi^*)$ \\
\hline
\end{tabular}}}
\label{tab:10dec}

\end{table}
\normalsize

\subsubsection{Formalism}
In our calculations, we assume that the Breit-Wigner width of all the resonances is energy-dependent and that the coupling constants for the decay into $8\oplus8$ are independent of the decaying particle's hypercharge/strangeness. This enables the total width of the decaying state to be expressed as
\begin{equation}
\Gamma_{tot}=\Gamma_8+\Gamma_{10},
\end{equation}
with $\Gamma_8$ corresponding to the partial width for hexaquark decay into $8\oplus8$ and $\Gamma_{10}$ for $10\oplus10$.
In case of $8\oplus8$ decay, the final state particles are stable against strong decay so the $\Gamma_8$ decay width can be expressed as 
\begin{equation}
    \Gamma_8=g_8^2p^{2L+1}F_{8}(p)
\end{equation}
with
\begin{equation}
    F_{8}(p)=\frac{R^{2L}}{1+R^{2L}p^{2L}}
\end{equation}
where $g_8$ is the coupling constant of the hexaquark decay into $8\oplus8$, $p$ is the momentum of the ejectile in the hexaquark rest frame, $L$ is the angular momentum in the system and $F(p)$ is a Form-Factor, usually introduced to account for potential barriers, similar to Ref.~\cite{PBC,Kuk2}. For a $J^p=3^+$ particle decaying into two $J^p=1/2^+$ baryons there are two possible scenarios - a $^3D_3$ partial wave with $L=2$ and $^3G_3$ partial wave with $L=4$. From partial wave analysis of the $d^*(2380)\to pn$ reaction we know that the majority ($\sim 90\%$) of the $8\oplus8$ decay proceeds through the $^3D_3$ partial wave~\cite{MBE1,MBE2,TSE1}. In our calculation, we will assume that all $8\oplus8$ decays proceed with $L=2$. This assumption leads to minor corrections but allows a significant reduction in the required number of coupling constants. Since all the $3^+$-hexaquarks lie far above the $8\oplus8$ threshold the $\Gamma_8$ width is expected to be nearly constant throughout the resonance. The $g_8$ constant is the same for all members of the antidecuplet. It is fixed to the value extracted from $\Gamma(d^*\to pn) = 8$~MeV.  

The $10\oplus10$ channel is much more challenging since the baryon decuplet contains resonant states with a rather sizeable width and an associated strong energy dependence. The $\Gamma_{10}$ width is expected to vary significantly between different resonances. We have calculated the $\Gamma_{10}$ as follows

\begin{equation}
        \Gamma_{10}=\gamma_{10}\int dm_1^2dm_2^2F^2(q_{10})|D_{D_{1}}(m_1^2)D_{D_{2}}(m_2^2)|^2  
        \end{equation}
        \begin{equation}
        F(q_{10})=\frac{\Lambda^2}{\Lambda^2+q^2_{10}/4},\;\;\; \Lambda=0.16~GeV/c \\
        \end{equation}
        \begin{equation}
        D_{D}=\frac{\sqrt{m_{D}\Gamma_{D}(q_{M})/q_{M}}}{M_{BM}^2-m_{D}^2+im_{D}\Gamma_{D}^{tot}(q_{M})} \\
        \end{equation}
        \begin{equation}
        \Gamma_{D}=\gamma(q_{M})^3\frac{R^2}{1+R^2(q_M)^2} 
        \end{equation}
        

here $F(q_{10})$ is a Form-Factor which depends on the relative momentum ($q_{10}$) between the two decuplet baryons in the hexaquark decay. We follow the prescription of Ref.~\cite{PBC}, and parameterise this dependence in a monopole form with a cut-off parameter $\Lambda$~\footnote{the appropriateness of the value for this cutoff parameter is explored in Ref~\cite{MB}, where it was shown to provide agreement with measured invariant mass distributions in the region of the so-called ABC-effect.}. $D_{D_1}/D_{D_2}$ are the propagators for the decuplet of baryons, with $m_1/m_2$ the baryon masses (or correspondingly the invariant mass of the meson-baryon system $M_{BM}$ from the decuplet baryon decay into the octet of mesons and the octet of baryons) and $m_{D}$ being their nominal Breit-Wigner masses. The energy-dependent width of the baryon decuplet decay, $\Gamma_D$, is parameterised in a standard form taking a $P$-wave decay resonance with Blatt-Weisskopf barrier factors of $R=6.3$~GeV/c and $\gamma = 0.74, 0.28, 0.13$ for the $\Delta,\Sigma^*$ and $\Xi^*$ respectively~\cite{Gal2}. $\Gamma_D^{tot}$ is the sum of all partial widths. The width of the $\Omega$ is considered to be zero since it is stable with respect to strong decays. The Form-Factor parameters are fixed based on the $d^*\to \Delta\Delta\to d\pi\pi$ invariant mass distributions of Ref.~\cite{MB}. The $\gamma_{10}$ is taken as a normalisation factor and is fixed to reproduce the width at zero binding energy, e.g $\Gamma(d^* \to \Delta\Delta) = 2\cdot \Gamma_{\Delta}$ for $M_{d^*}=2\cdot M_{\Delta}$.

\begin{equation}
\Gamma_{10}(B=0)=\Gamma_{D_1}+\Gamma_{D_2}     
\end{equation}

\subsubsection{$d^*(2380)$ multiplet decays in a molecular picture}

If one would consider the $d^*$ multiplet to be a purely molecular state one can expect some difference compared to the calculations above. The $10\oplus 10$ decay would stay unchanged since it is a fall-apart decay and it is only driven by available phase space and the energy dependence of the widths. However, to get a $8\oplus8$ decay, one needs to get a quark rearrangement, so this decay would be, besides other things, dependent on the wave function overlap. In the calculations above, the $8\oplus8$ decay width stayed essentially constant over the large range of binding energies, while in the molecular picture the wave function overlap at $B=0$ should be zero, hence the $8\oplus8$ decay width should be also zero. To account for this effect we have modified Eq. 2 with

\begin{equation}
    \Gamma_8=\tilde{g_8}^2p^{2L+1}F_{8}(p)\cdot P(B)
\end{equation}

here $P(B)$ is a wave function overlap and $\tilde{g_8}$ is a modified $g_8$ constant to reproduce the $d^*(2380)$ width.

We have evaluated $P(B)$ in the following assumptions: i) all particles assumed to be spherical with a distribution, similar to the proton charge distribution $\rho(r)=exp(-a\cdot r)$, with a standard $a=4.27 fm^{-1}$ (a coordinate space analogue of the proton dipole Form-Factor with a cut-off parameter $\Lambda=0.72 GeV/c$). The distance between molecular components was taken as

\begin{equation}
    d=\frac{1}{\sqrt{2\cdot M_{Red}|B|}}
\end{equation}

with $M_{Red}$  being reduced mass and $B$ is the binding energy. The wave function overlap $P(B)$ derived under these assumptions is shown on the figure below, Fig.~\ref{overlap_fig}. The $\tilde{g_8}$ is again fixed to reproduce the value extracted from $\Gamma(d^*\to pn) = 8$~MeV.

 \begin{figure}[!h]
\begin{center}
        \includegraphics[width=0.4\textwidth,angle=0]{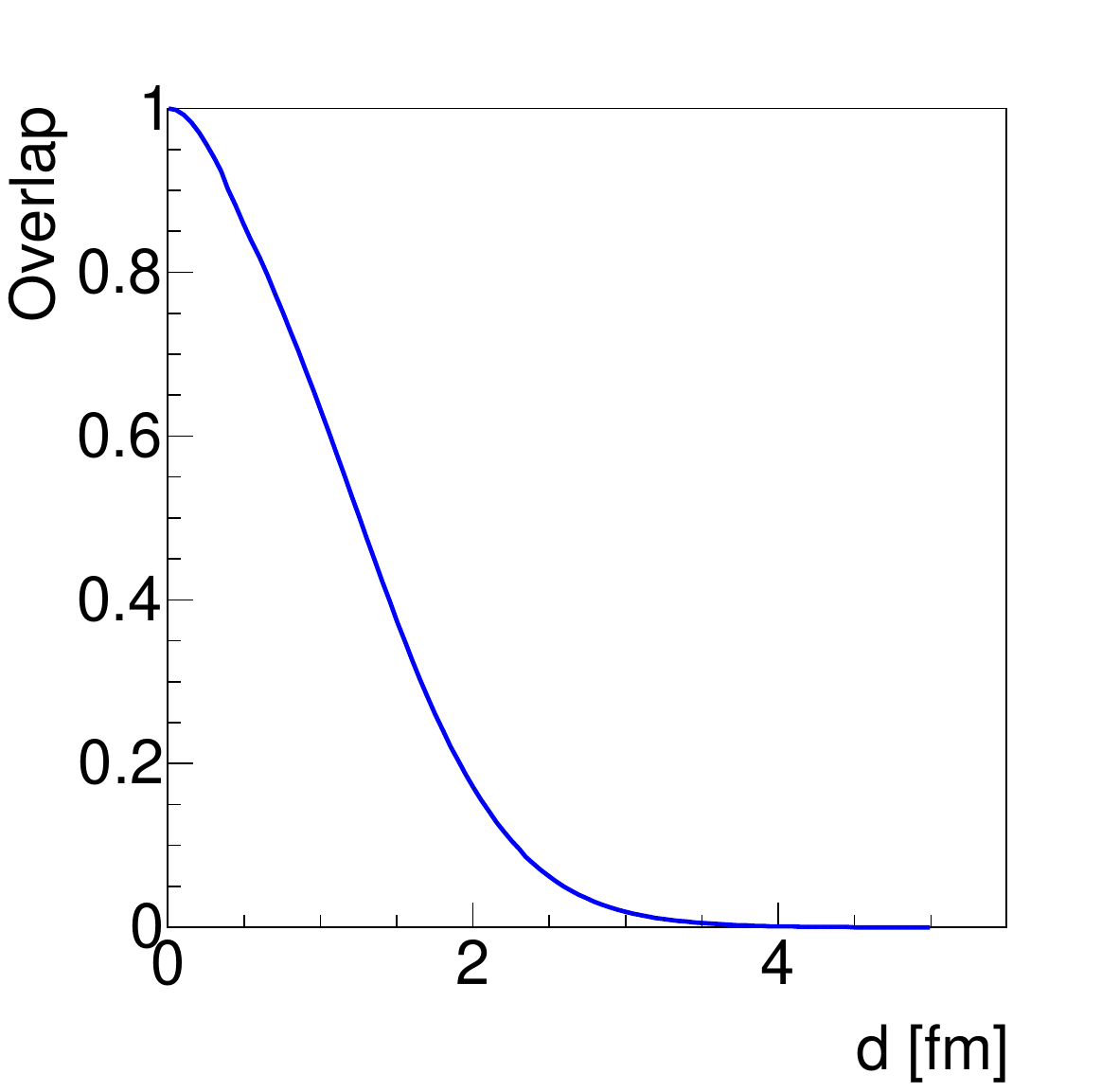}
        \includegraphics[width=0.4\textwidth,angle=0]{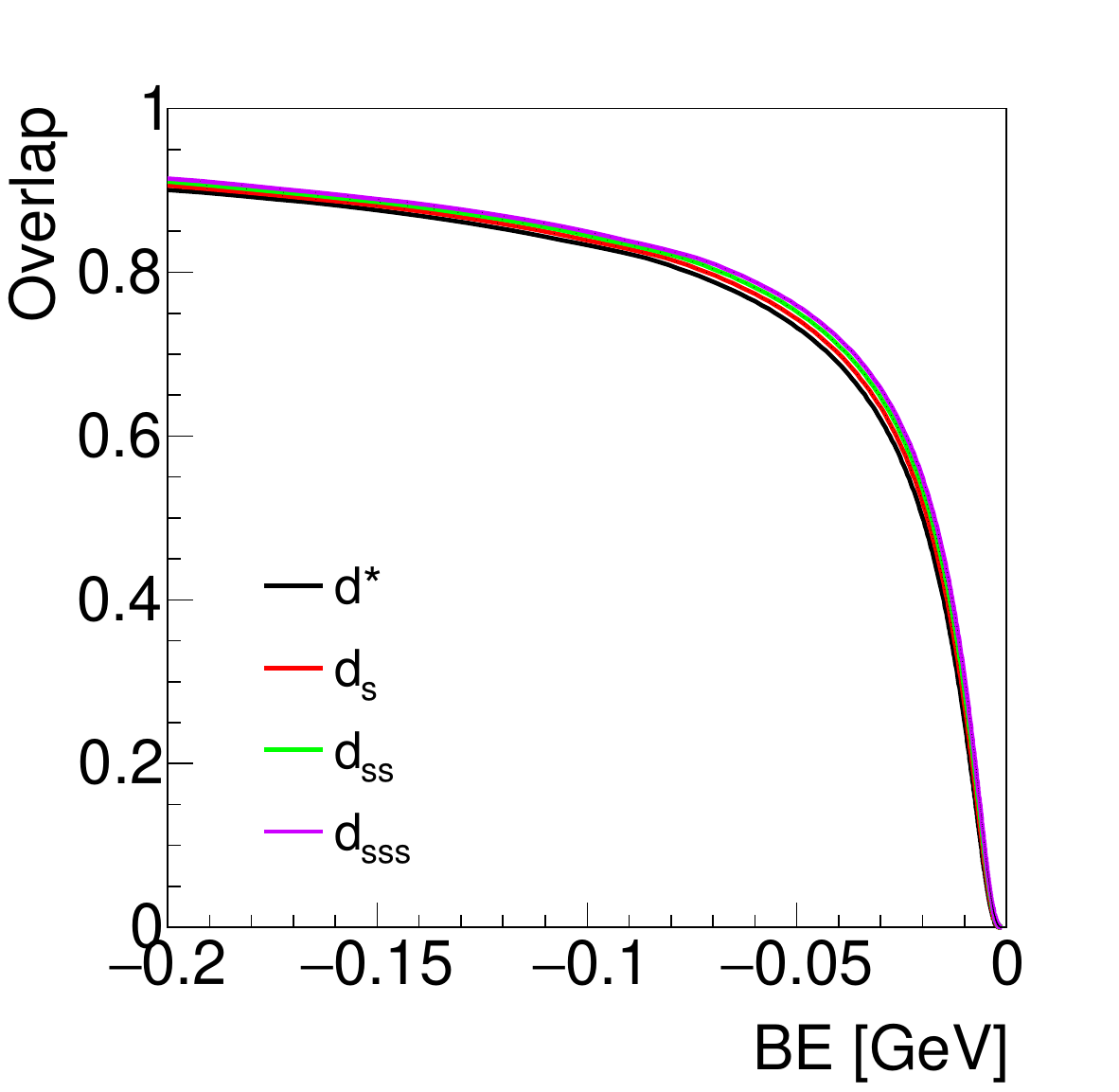}
\end{center}

\captionsetup{justification=centering,margin=2cm}
\caption{The density overlap for two diffuse balls in a molecule as a function of the distance(left) and binding energy(right).}
\label{overlap_fig}
\end{figure}

With these modified $8\oplus8$ decay width both the branching ratios and decay widths would change.

\subsubsection{Results for the d*(2380) decuplet under assumption of a genuine hexaquark}

\begin{figure}[!h]
\begin{center}
        \includegraphics[width=0.4\textwidth,angle=0]{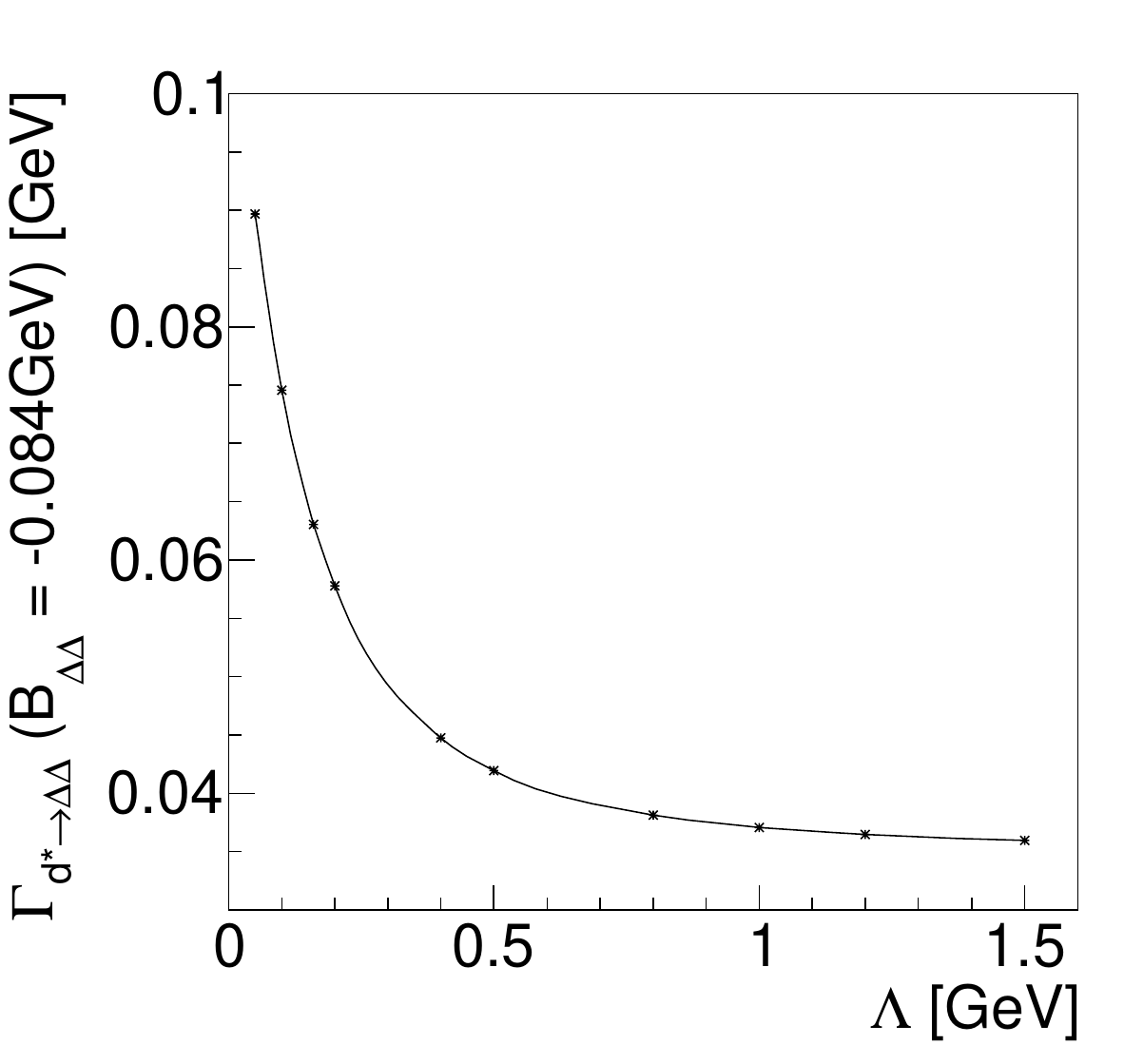}
        \includegraphics[width=0.4\textwidth,angle=0]{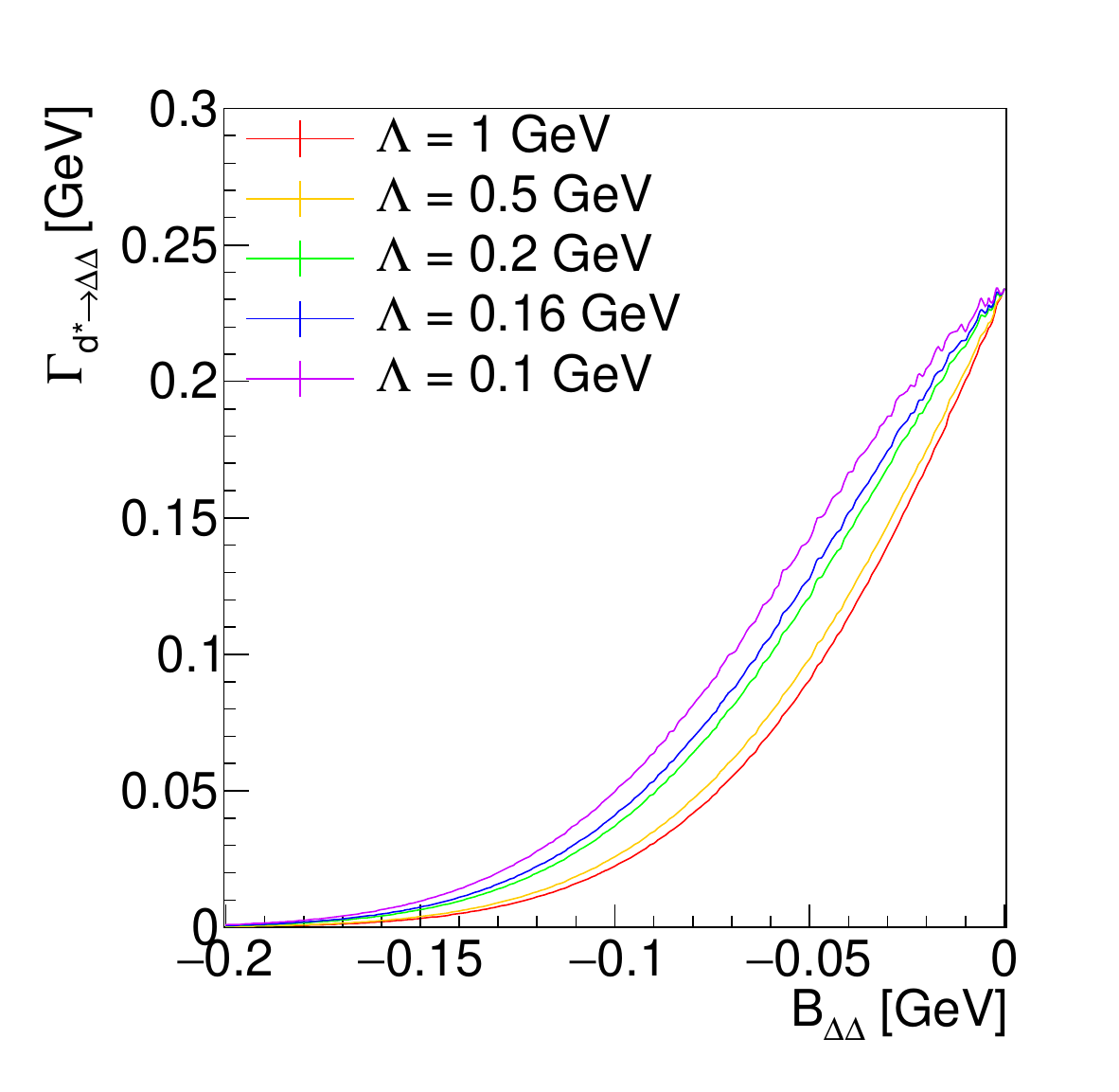}
\end{center}

\captionsetup{justification=centering,margin=1cm}
\caption{$d^*(2380)$ width as a function of the Form Factor cut-off parameter $\Lambda$ (left) and partial width $\Gamma_{d^*\to\Delta\Delta}$ as a function of binding energy (right) for various $\Lambda$ values, $\Lambda = 0.1, 0.16, 0.2, 0.5, 1$~GeV/c in rainbow order from violet to red.}
\label{d_width}
\end{figure}

We first explored the validity of the adopted cut-off parameter $\Lambda =0.16$~GeV/c in the model.
The Form-Factor of the form Eq. 5. with a cut-off parameter $\Lambda =0.16$~GeV/c in the $d^*\to\Delta\Delta\to d\pi\pi$ was first introduced in a Ref.~\cite{MB} to explain the so-called ABC-effect, an enhancement in $M_{\pi\pi}$ close to the threshold. Indeed, for the case where the nucleons in the deuteron have a very small relative momentum, there is a correspondence in the relative momentum between the $\Delta$'s and the relative momentum between the pions (and hence with the pion invariant mass). It was later speculated by A. Gal, that reduction of the $\Delta\Delta$ system size within the $d^*$  can lead to a further reduction of the $d^*$ width~\cite{Gal2}. To clarify the situation we have studied the predicted $d^*\to\Delta\Delta$ width dependence as a function of the cut of parameter $\Lambda$, Fig.~\ref{d_width}. The results indicate that the adopted $\Lambda =0.16$~GeV/c can not only reproduce the ABC effect but also gives agreement with the measured $\Gamma(d^*\to\Delta\Delta)=62$~MeV. As evident in Fig~\ref{d_width}  we also reproduce the trend in which a higher value adopted for the cut-off parameter leads to a smaller width. 

\begin{figure}[!h]
\begin{center}
        \includegraphics[width=0.24\textwidth,angle=0]{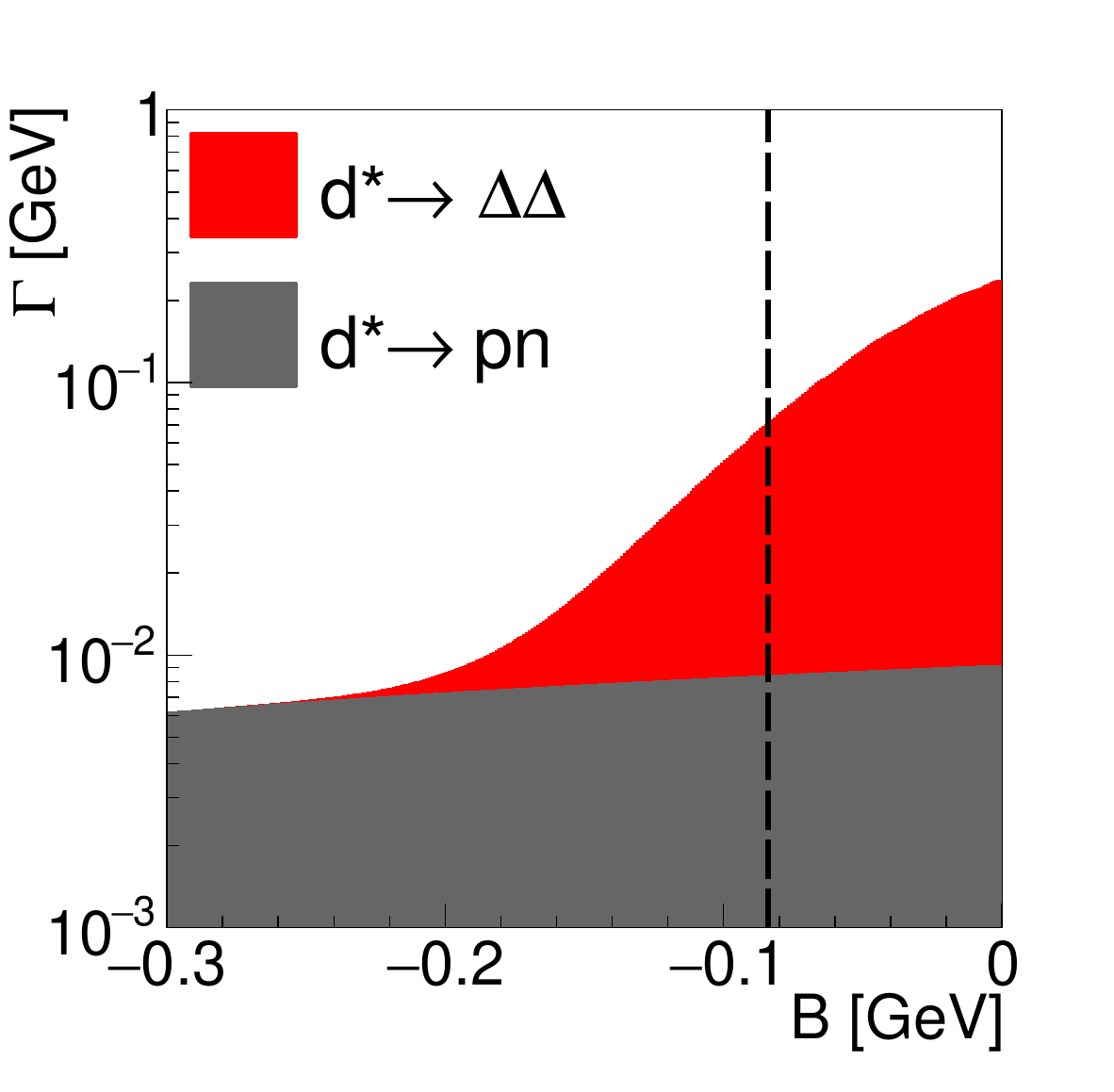}
        \includegraphics[width=0.24\textwidth,angle=0]{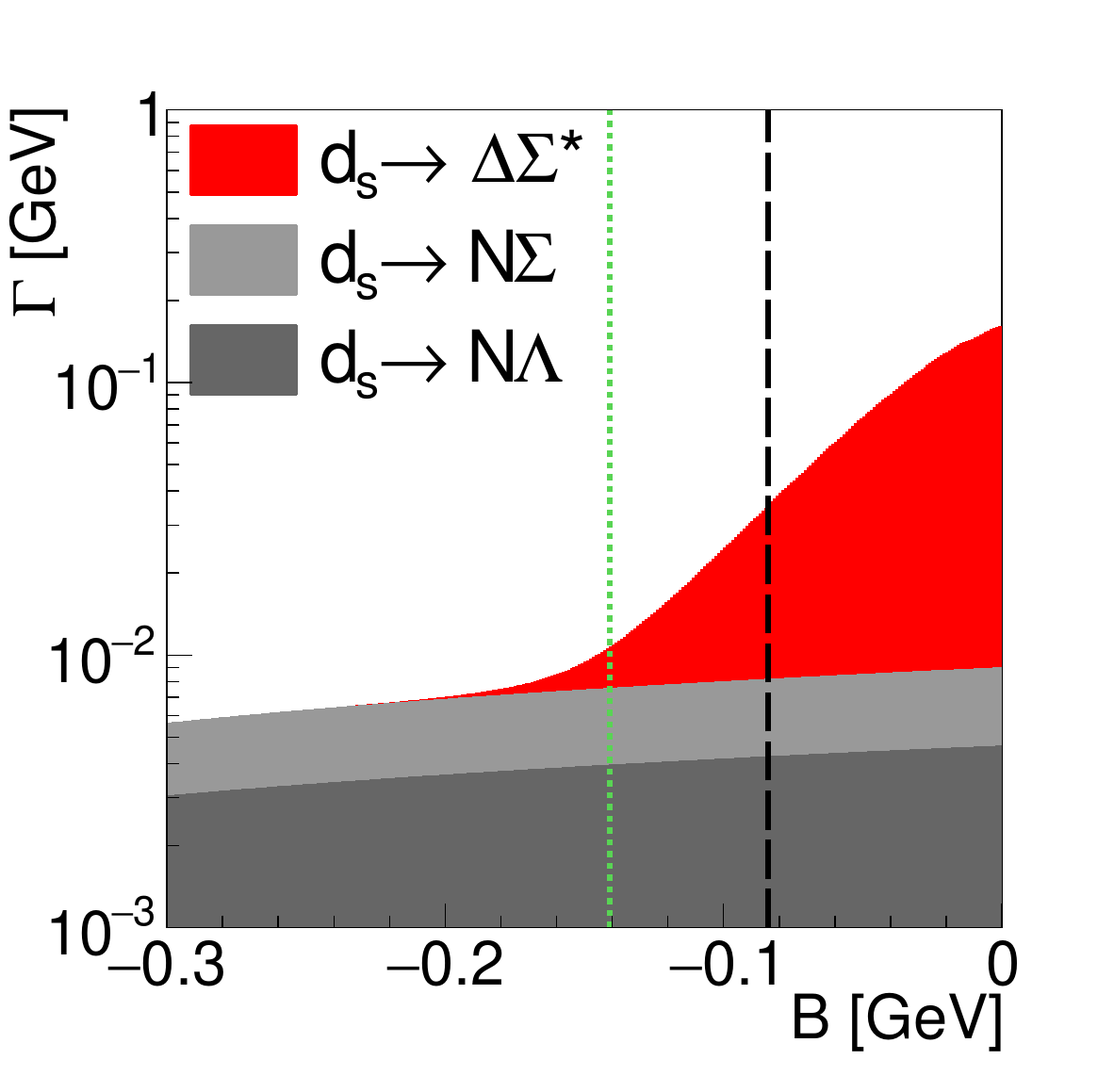}
        \includegraphics[width=0.24\textwidth,angle=0]{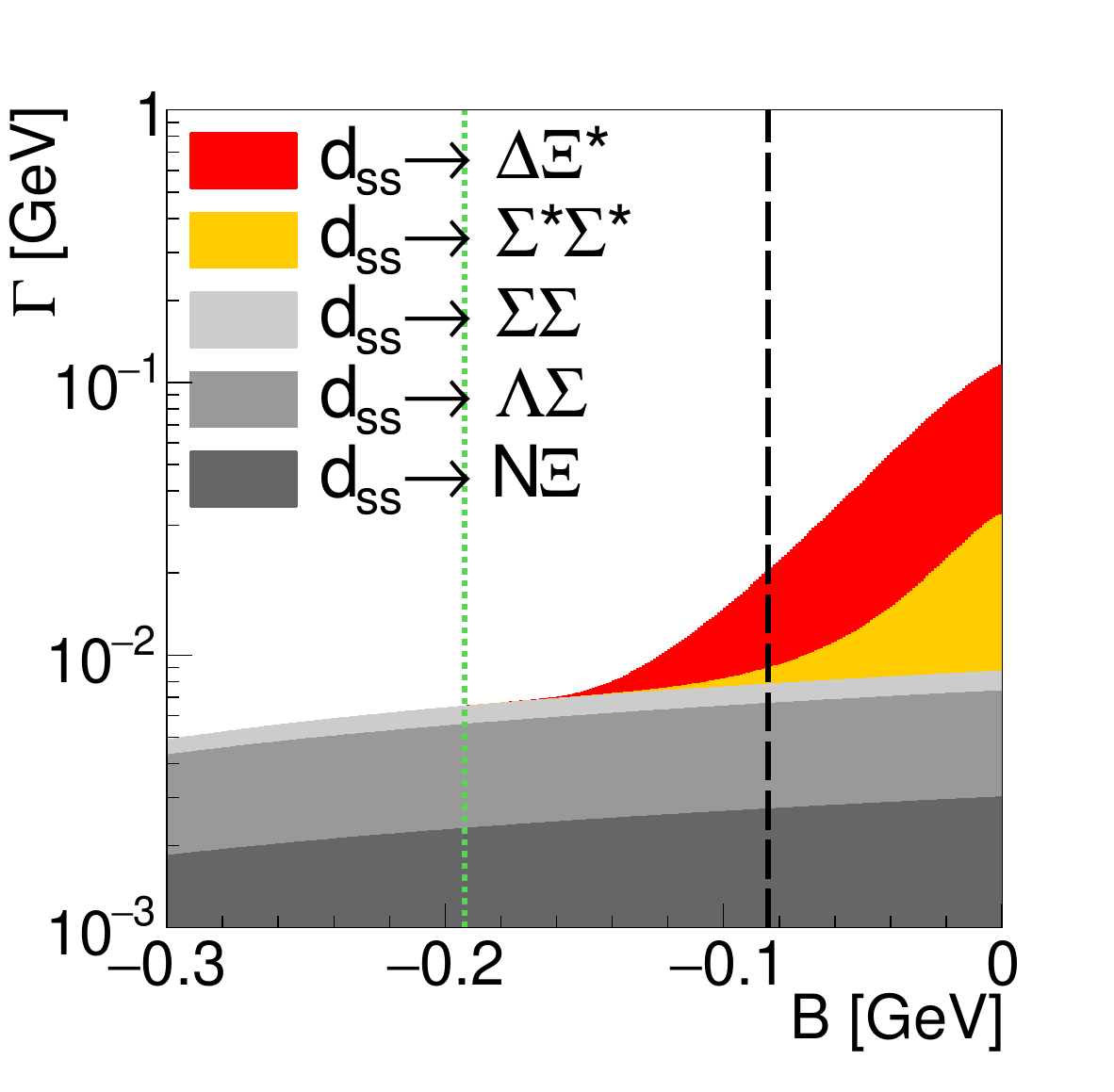}        \includegraphics[width=0.24\textwidth,angle=0]{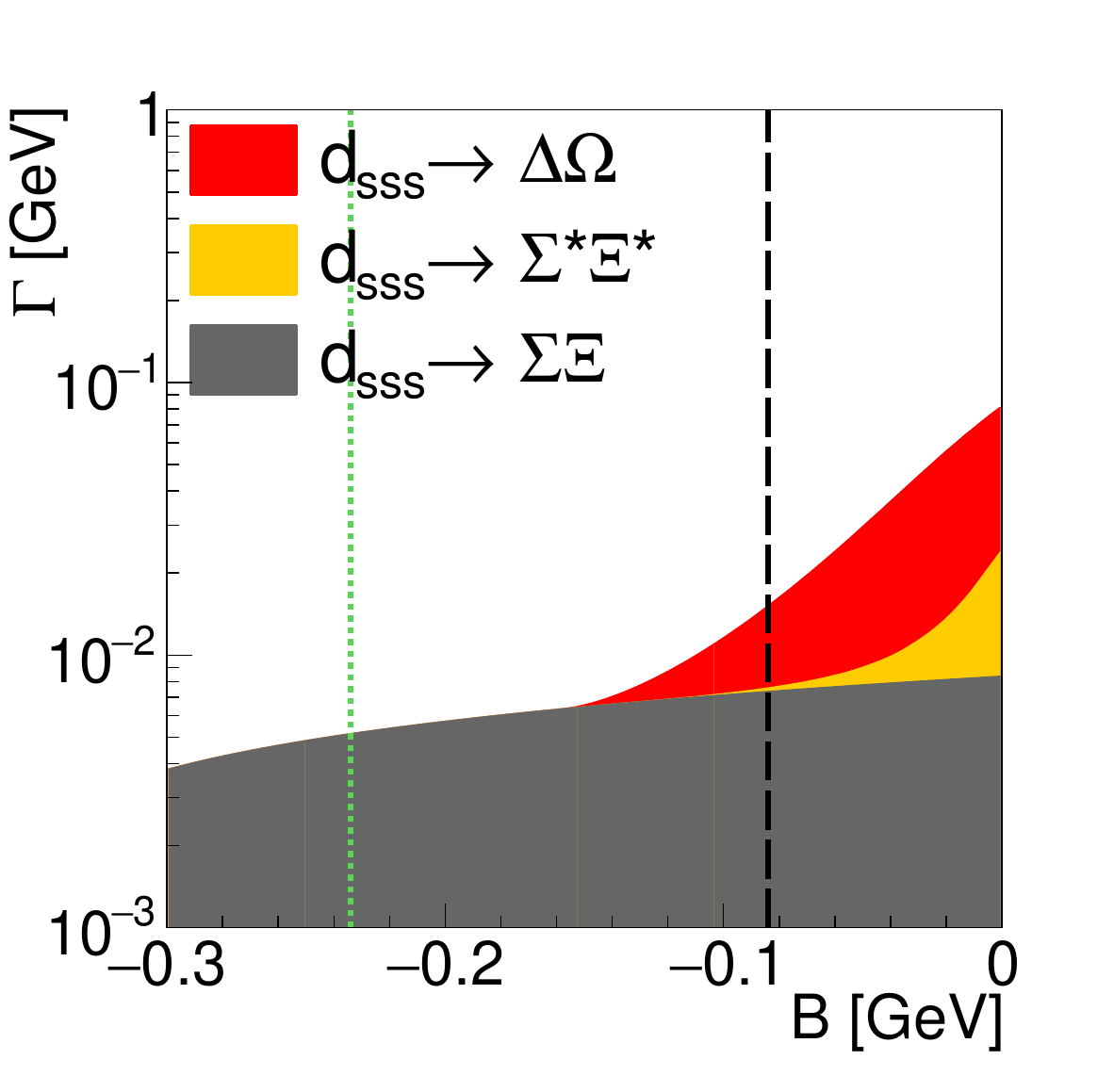}
\end{center}

\captionsetup{justification=centering,margin=0.5cm}
\caption{$d^*$ multiplet total width as a function of binding energy (relative to the lightest member of the Decuplet-Decuplet pole) for the $d^*,d_{s},d_{ss},d_{sss}$ from left to right split into major decay branches (note the log scale). The vertical line (common to all figures) shows the nominal expected mass, obtained under the assumption of the same binding (84 MeV) for all multiplet members(black dashed) and for specific binding (green dotted) as specified in Table.~\ref{DMass1Tab}.}
\label{d_all_width}
\end{figure}

\begin{figure}[!h]
\begin{center}
        \includegraphics[width=0.24\textwidth,angle=0]{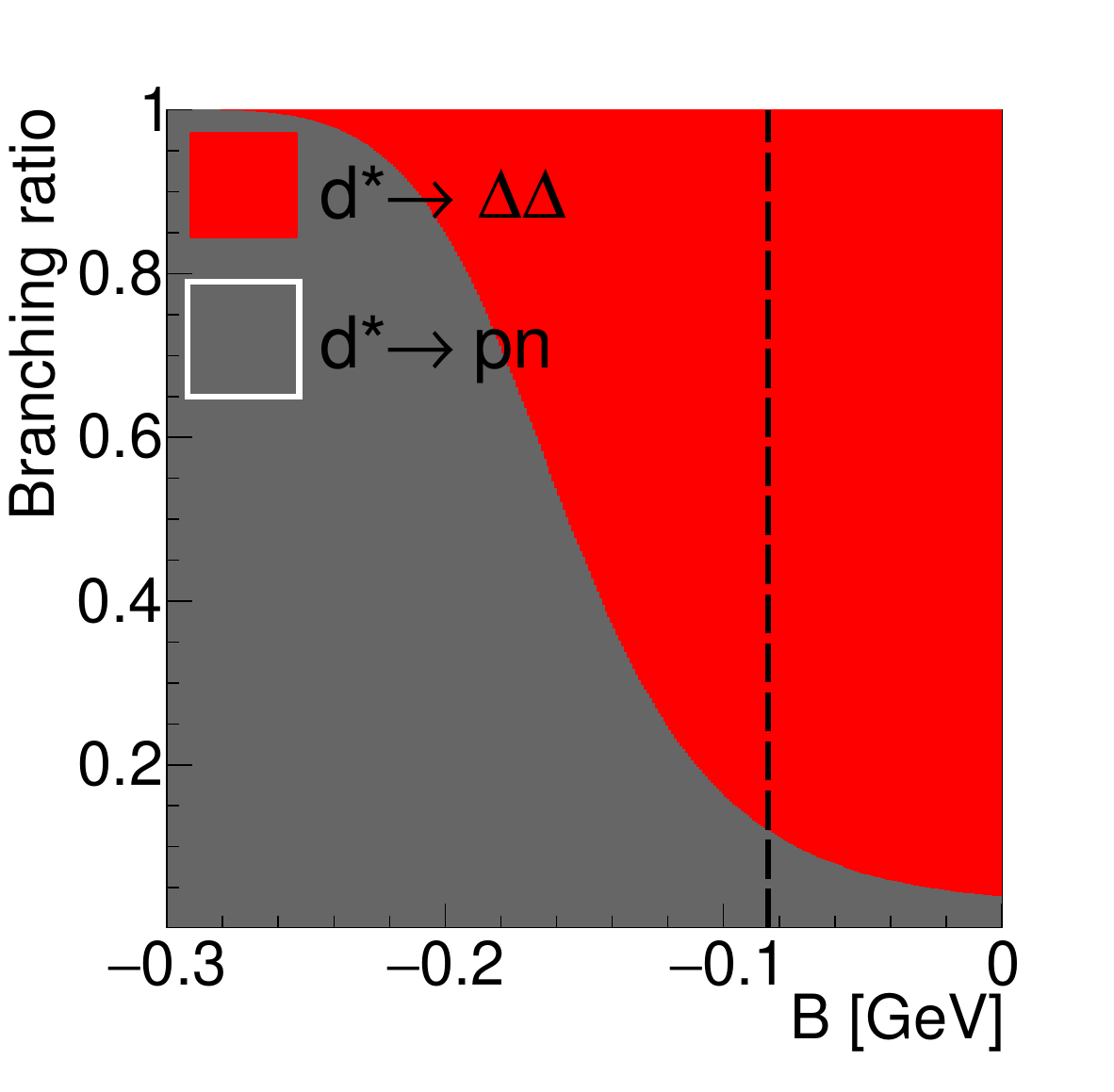}
        \includegraphics[width=0.24\textwidth,angle=0]{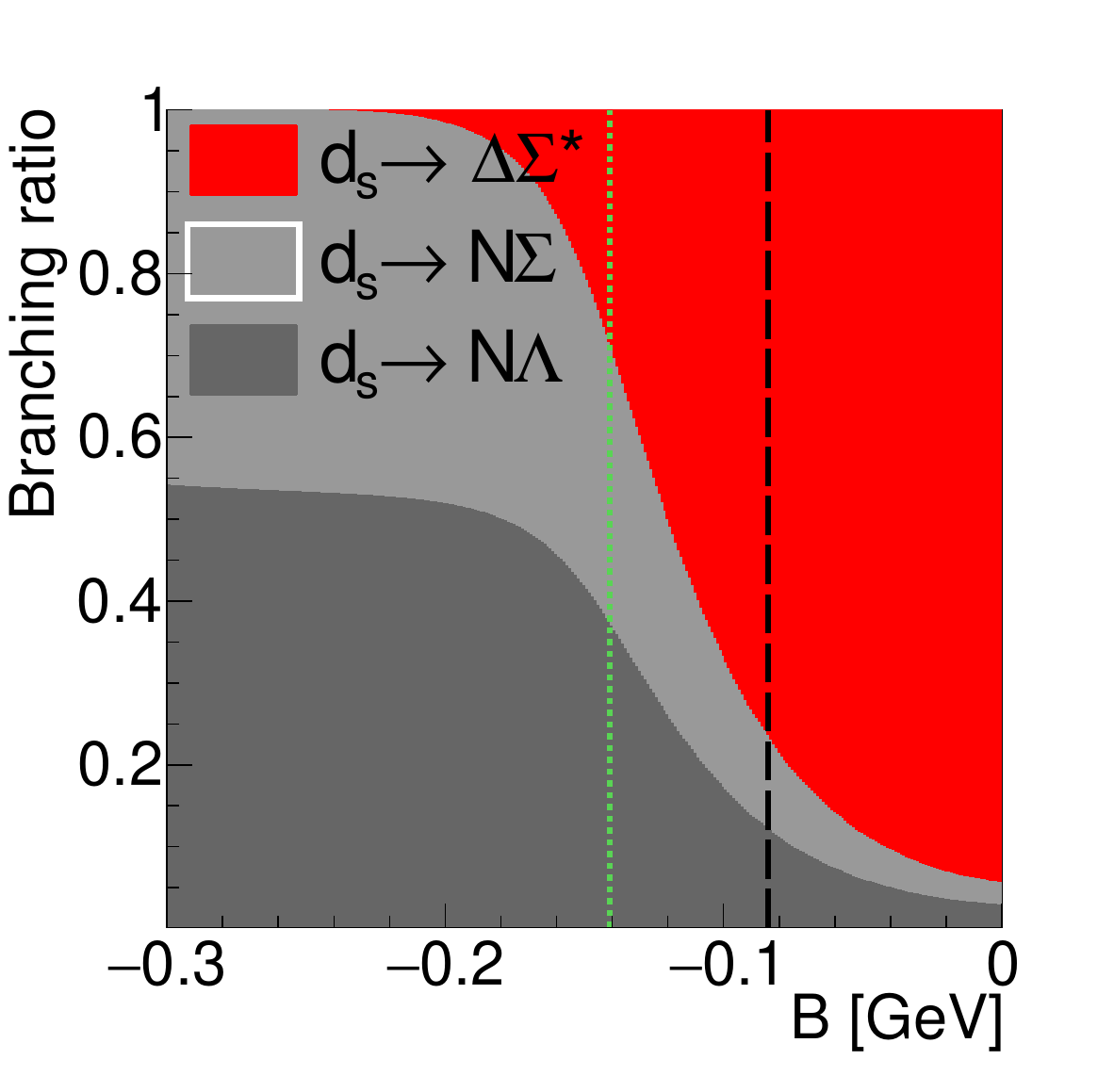}
        \includegraphics[width=0.24\textwidth,angle=0]{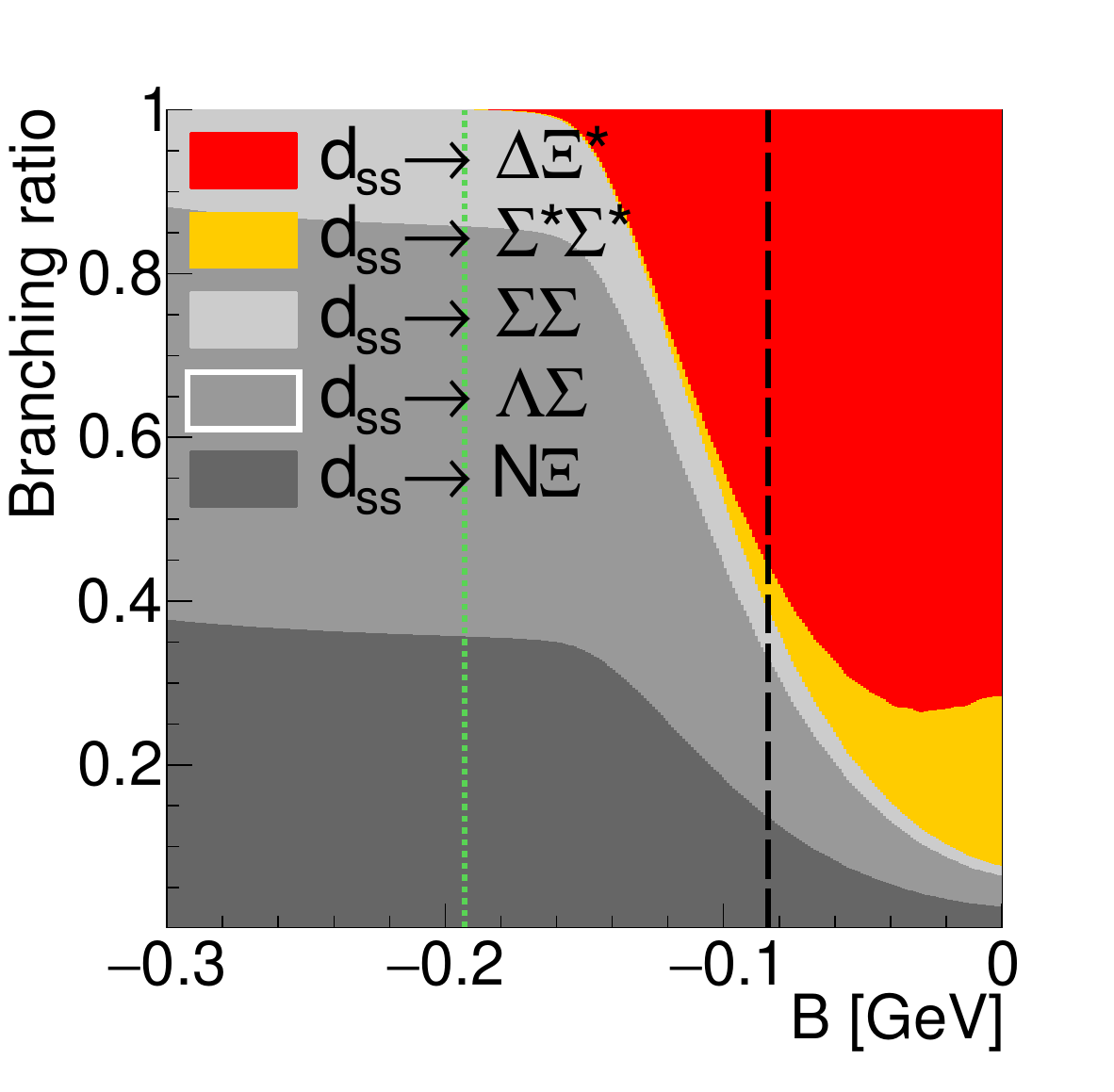}        
        \includegraphics[width=0.24\textwidth,angle=0]{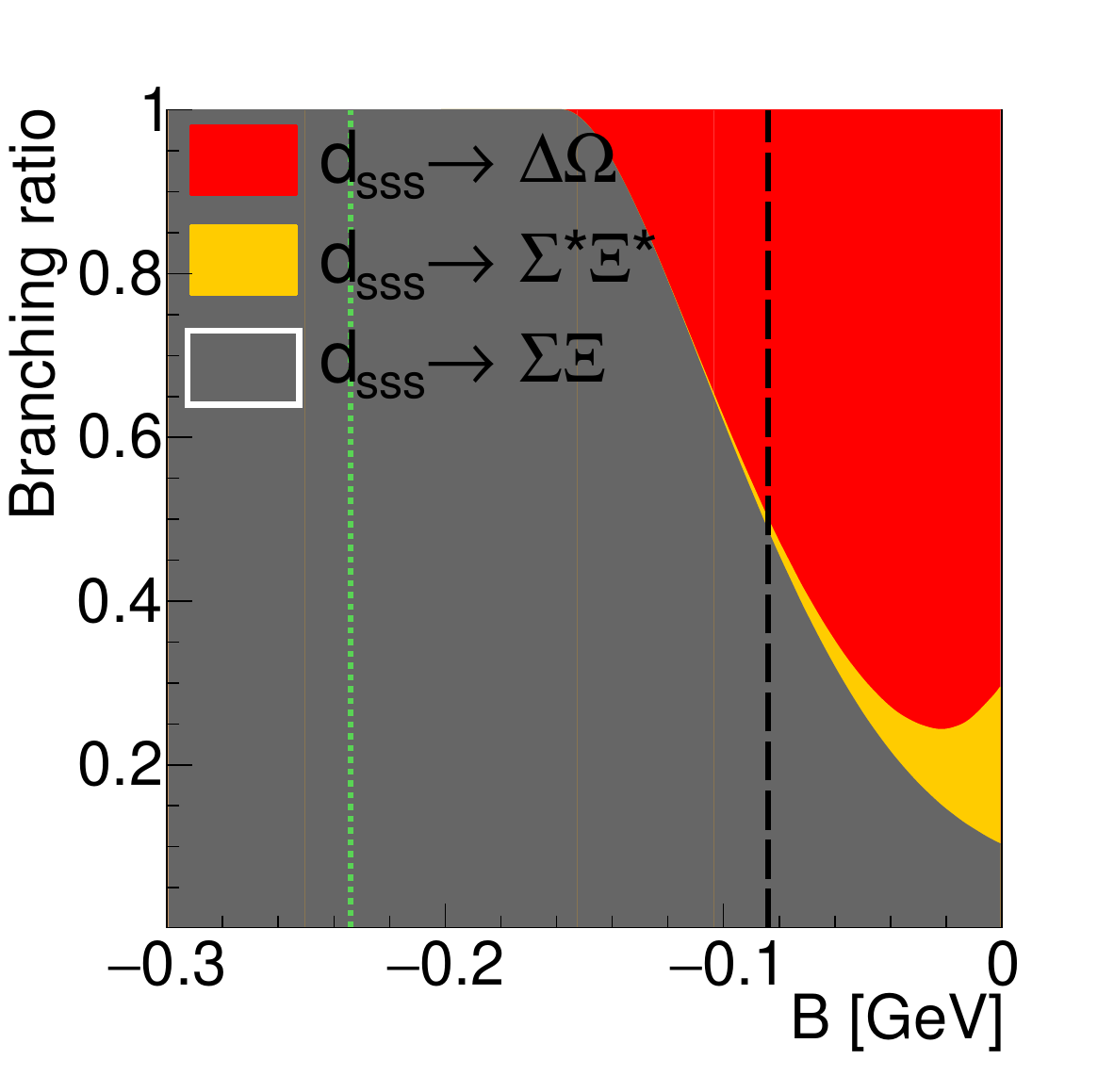}
\end{center}

\captionsetup{justification=centering,margin=0.2cm}
\caption{$d^*$ multiplet branching ratio as a function of binding energy (relative to Decuplet-Decuplet pole) for the $d^*,d_{s},d_{ss},d_{sss}$ from left to right. The vertical dashed line (common to all figures) shows the nominal expected mass, obtained under the assumption of the same binding (84 MeV) for all multiplet members(black dashed) and for specific binding (green dotted) as specified in Table.~\ref{DMass1Tab}.}
\label{d_all_Br}
\end{figure}

We, therefore, adopt $\Lambda =0.16$~GeV/c for subsequent calculations. The predicted width for all decuplet members is shown in Fig.~\ref{d_all_width}. In Fig.~\ref{d_all_Br} we show the predicted branching ratios as a function of binding energy. The results are summarised in Tables.~\ref{Tab_width_d84},~\ref{Tab_width}.

One can clearly see that the $8\oplus 8$ decays are predicted to be increasingly  important for the higher strangeness states, while both partial and total widths reduce substantially. For all hexaquark members only $10\oplus 10$ with $\Delta$ in a final state are important. For the $d_{ss}$ non-$\Delta$ channels ($\Sigma^*\Sigma^*\to$~anything) covers only 6\%. For the $d_{sss}$ the non-$\Delta$ ($\Xi^*\Sigma^*$) decay has a 2\% branching ratio. It is interesting to note that the semi-electromagnetic decay branch $d_{sss}\to \Omega\Delta\to \Omega \textrm{N}\gamma$ is predicted to have a higher probability than the purely hadronic $d_{sss}\to \Xi^*\Sigma^* \to \Xi\Sigma\pi\pi$ decay. 
\begin{table}[ht]
\centering \protect\caption{$d^*$ multiplet width results for the $B= -84$~MeV. }
\vspace{2mm}
{%
\resizebox{\textwidth}{!}{
\begin{tabular}{lrr|lrr|lrr|lrr}
\hline
\multicolumn{3}{|c|}{$d^*$(2380)} & \multicolumn{3}{|c|}{$d_s$(2531)} & \multicolumn{3}{|c|}{$d_{ss}$(2670)} & \multicolumn{3}{|c|}{$d_{sss}$(2820)} \\
\hline

decay & $\Gamma$, [MeV]  & BR [\%] & decay & $\Gamma$, [MeV]  & BR [\%] & decay & $\Gamma$, [MeV]  & BR [\%] & decay & $\Gamma$, [MeV]  & BR [\%]\\
$\Delta\Delta$ & 62.3 & 88 & $\Delta\Sigma^*$ & 26.9 & 77 & $\Delta\Xi^*$ & 11.3 & 56 & $\Delta\Omega$ & 7.6 & 50 \\
 &  &  &   &  &  & $\Sigma^*\Sigma^*$  & 1.1 & 6 & $\Sigma^*\Xi^*$ & 0.2 & 2 \\
 \hline
$pn$ & 8.4 & 12 &  $\textrm{N}\Lambda$ & 4.2 & 12 & $\textrm{N}\Xi$ & 2.7 & 13 & $\Sigma\Xi$ & 7.4 & 48 \\
  &  &  & $\textrm{N}\Sigma$ & 3.9 & 11 & $\Lambda\Sigma$ & 3.9 & 19 &  &  &  \\
 
&  &  &  & & & $\Sigma\Sigma$ & 1.2 & 6 & & &  \\
\hline
\hline
total & {\bf 70.7} &  & total  & {\bf 35.0} & & total & {\bf 20.2} & & total &  {\bf 15.2} &  \\
\hline
\end{tabular}} \label{Tab_width_d84}
}
\end{table}

Hexaquarks are expected to be produced copiously in heavy ion collisions, however, our estimations indicate the width for all $d^*$ multiplet states is rather large. Unfortunately, this indicates their clean identification in the tough background conditions in typical heavy ion collisions may be challenging. The most feasible channel for such studies, having both a large partial width and a convenient isospin $3/2$ separation, is the $\Xi\Sigma$ branch in $d_{sss}$ decay. This could potentially be tested using $\Xi-\Sigma$ correlation functions.
\begin{table}[ht]
\centering \protect\caption{$d^*$ multiplet width results for the expected binding energies from Table.~\Ref{DMass1Tab}. }
\vspace{2mm}
{%
\resizebox{\textwidth}{!}{
\begin{tabular}{lrr|lrr|lrr|lrr}
\hline
\multicolumn{3}{|c|}{$d^*$(2380), B=84~MeV} & \multicolumn{3}{|c|}{$d_s$(2474), B=141~MeV} & \multicolumn{3}{|c|}{$d_{ss}$(2571), B=193~MeV} & \multicolumn{3}{|c|}{$d_{sss}$(2670), B=234~MeV} \\
\hline

decay & $\Gamma$, [MeV]  & BR [\%] & decay & $\Gamma$, [MeV]  & BR [\%] & decay & $\Gamma$, [MeV]  & BR [\%] & decay & $\Gamma$, [MeV]  & BR [\%]\\
$\Delta\Delta$ & 62.3 & 88 & $\Delta\Sigma^*$ & 3.1 & 29 & $\Delta\Xi^*$ & $<0.1$ & $<1$ & $\Delta\Omega$ & 0 & 0 \\
 &  & &   &  &  & $\Sigma^*\Sigma^*$  & $<0.1$ & $<1$ & $\Sigma^*\Xi^*$ & 0 & 0 \\
 \hline
$pn$ & 8.4 & 12 &  $\textrm{N}\Lambda$ & 3.9 & 37 & $\textrm{N}\Xi$ & 2.3 & 36 & $\Sigma\Xi$ & 5.1 & 100 \\
  &  &  & $\textrm{N}\Sigma$ & 3.6 & 34 & $\Lambda\Sigma$ & 3.3 & 50 &  &  &  \\
  &  &  &  &  & 
  & $\Sigma\Sigma$ & 0.9 & 14 &  &  &  \\
  
\hline
\hline
total & {\bf 70.7} &  & total  & {\bf10.6 } & & total & {\bf 6.5} & & total &  {\bf 5.1} &  \\
\hline
\end{tabular}} \label{Tab_width}
}
\end{table}


For all channels the most prominent final states have high particle multiplicity, indicating hermetic detector apparatus and exclusivity conditions would need to be established. The $d_{sss}$ is the only member with a dominant octet-octet decay, however, the necessity of associated kaon production to conserve strangeness in the production mechanism also provides challenges. The requirements may be more easily achieved using beams containing intrinsic strangeness (e.g. Kaon beams as proposed in Ref.~\cite{KLF}).

\subsubsection{Results for d*(238) decuplet in a molecular picture}

\begin{figure}[!h]
\begin{center}
        \includegraphics[width=0.24\textwidth,angle=0]{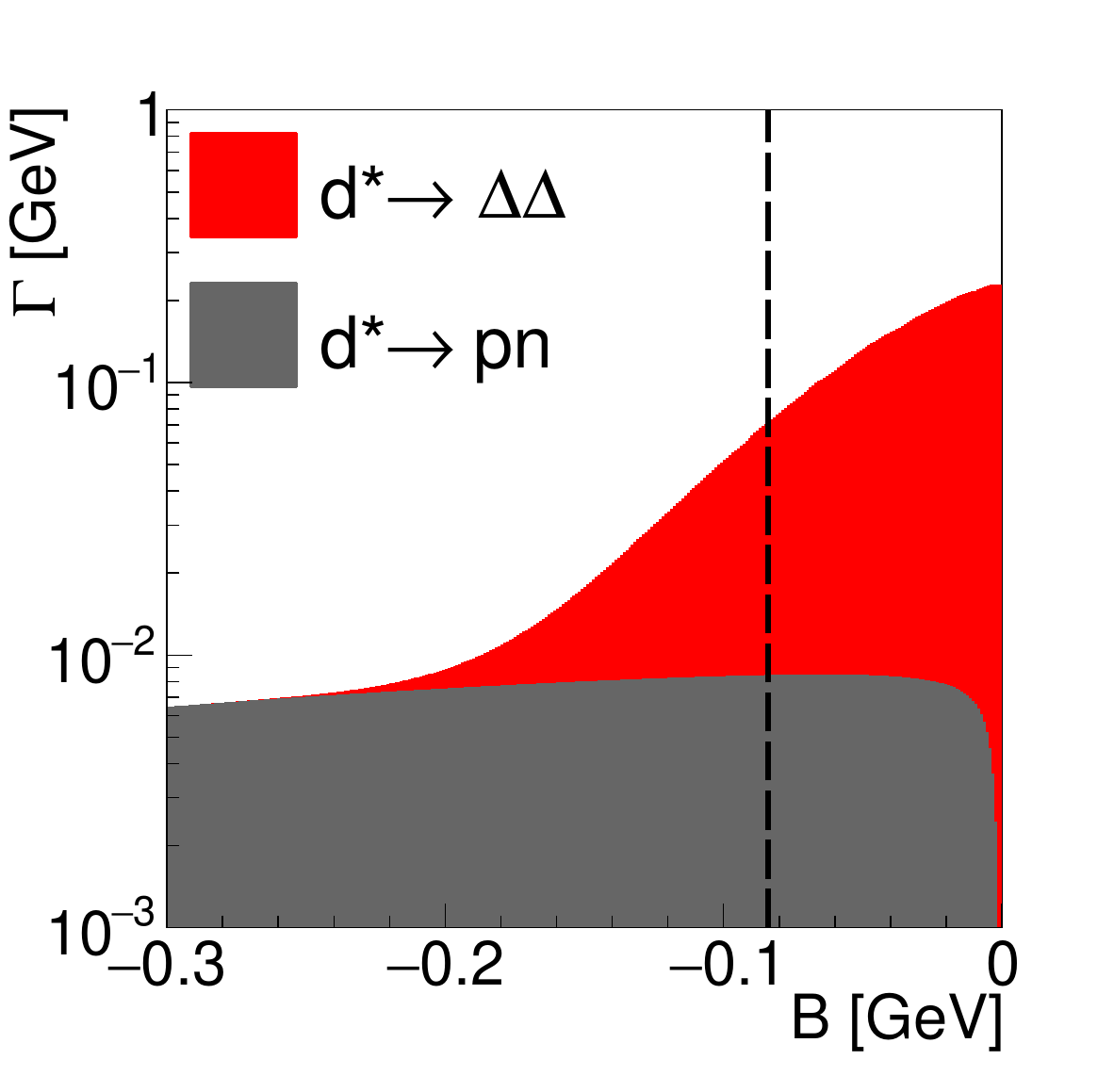}
        \includegraphics[width=0.24\textwidth,angle=0]{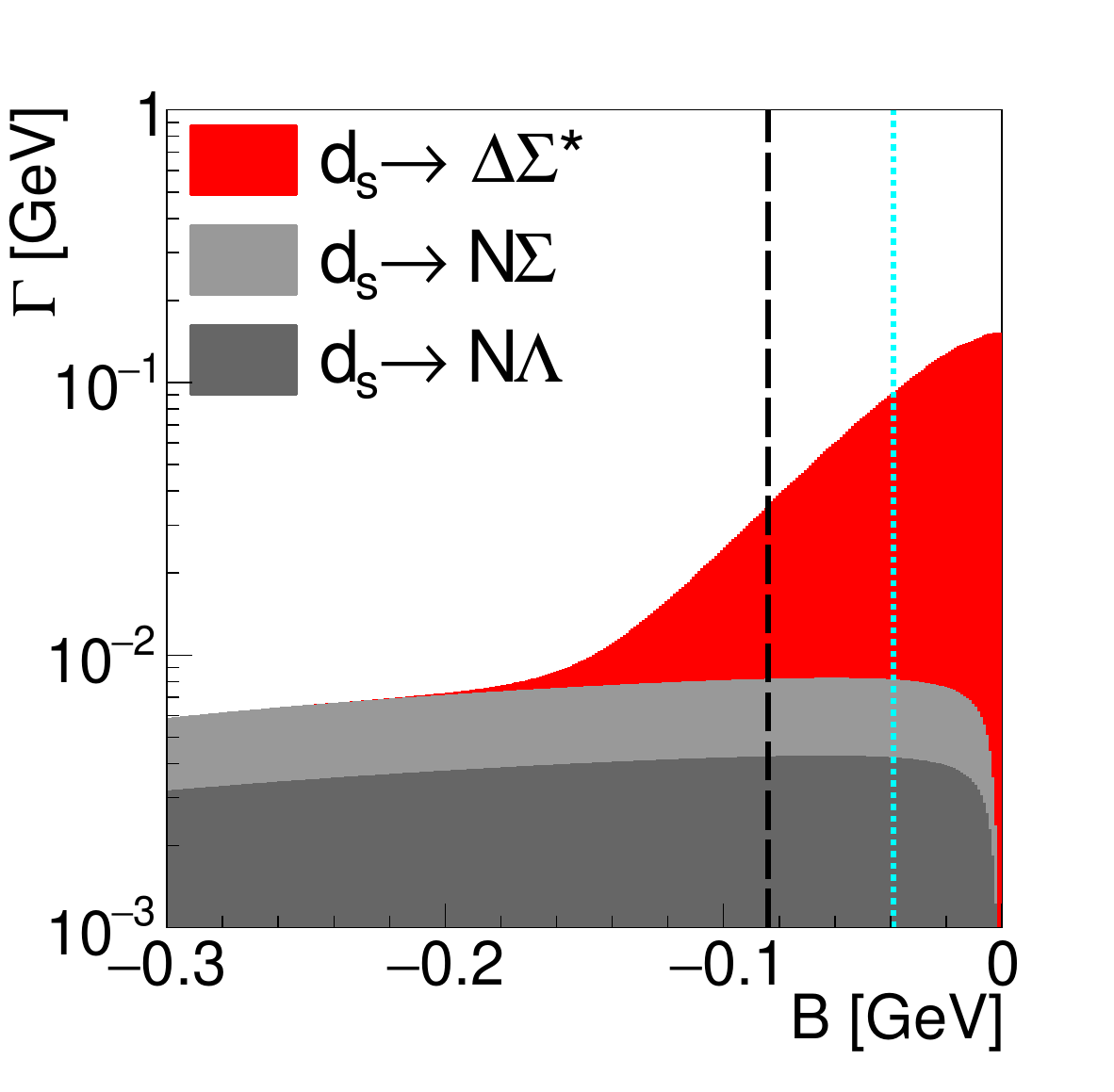}
        \includegraphics[width=0.24\textwidth,angle=0]{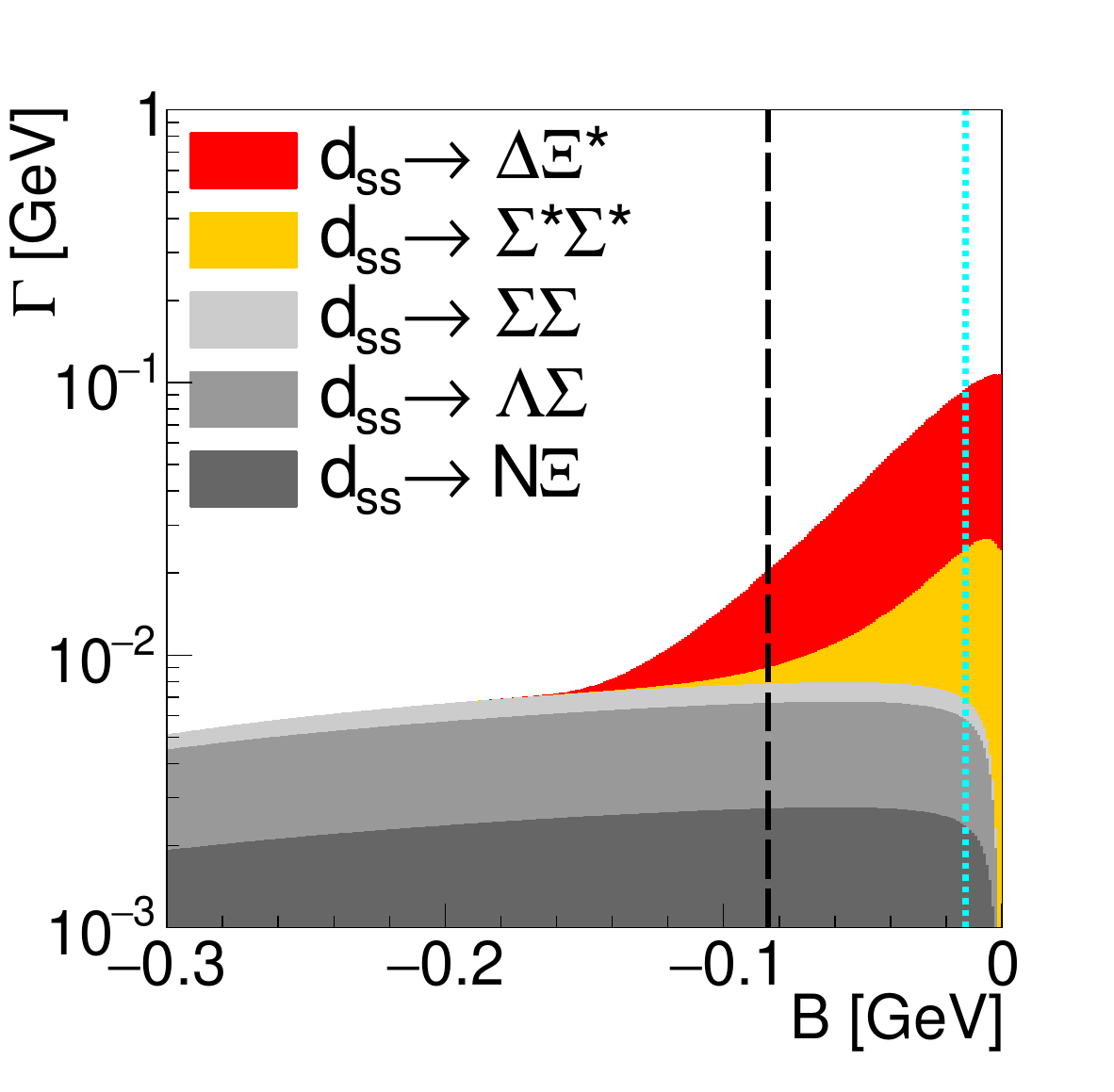}        \includegraphics[width=0.24\textwidth,angle=0]{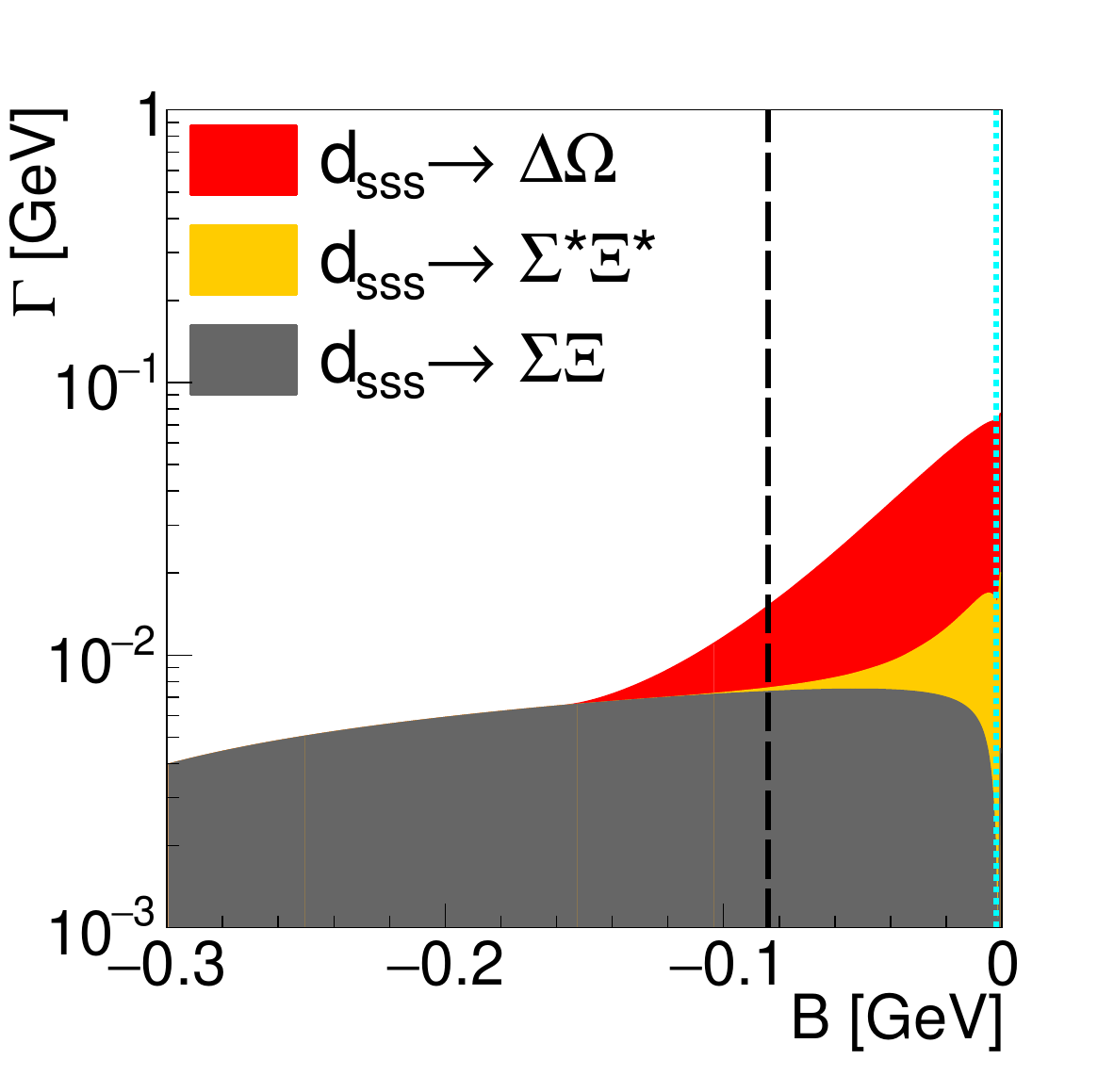}
\end{center}

\captionsetup{justification=centering,margin=0.4cm}
\caption{Same as Fig.~\ref{d_all_width}, but for a molecular picture with nominal molecular masses shown by the cyan dotted line as in Table.~\ref{DMass2Tab}.}
\label{d_all_width_M}
\end{figure}

\begin{figure}[!h]
\begin{center}
        \includegraphics[width=0.24\textwidth,angle=0]{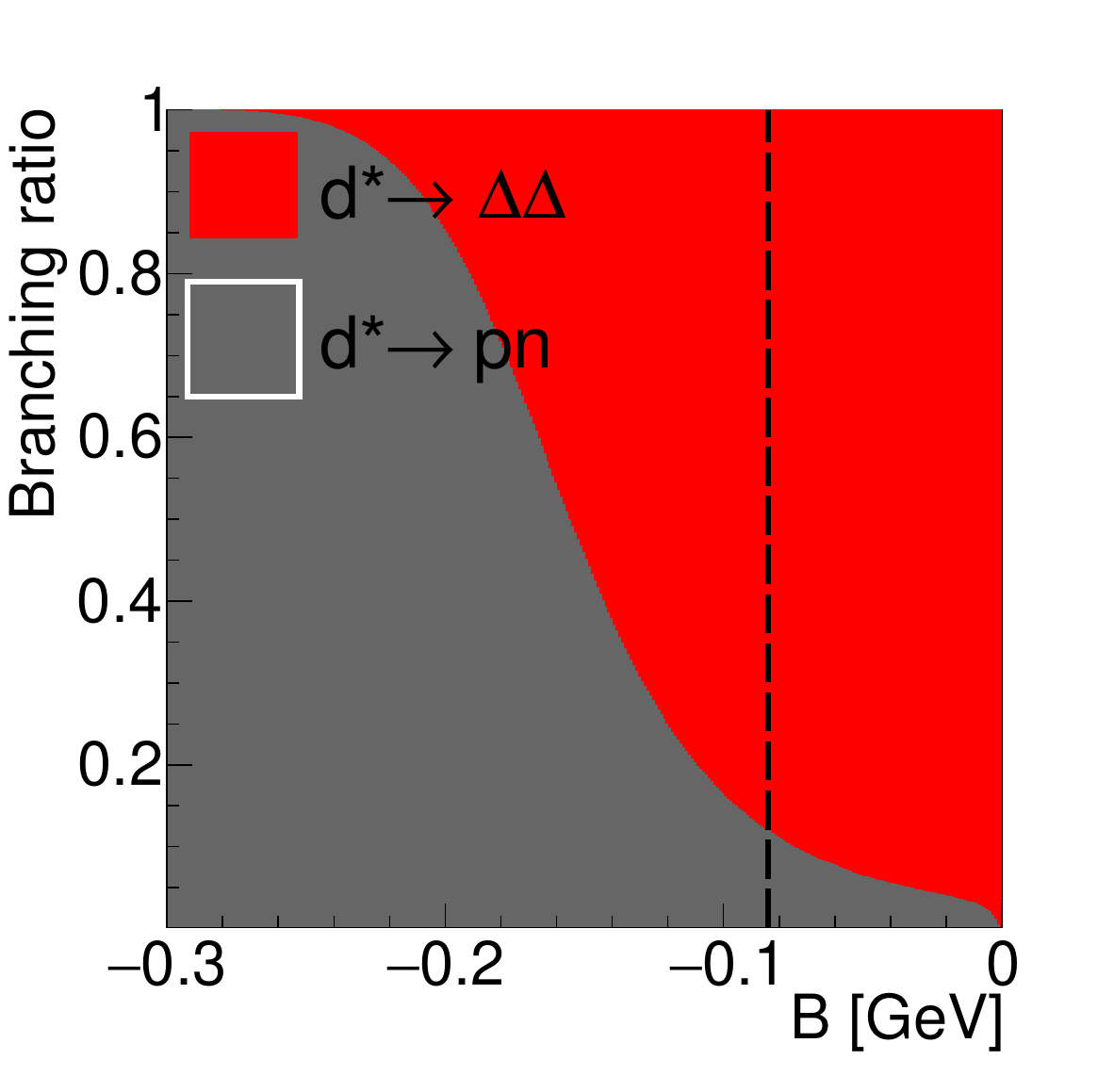}
        \includegraphics[width=0.24\textwidth,angle=0]{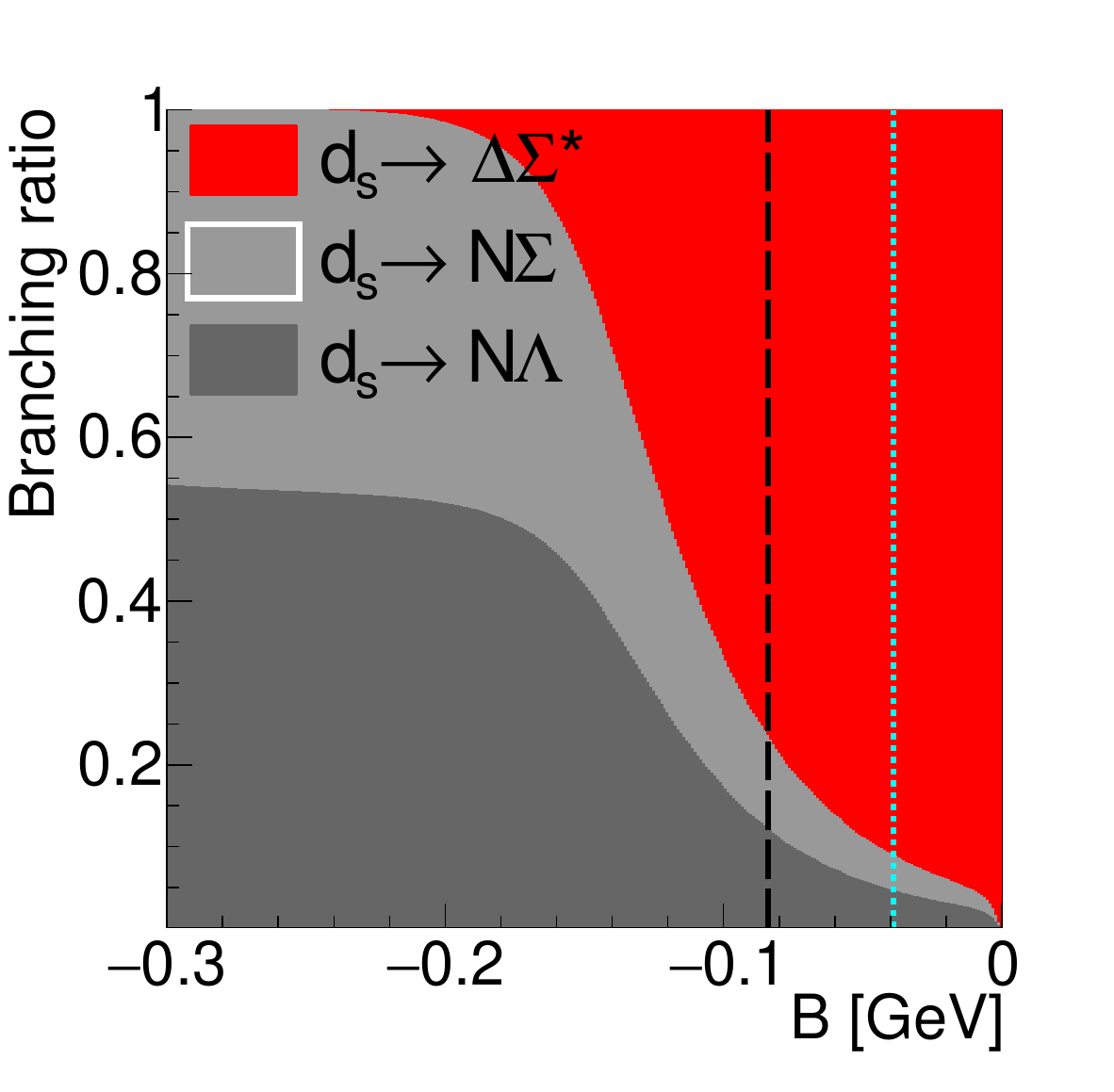}
        \includegraphics[width=0.24\textwidth,angle=0]{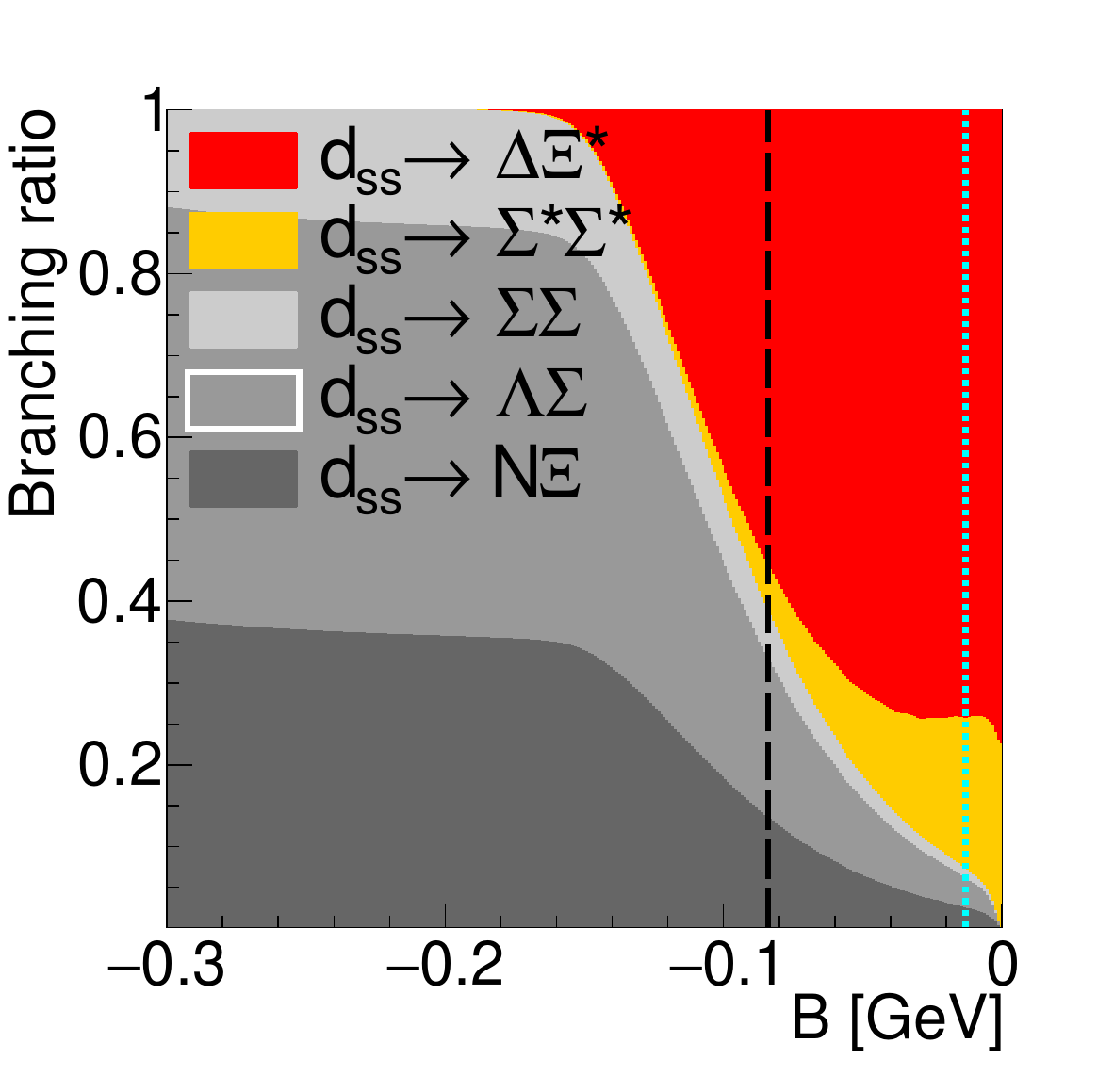}        \includegraphics[width=0.24\textwidth,angle=0]{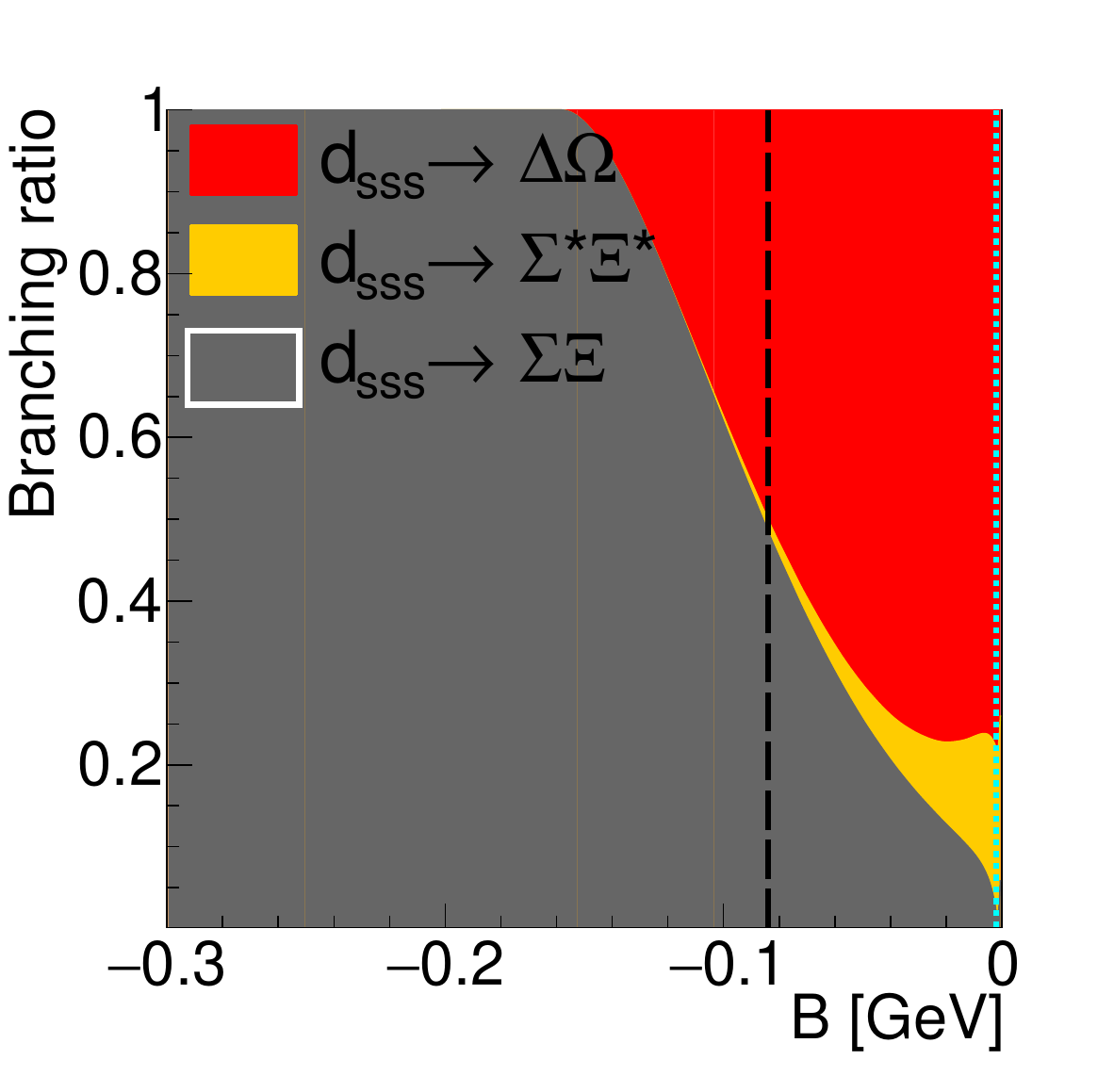}
\end{center}

\captionsetup{justification=centering,margin=0.4cm}
\caption{same as Fig.~\ref{d_all_Br}, but for a molecular picture with nominal molecular masses shown by cyan dotted line as in Table.~\ref{DMass2Tab}.}
\label{d_all_Br_M}
\end{figure}

The strong variation of the $8\oplus8$ decay in the molecular picture leads to very different behaviour in total width and branching ratios, see Table.~\ref{Tab_width_M}, Fig.~\ref{d_all_width_M},~\ref{d_all_Br_M}. One also should keep in mind that for this case we generally would expect much smaller binding energies for all strange members than $B=-84~MeV$ of a $d^*(2380)$. 
While at $B=-84~MeV$ binding the decay branchings are nearly identical for the two scenarios,  in a purely molecular picture with reduced binding the importance of the $8\oplus8$ decays is also reduced. For the $d_{sss}$ case with a tiny ($B=2$~MeV) binding the $8\oplus8$, $\Sigma\Xi$ decay is $\Gamma(\Sigma\Xi)=1.5$~MeV, or only 2\% of the total decay width. It is also interesting to note that in a molecular picture, the state with zero strangeness, $d^*(2380)$ should have the smallest width. One can see that even the stability of the $\Omega$ baryon for the case of $d_{sss}$ state cannot compensate the massive reduction of the binding energy, which leads to the prediction of a much larger width for  $d_{sss}$ compared to the $d^*(2380)$. 




 

\begin{table}[ht]

\captionsetup{justification=centering,margin=0.5cm}
\centering \protect\caption{$d^*$ multiplet width results for the expected binding energies from Table\Ref{DMass2Tab}. (molecule) }
\vspace{2mm}
{%
\resizebox{\textwidth}{!}{
\begin{tabular}{lrr|lrr|lrr|lrr}
\hline
\multicolumn{3}{|c|}{$d^*$(2380)} & \multicolumn{3}{|c|}{$d_s$(2576), B=39~MeV} & \multicolumn{3}{|c|}{$d_{ss}$(2751),B=13~MeV} & \multicolumn{3}{|c|}{$d_{sss}$(2902), B=2~MeV} \\
\hline

decay & $\Gamma$, [MeV]  & BR [\%] & decay & $\Gamma$, [MeV]  & BR [\%] & decay & $\Gamma$, [MeV]  & BR [\%] & decay & $\Gamma$, [MeV]  & BR [\%]\\
$\Delta\Delta$ & 62.3 & 88 & $\Delta\Sigma^*$ & 82.9 & 91 & $\Delta\Xi^*$ & 69.9 & 74 & $\Delta\Omega$ & 56.3 & 78 \\
 &  &  &   &  &  & $\Sigma^*\Sigma^*$  & 17.3 & 18 & $\Sigma^*\Xi^*$ & 14.4 & 20 \\
 \hline
$pn$ & 8.4 & 12 &  $\textrm{N}\Lambda$ & 4.2 & 5 & $\textrm{N}\Xi$ & 2.3 & 3 & $\Sigma\Xi$ & 1.5 & 2 \\
  &  &  & $\textrm{N}\Sigma$ & 3.9 & 4 & $\Lambda\Sigma$ & 3.3 & 4 &  &  &  \\
 
&  &  &  & & & $\Sigma\Sigma$ & 1.0 & 1 & & &  \\
\hline
\hline
total & {\bf 70.7} &  & total  & {\bf 91.0} & & total & {\bf 94.0} & & total &  {\bf 72.2} &  \\
\hline
\end{tabular}} \label{Tab_width_M}
}
\end{table}

\section{$\Delta \Delta$ 28-plet}\label{sec:DD}
\begin{figure}[!h]
\begin{center}
\includegraphics[angle=0,width=0.5\textwidth]{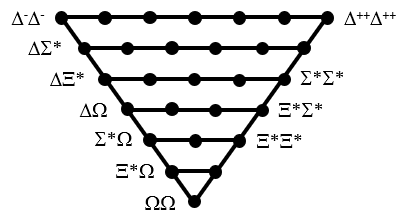}
\end{center}
\caption{$\Delta\Delta$ 28-plet $10\oplus10$}
\label{28_deldel_multiplet}
\end{figure}

The $\Delta\Delta$ 28-plet(Fig.~\ref{28_deldel_multiplet}) couples only to a $10\oplus10$, hence it is extremely difficult to access it even in a low strangeness state, see Table~\ref{Tab1}. Indeed, the zero strangeness state has isospin equal to 3. Due to isospin conservation, it does not couple to $\textrm{N}\textrm{N}$, hence cannot be directly accessed in nucleon-nucleon collisions. It does not couple to $\textrm{N}\textrm{N}\pi$, hence cannot be probed with pion beams on nuclei. (One can still use pion double-charge exchange reactions on nuclei, e.g. $\pi^+A\to\pi^- X$, but not with direct $S$-channel production).

\begin{table}[ht]
\footnotesize
\centering \protect\caption{Expected decay branches of the spin $J=0$  SU(3) 28-plet}
\vspace{2mm}
\scalebox{0.85}{%
\setlength{\tabcolsep}{10pt}
\renewcommand{\arraystretch}{1.5}
\begin{tabular}{|l|c|c|}
\hline
Strangeness  & $10\oplus10$ & Mass [MeV] \\
\hline
\ 0     & $\Delta\Delta$ & $\Delta\Delta(2464)$ \\
-1    & $\Delta\Sigma^* $ & $\Delta\Sigma^*(2615)$   \\
-2    & $\frac{1}{\sqrt{5}}(\sqrt{2}\Delta\Xi^* + \sqrt{3}\Sigma^*\Sigma^*)$ & $\Sigma^*\Sigma^*(2766)$ $\Delta\Xi^*(2764)$ \\
-3    & $\frac{1}{\sqrt{10}}(\Delta\Omega + 3\Sigma^*\Xi^*)$ & $\Sigma^*\Xi^*(2915)$ $\Delta\Omega(2904)$  \\
-4    & $\frac{1}{\sqrt{5}}(\sqrt{2}\Sigma^*\Omega + \sqrt{3}\Xi^* \Xi^*)$ & $\Xi^* \Xi^*(3064)$ $\Sigma^*\Omega(3055)$  \\
-5    & $\Xi^*\Omega$ & $\Xi^*\Omega(3204)$ \\
-6    & $\Omega\Omega$ & $\Omega\Omega(3344)$ \\
\hline
\end{tabular}} \label{Tab1}
\end{table}
\normalsize

The simplest way to produce even the lowest lying $\Delta\Delta$ state is a nucleon-nucleon collision with two associated pions, $pp\to (\pi\pi)(\Delta\Delta)$. A six-body final state and rather high energy required for this reaction make it extremely challenging to predict the background from conventional reactions, which can also potentially produce structures in the measured cross sections. As a result bump-hunting techniques become extremely unreliable. If the $\Delta\Delta$ 28-plet would be strongly bound ($\sim 80$~MeV or higher, as the $d^*$ antidecuplet) it members  would have an even smaller width than corresponding $d^*$ multiplet particles, since the $8\oplus8$ decay is not allowed here. Unfortunately, in the only measurement from Ref.~\cite{iso3} , where the authors tried to extract the maximum isospin projection state, $pp\to \pi^-\pi^-\Delta^{++}\Delta^{++}$, it was demonstrated that this multiplet is either loosely bound or loosely unbound. That unavoidably means that the widths of all states would be similar to the combined width of constituent particles, e.g. in $\Delta\Delta$ case it would be $\Gamma\sim 240$~MeV. Under such conditions, the only state which can be experimentally accessed is the strong-decay-free $\Omega\Omega$ dibaryon. There is a hope to get the $\Omega\Omega$ scattering length from heavy ion collisions. 
Lattice calculations also support the expectation of loosely bound 28-plet. For example, Ref.~\cite{Omega2LQCD} predict $\Omega\Omega$ state to be either bound by less than an MeV or unbound. 

 \begin{figure}[!h]
\begin{center}
        \includegraphics[width=0.3\textwidth,angle=0]{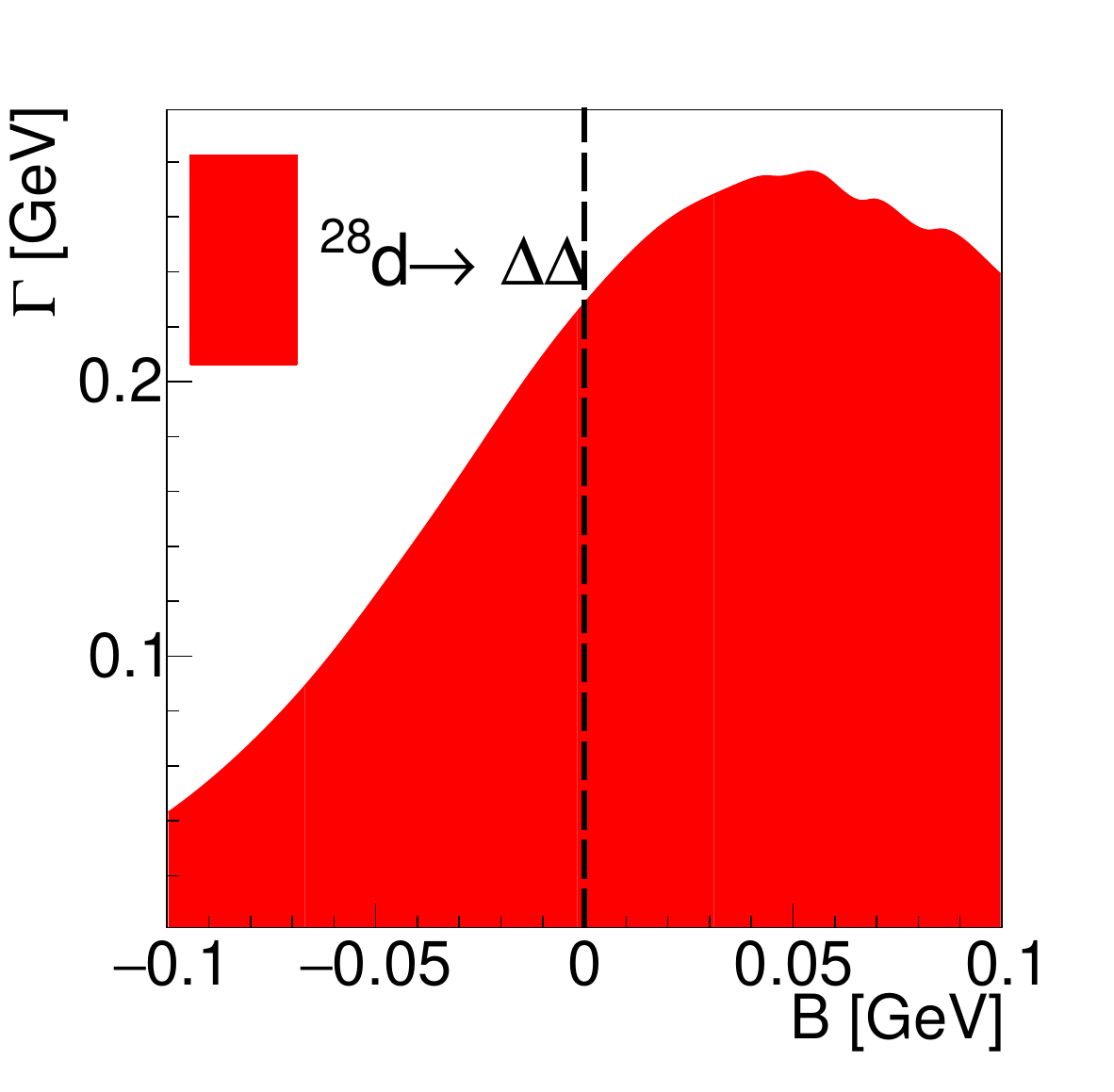}
        \includegraphics[width=0.3\textwidth,angle=0]{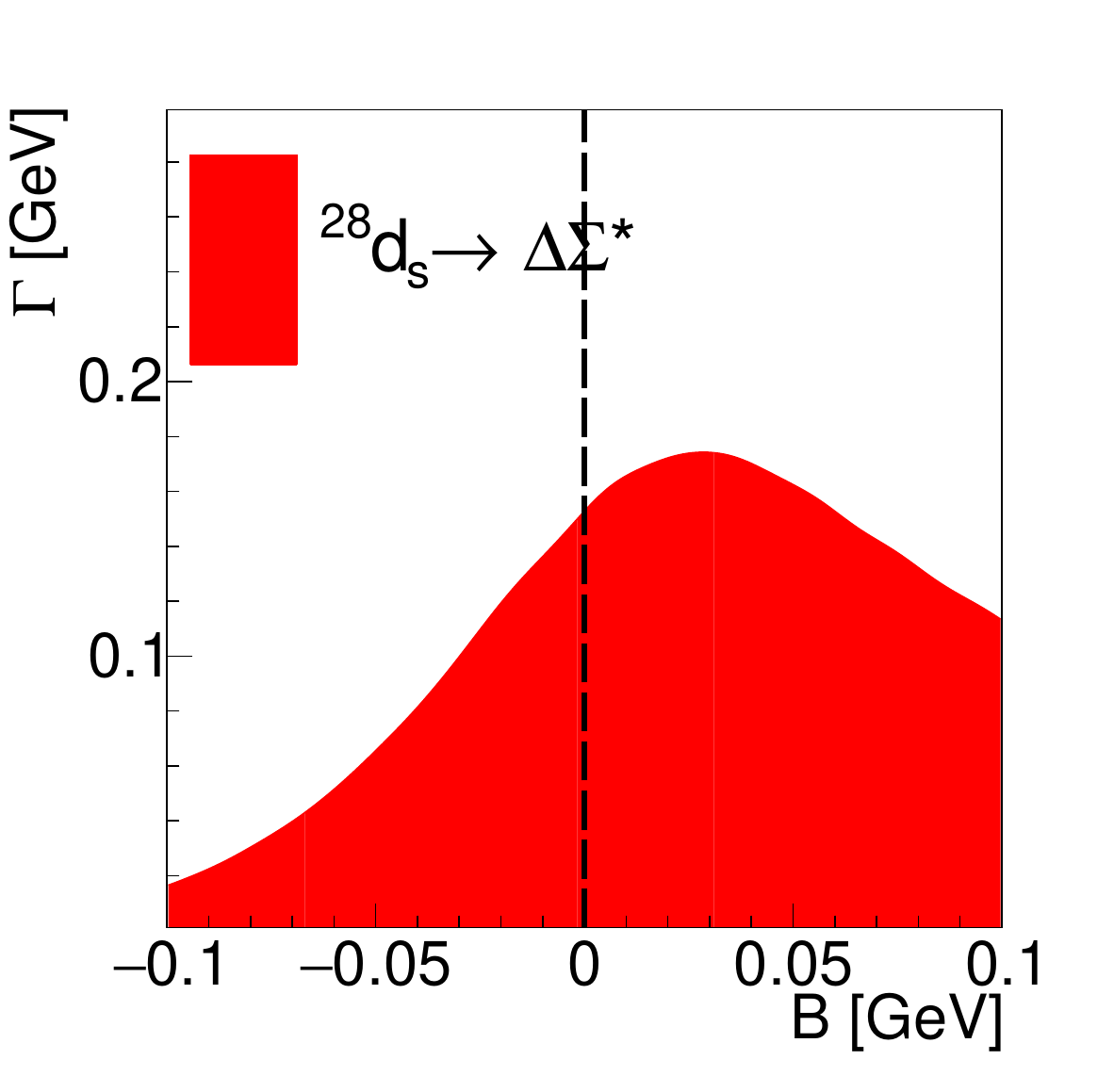}
        \includegraphics[width=0.3\textwidth,angle=0]{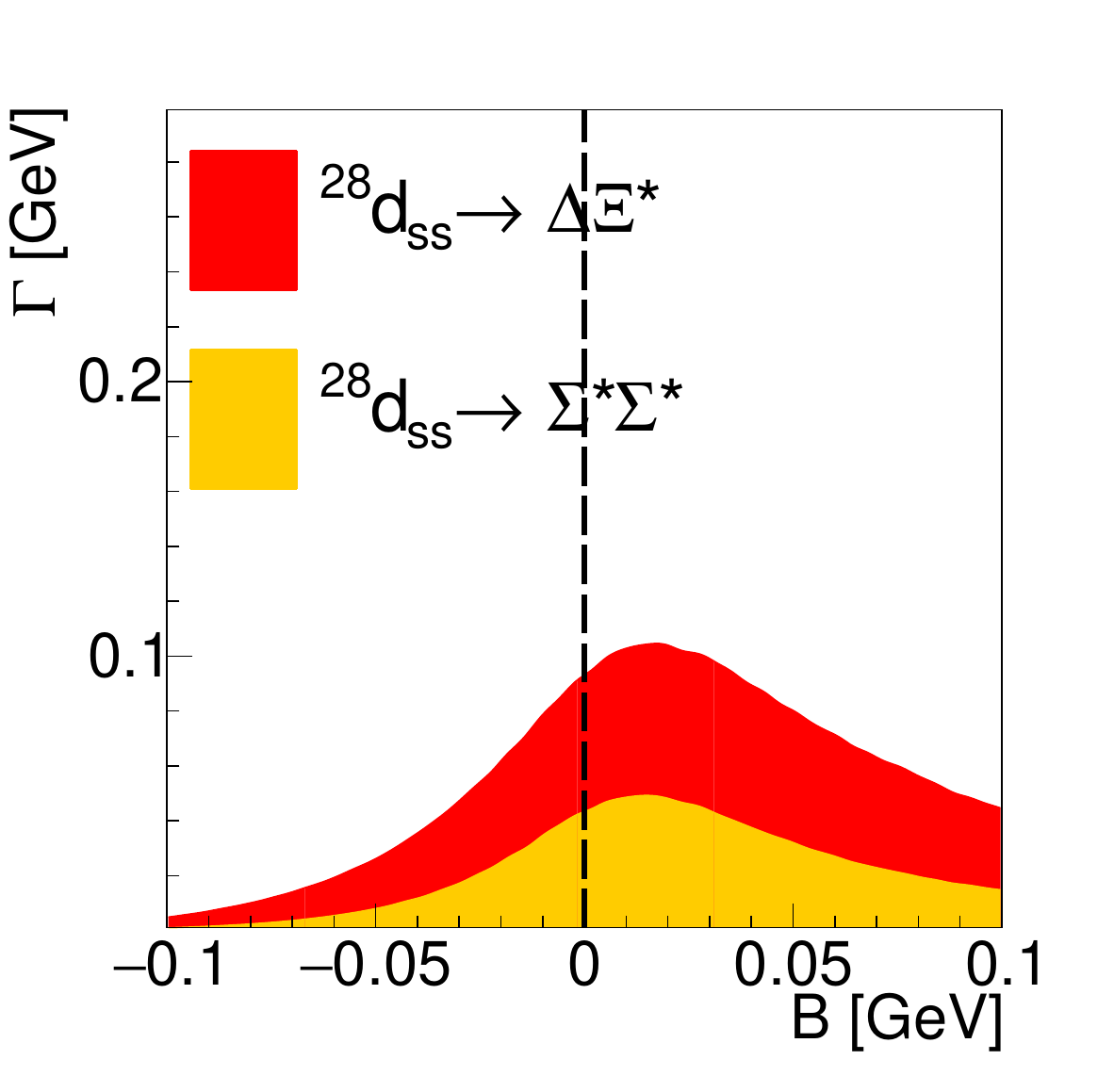}        \includegraphics[width=0.3\textwidth,angle=0]{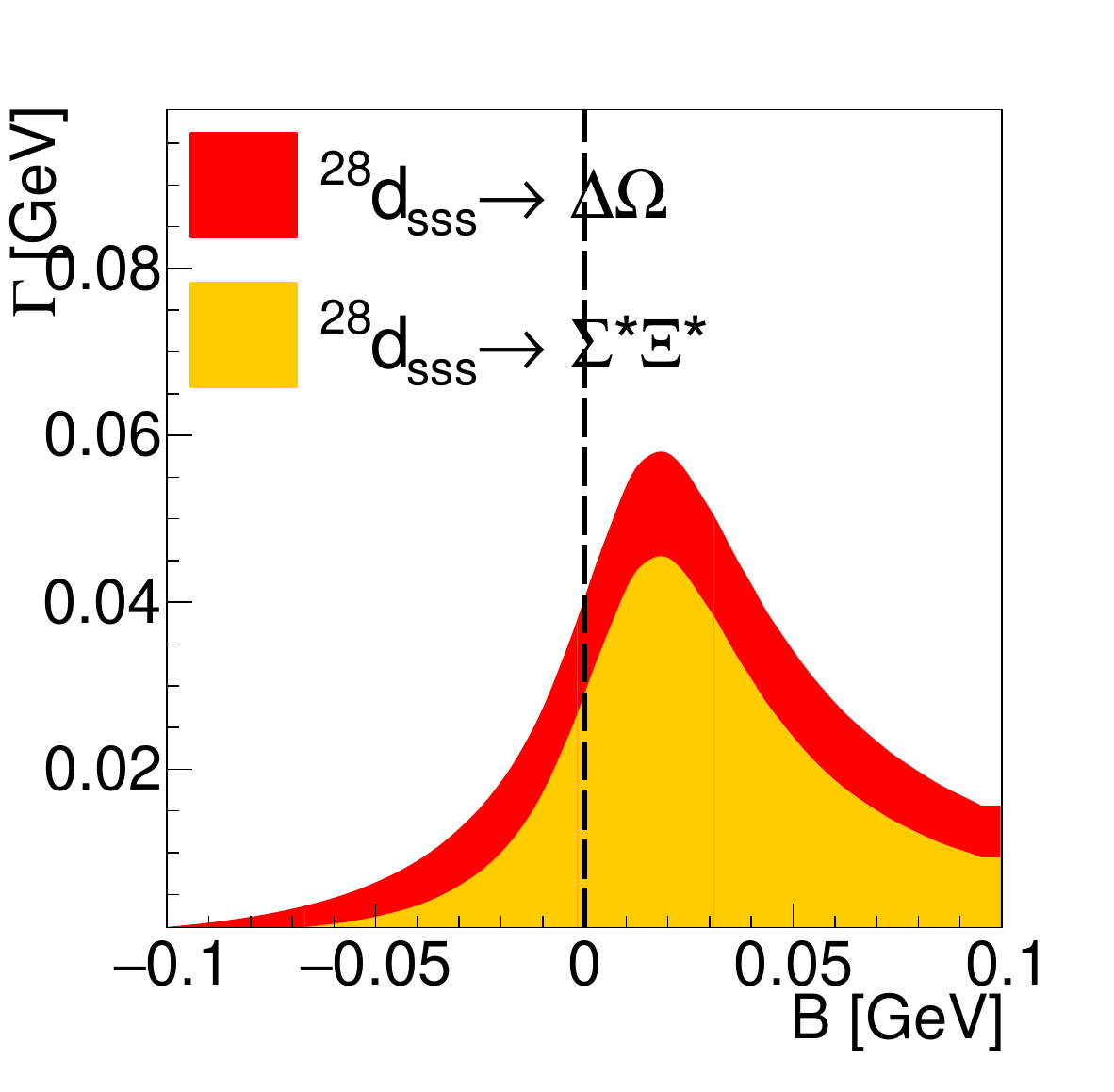}
        \includegraphics[width=0.3\textwidth,angle=0]{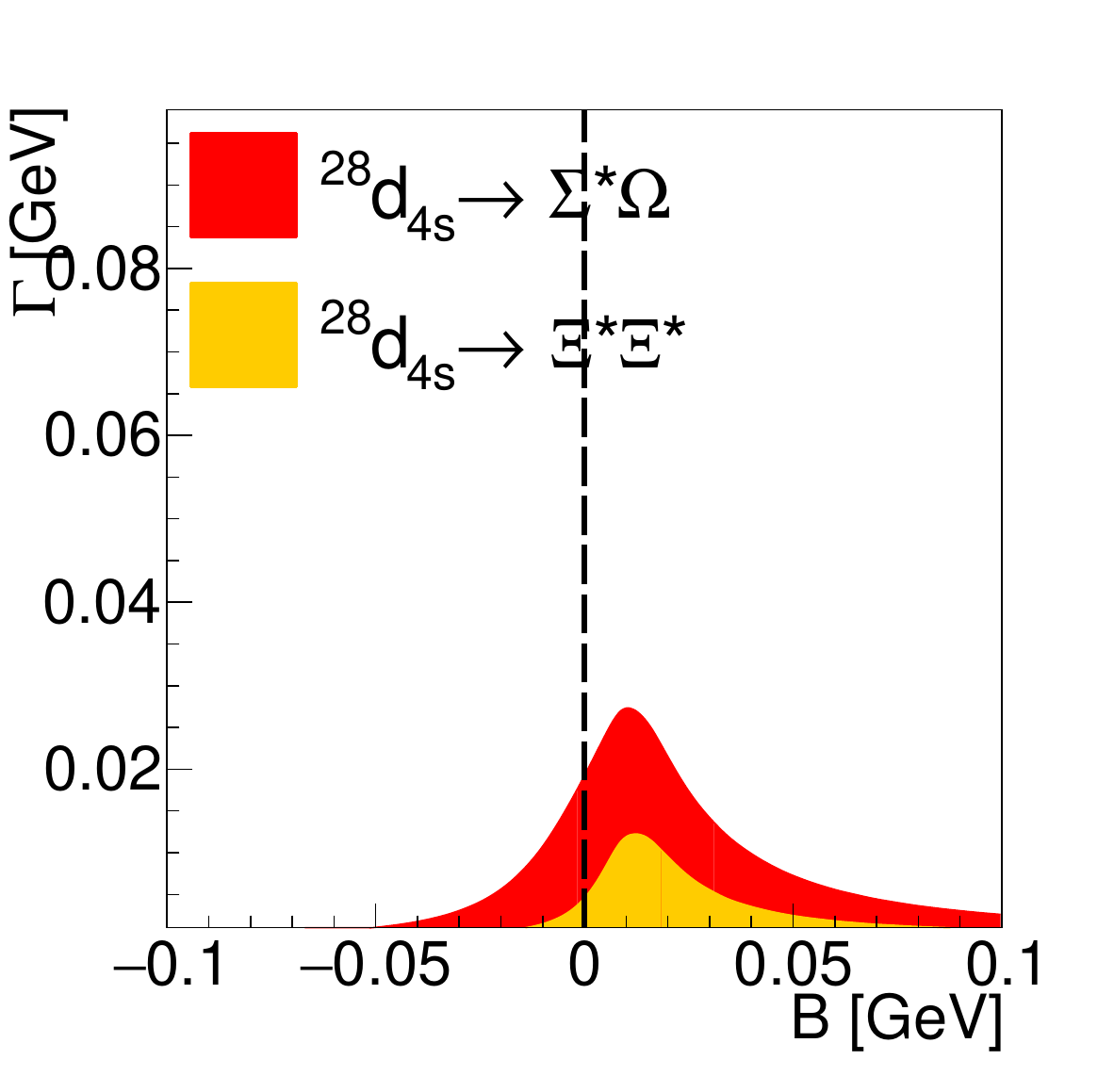}
        \includegraphics[width=0.3\textwidth,angle=0]{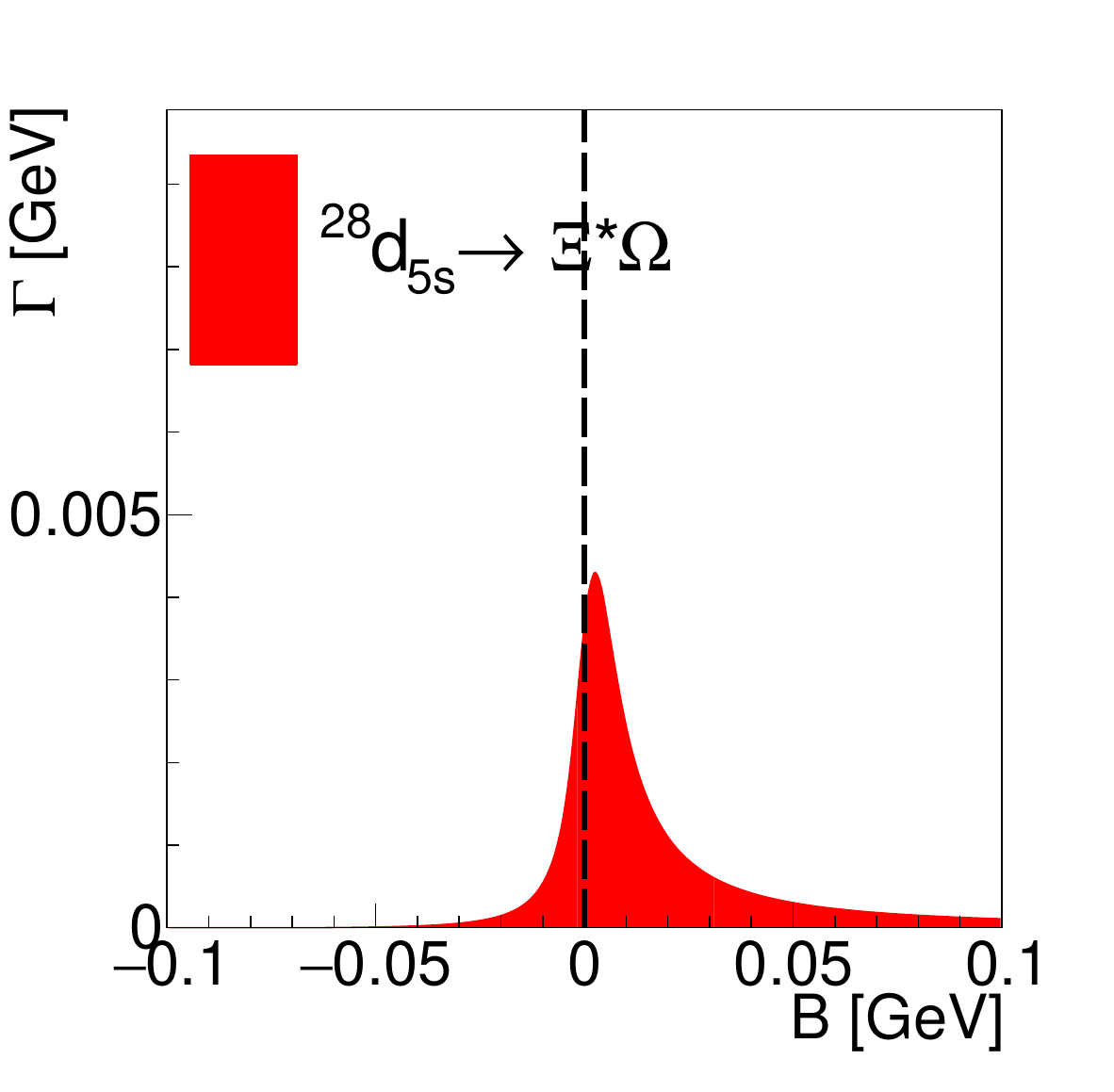}
\end{center}

\captionsetup{justification=centering,margin=0.2cm}
\caption{28-plet total width as a function of binding energy (relative to the Decuplet-Decuplet pole) for the various states with increased strangeness from top left ($S=0$) to bottom right ($S=-5$) split into major decay branches.}
\label{28_width}
\end{figure}

As a first approximation, the 28-plet decay widths can be evaluated by taking the $10\oplus10$ decay width from the $d^*$ antidecuplet. The only difference is the binding energy dependence of the width for a 28-plet compared to an antidecuplet would be the size of a cut-off parameter $\Lambda$. As one can see in Fig.~\ref{d_width}, for a given binding energy the width gets larger with an increase of the cut-off and for a small binding energy, close to $B=0~MeV$ the effect of cut-off is minimal due to the normalisation condition. Since we expect the 28-plet $\Lambda$ to be larger than for the antidecuplet, the use of $\Lambda_{\bar{10}}\sim0.16~GeV$ give us an upper limit for an expected width, Fig.~\ref{28_width}. The $\Omega\Omega$ state is stable against hadronic decays if bound so is therefore not shown in Fig.~\ref{28_width}.

\section{$\textrm{N} \Delta$ 27-plet}\label{sec:ND27}

\begin{figure}[!h]
\begin{center}
\includegraphics[angle=0,width=0.4\textwidth]{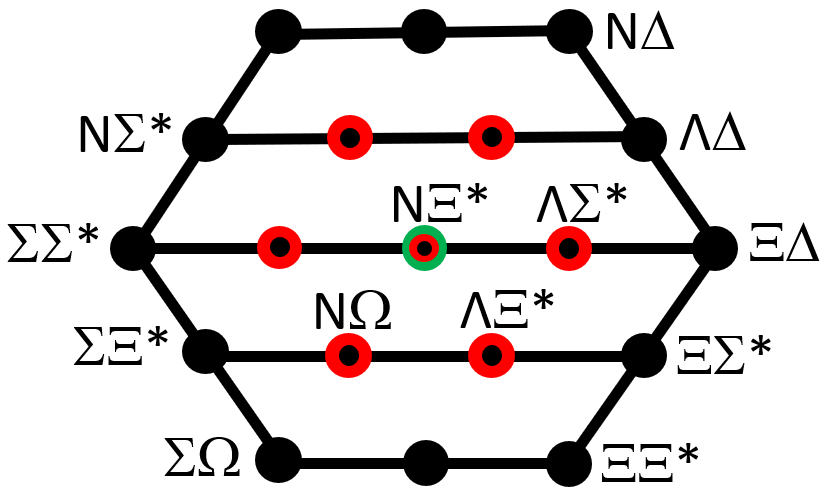}
\end{center}
\caption{$\textrm{N}\Delta$ 27-plet $8\oplus10$}
\label{27_n_delta_multiplet}
\end{figure}

This multiplet (Fig.~\ref{27_n_delta_multiplet}) is very interesting. The zero strangeness $\textrm{N}\Delta$ state is known since the '60s and is often referred to as an R. Arndt resonance~\cite{Arndt1,Arndt2}. The best way to observe it is a $\pi d \to pp$ or $pp \to \pi d$ reaction, where it reveals itself as a perfect loop on the Argand plot in a $^1D_2$ partial wave. There are plenty of data collected on this resonance experimentally - including differential cross-sections, single- and double-polarisation observables~\cite{Arndt2}. There are even some data on deuteron tensor polarisation for this reaction~\cite{dpiTensor}. While experimental evidence for this state is settled, the theoretical interpretation of these data is challenging. We have a very long list of proposed explanations, including  $\textrm{N}\Delta$-FSI, extra attraction due to $\rho$-meson exchange, the effect of box diagrams and it being a genuine dibaryon state. The latest analysis of Hoshizaki tends to explain it with a dibaryon state~\cite{Hoshizaki1,Hoshizaki2}. Also, modern Fadeev calculations from Gal\&Garcilazo tend to explain this state with $\textrm{N}\Delta$ molecule~\cite{Gal1}. Very recently Niskanen put a paper explaining this state by scattering dynamics in presence of attractive $N-\Delta$ potential~\cite{Niskanen}. So after 50 years of research the field is not settled. It is also unclear if its various descriptions derive from semantics due to the different theoretical approaches or if we are really discussing different phenomena.  Regardless, under the assumption of a bound state with 20 MeV binding energy, it is hard to expect anything other than a loosely bound molecules, inline with the prediction of Ref.~\cite{Gal1}. We therefore take this as an assumption in our theoretical analysis. 

There are two important questions to be addressed - what are the expected binding energies for the other states and what are their branching ratios? Flavour configurations for all members of multiplet for both $8\oplus10$ and $8\oplus8$ decays together with their nominal masses are shown in Tables~\ref{Tab27_810},~\ref{Tab27_88},~\ref{Tab27_810a}. Making parallels to the deuteron antidecuplet, where the first state (deuteron) is bound by 2.2~MeV and the second($\textrm{N}\Lambda$) is unbound by 166keV only, we can expect for the 27-plet with the first state bound by 20 MeV that higher strangeness bound states might also exist. Since the mass splitting is comparable or even larger than expected binding energies, the use of SU(6) formalism from Section~\ref{sec:dStar.mol} is not applicable. However, we can use similar arguments to make qualitative evaluations. One can say that the binding should be proportional to the number of light quarks squared, so if a state with all 6 light quarks has a binding energy of 20 MeV, the state with 3 light and 3 strange quarks should have a binding energy of $\sim 1.3$~MeV~(Table \ref{D27MassTab}). The latter state was recently explored both theoretically and experimentally with the observation of $p\Omega$ state in heavy ion collisions~\cite{NOmega} and in lattice QCD calculations~\cite{NOmegaLQCD}, which predicted it to be bound by 2.5 MeV (1.5 MeV strong binding plus 1.0MeV electromagnetic binding). Hence we can expect a single strangeness state to be bound by about 10 MeV and a double strangeness state by 3.2 MeV\footnote{the strangeness 4 state is likely to be unbound}. 

In all states with a $\Delta$ component, the width is largely dominated by the width of the $\Delta$ and resulting states appeared to have similar widths e.g. $\Gamma(^{27}d)=111$MeV and $\Gamma(^{27}d_{ss})=96$~MeV. All of these states show interesting, and rather counterintuitive behaviour. An increase of binding energy first leads to an increase of the width. However, the increase of the binding leads to a decrease of particle separation in the state and therefore an increase of the wave function overlap. This unavoidably means an increase of the phase Space available for the octet-octet decay branch. That is why all members of this multiplet have wider widths than free states.

\begin{table}[ht]

\centering \protect\caption{Expected $8\oplus10$ decay branches of the  spin $J=2$ 27-plet}
\vspace{2mm}
{%
\tiny
\resizebox{\textwidth}{!}{%
\setlength{\tabcolsep}{10pt}
\renewcommand{\arraystretch}{1.5}
\begin{tabular}{|l|c|c|c|c|}
\hline
$S$  & Max Isospin & Med Isospin & Min Isospin  \\
\hline
\ 0     & $\textrm{N}\Delta$ & & \\
-1    & $\frac{1}{4}(\sqrt{5}\Sigma\Delta+3\Lambda\Delta-\sqrt{2}\textrm{N}\Sigma^*)$ & $\frac{1}{\sqrt{5}}(2\textrm{N}\Sigma^{*}-\Sigma\Delta)$   & \\
-2    & $\frac{1}{2}( \sqrt{3}\Xi\Delta-\Sigma\Sigma^{*})$ & $\frac{\sqrt{5}}{10}(3\Sigma\Sigma^*+\sqrt{6}\Lambda\Sigma^*-2\textrm{N}\Xi^*-\Xi\Delta)$ & $\frac{1}{\sqrt{5}}( \sqrt{3}\textrm{N}\Xi^{*}-\sqrt{2}\Sigma\Sigma^{*})$ \\
-3    & $\frac{1}{\sqrt{2}}(\Xi\Sigma^{*}-\Sigma\Xi^{*})$ & $\frac{\sqrt{5}}{20}(7\Sigma\Xi^*+3\Lambda\Xi^*-3\sqrt{2}\textrm{N}\Omega-2\Xi\Sigma^*)$ &  \\
-4    & $\frac{1}{2}(\sqrt{3}\Sigma\Omega - \Xi\Xi^*)$ & & \\

\hline
\end{tabular}} \label{Tab27_810}
}
\end{table}


\begin{table}[ht]

\centering \protect\caption{Expected $8\oplus8$ decay branches of the  spin $J=2$ 27-plet}
\vspace{2mm}
{%
\tiny
\scalebox{1.2}{
\setlength{\tabcolsep}{10pt}
\renewcommand{\arraystretch}{1.5}
\begin{tabular}{|l|c|c|c|}
\hline
$S$  & Max Isospin & Med Isospin & Min Isospin  \\
\hline
\ 0     & $\textrm{NN}$ & & \\
-1    & $\textrm{N}\Sigma$ & $\frac{1}{\sqrt{10}}(3\textrm{N}\Lambda$ - $\textrm{N}\Sigma)$ &  \\
-2    & $\Sigma\Sigma$ & $\frac{1}{\sqrt{10}}(2\textrm{N}\Xi+\sqrt{6}\Sigma\Lambda)$ & $\frac{1}{\sqrt{10}}(\frac{3\sqrt{3}}{2}\Lambda\Lambda-\sqrt{3}\textrm{N}\Xi- \frac{1}{2}\Sigma\Sigma)$ \\
-3    & $\Sigma\Xi$ & $\frac{1}{\sqrt{10}}(3\Lambda\Xi-\Sigma\Xi)$ & \\
-4    & $\Xi\Xi$ & & \\

\hline
\end{tabular}} \label{Tab27_88}
}
\end{table}


\begin{table}[ht]

\centering \protect\caption{Masses of $8\oplus10$ and $8\oplus8$ decay branches of the  spin $J=2$ 27-plet}
\vspace{2mm}
{%
\tiny
\scalebox{1.2}{
\setlength{\tabcolsep}{10pt}
\renewcommand{\arraystretch}{1.5}
\begin{tabular}{|c|c|c|}

\hline
$S$  & $8\oplus10$ Mass [MeV] & $8\oplus8$ Mass [MeV] \\
\hline
\ 0     & $\textrm{N}\Delta(2170)$ & $\textrm{N}\textrm{N}(1876)$\\
-1    & $\Sigma\Delta (2421)$ $\Lambda\Delta(2348)$ $\textrm{N}\Sigma^{*}(2321)$ & $\textrm{N}\Lambda(2054)$ $\textrm{N}\Sigma(2127)$\\
-2    & $\Sigma\Sigma^{*}(2572)$ $\Xi\Delta(2547)$ $\Lambda\Sigma^{*}(2499)$ $\textrm{N}\Xi^{*}(2470)$ & $\textrm{N}\Xi(2253)$ $\Lambda\Lambda(2232)$ $\Sigma\Sigma(2378)$ $\Sigma\Lambda(2305)$\\
-3    & $\Sigma\Xi^{*} (2721)$ $\Xi\Sigma^{*} (2698)$ $\Lambda\Xi^{*} (2648)$ $\textrm{N}\Omega (2610)$ & $\Lambda\Xi(2431)$ $\Sigma\Xi(2504)$\\
-4    &  $\Sigma\Omega(2861)$ $\Xi\Xi^*(2847)$ & $\Xi\Xi(2630)$\\

\hline
\end{tabular}} \label{Tab27_810a}
}
\end{table}

A $^{27}d_{s}$ state shows unusually small width of $\Gamma(^{27}d_s)\sim 40$~MeV. The lightest single strangeness component, $\textrm{N}\Sigma^*$ has a mass which is nearly 30~MeV lighter than the next lightest $\Lambda\Delta$ branch. An additional 10 MeV $^{27}d_{s}$ binding means that this state is bound by 40~MeV relative to $\Lambda\Delta$ pole. Such large binding leads to a sizeable reduction of a $\Delta$ width. On top of which, the $\Lambda\Delta$ branch is also suppressed by Clebsh-Gordan coefficients (see Tab.~\ref{Tab27_810}), making $^{27}d_{s}$ sufficiently narrow, but still wider than a free $\Sigma^*$ state. The experimental search of this state might be challenging. One could search for the $\Lambda\Delta$ branch, however the necessity of a partial wave analysis of a 4 body reaction($\Lambda,\textrm{N},\pi$ + an associated particle) makes this channel challenging from a theoretical point of view. An $8\oplus8$ decay, where the state can reveal itself as a Flatte~\cite{Flatte} shape with sub-threshold (relative to the $8\oplus10$ pole) peaking or as a cusp-like behaviour looks more appealing theoretically, but very inconvenient from an experimental point of view due to necessity to measure $\Sigma$ in the final state.

\begin{table}[ht]
\centering 

\captionsetup{justification=centering,margin=0.5cm}
\protect\caption{Expected masses of the  spin $J=2$ $d$-27 SU(3) multiplet in pure molecular picture}
\vspace{2mm}
{%
\resizebox{\textwidth}{!}{
\begin{tabular}{|l|c|c|r|}
\hline
Particle  &  Binding energy structure & Mass value [MeV] & Binding\footnote{relative to the lightest Octet-Decuplet pole} [MeV]\\
\hline
$^{27}d$    & $M_{Red}(\textrm{N}\Delta)(3f\cdot3f)^2$  &  2151 & 20.0\\
$^{27}d_s$    & $[\frac{2}{16}M_{Red}(\textrm{N}\Sigma^*)+\frac{9}{16}M_{Red}(\Delta\Lambda)+\frac{5}{16} M_{Red}(\Delta\Sigma)](3f\cdot2f)^2$  & 2311   &  10.0\\
$^{27}d_{ss}$    & $3/4M_{Red}(\Xi\Delta)(3f\cdot1f)^2+1/4M_{Red}(\Sigma\Sigma^*)(2f\cdot2f)^2$  & 2551 & 3.2\\
$^{27}d_{sss}$    &  $1/2[M_{Red}(\Sigma\Xi^*)+M_{Red}(\Xi\Sigma^*)](2f\cdot1f)^2$ & 2704 & 1.3  \\
$^{27}d_{4s}$    &  $3/4M_{Red}(\Sigma\Omega)(2f\cdot0f)^2+1/4M_{Red}(\Xi\Xi^*)(1f\cdot1f)^2$ & 2853 &  0.3 \\
\hline
\end{tabular}} \label{D27MassTab}
}
\end{table}

\begin{figure}[!h]
\begin{center}
        \includegraphics[width=0.19\textwidth,angle=0]{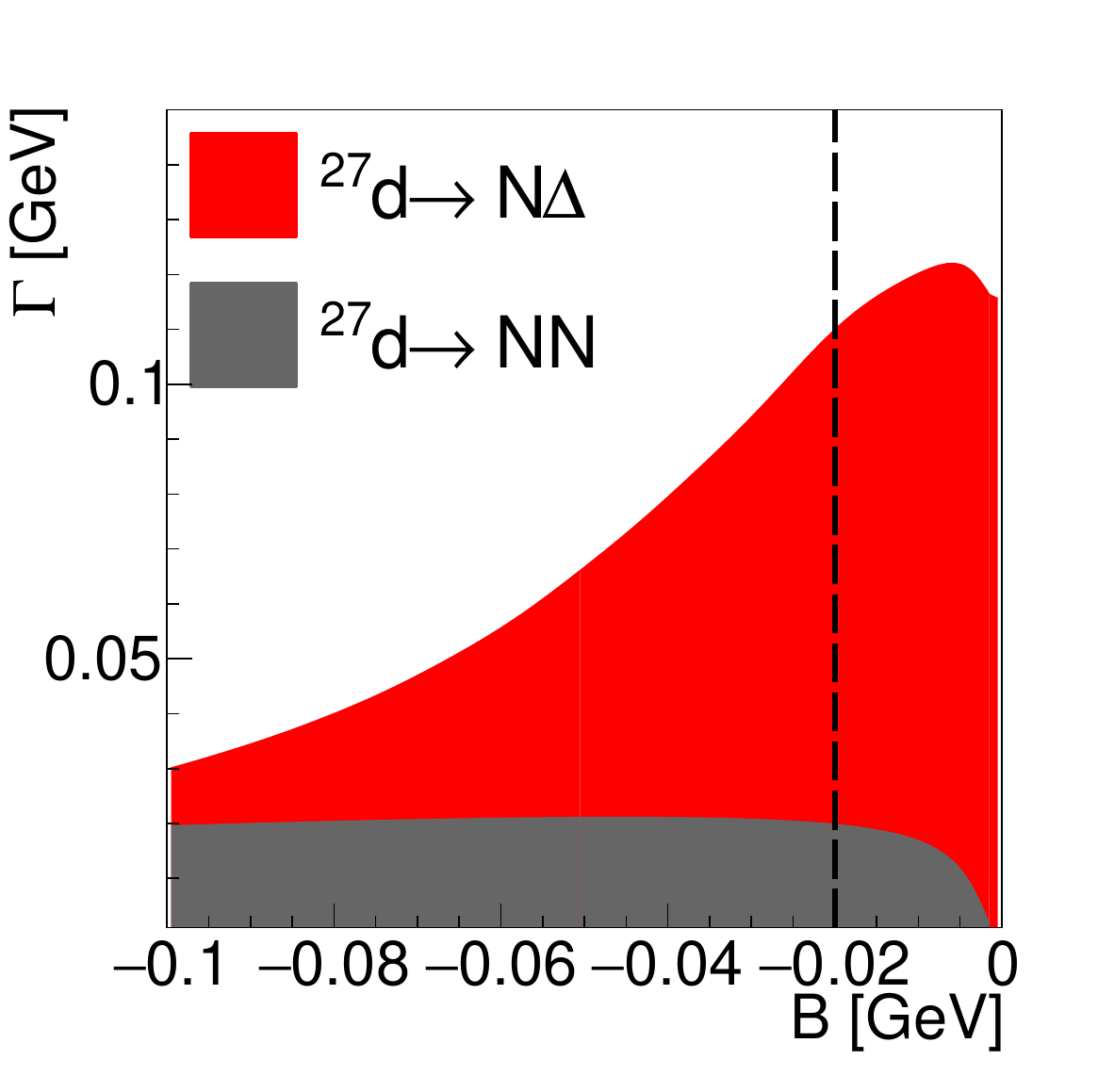}
        \includegraphics[width=0.19\textwidth,angle=0]{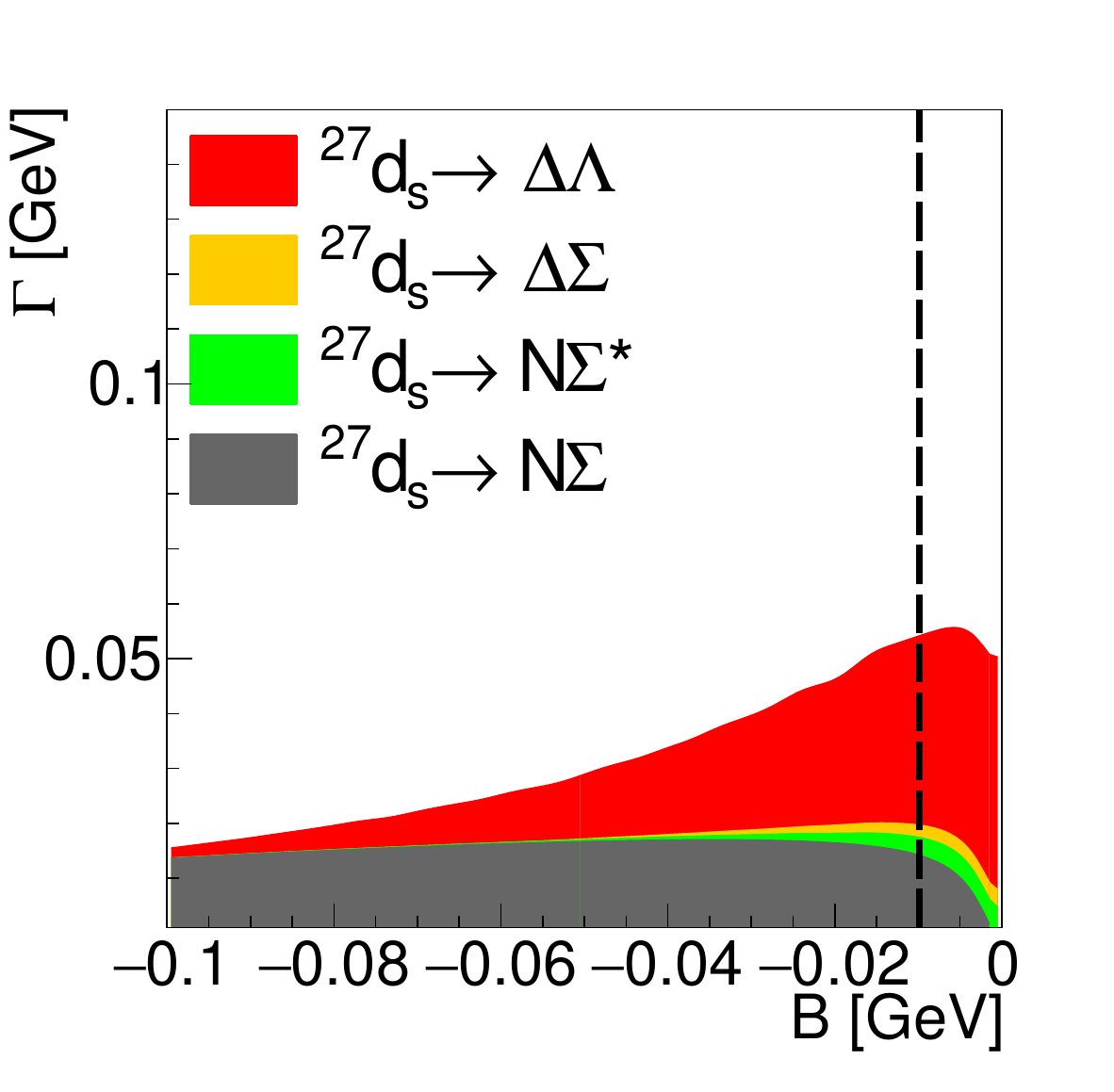}
        \includegraphics[width=0.19\textwidth,angle=0]{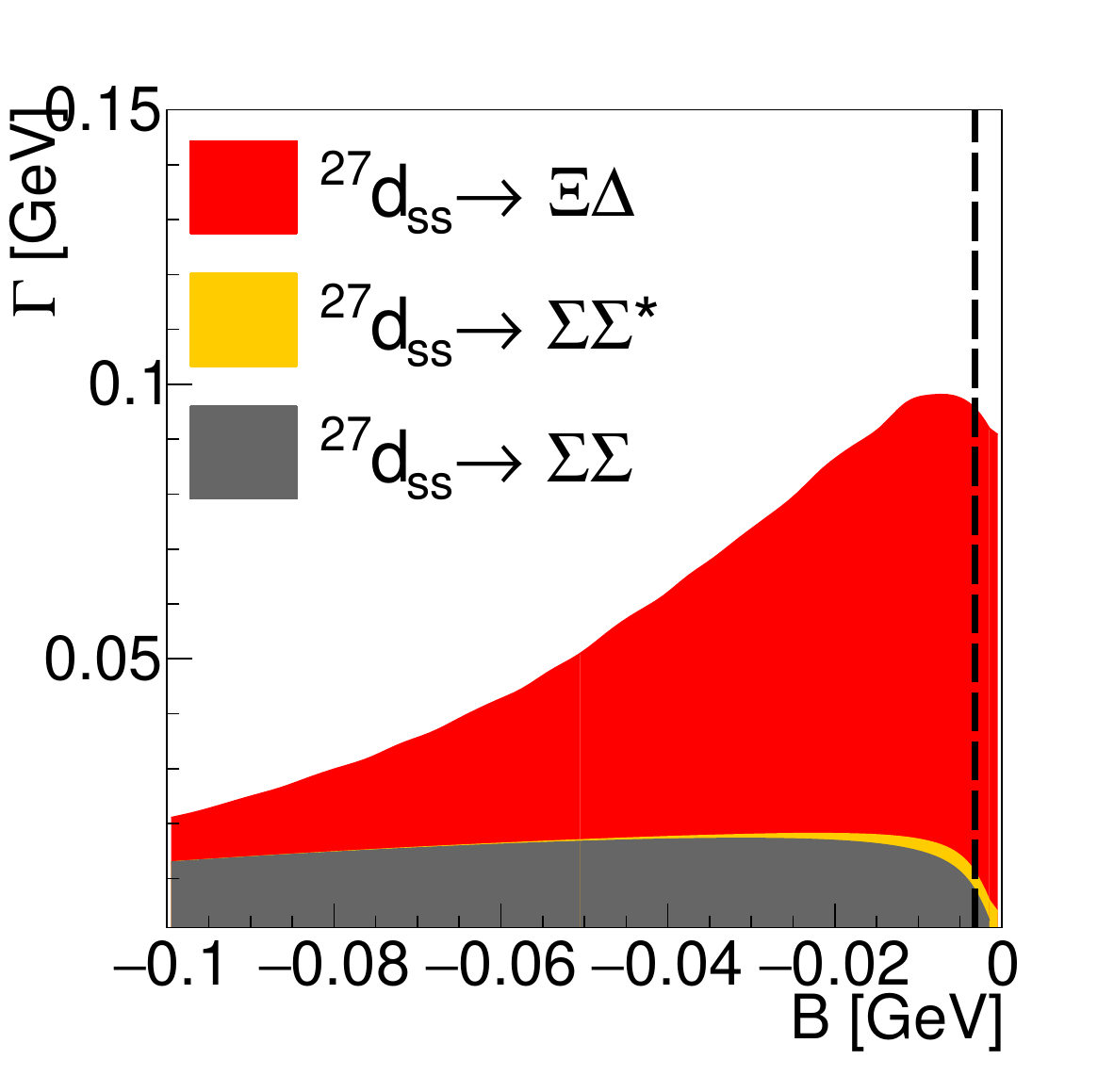}        \includegraphics[width=0.19\textwidth,angle=0]{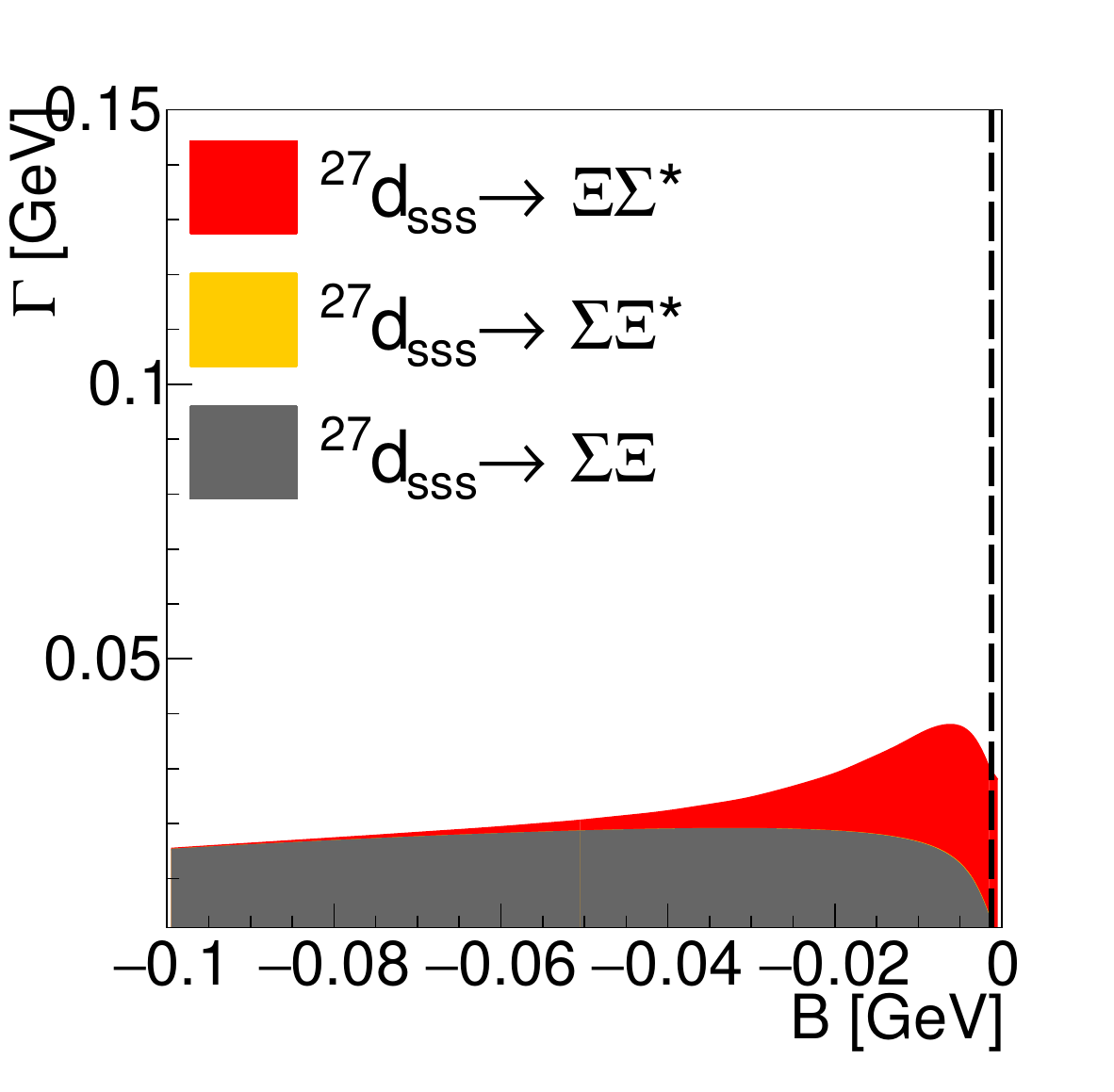}
        \includegraphics[width=0.19\textwidth,angle=0]{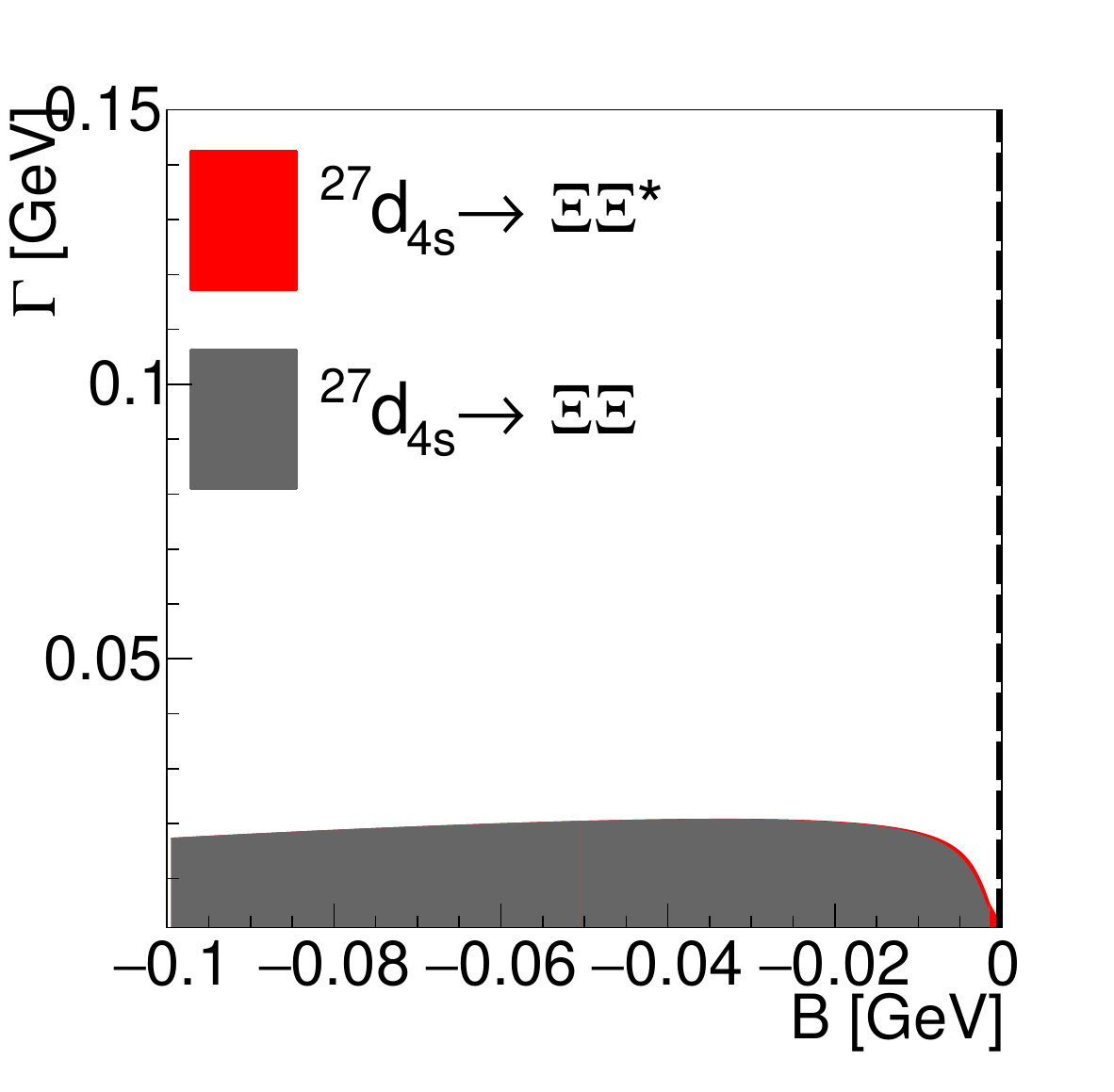}
\end{center}

\captionsetup{justification=centering,margin=0.2cm}
\caption{$d$-27 multiplet total width as a function of binding energy (relative to the lightest member of the Octet-Decuplet pole) for the states with various strangeness increasing from left to right split into major decay branches. The vertical dashed line shows the expected mass as specified in Table.~\ref{D27MassTab}}
\label{d27_all_width_M}
\end{figure}

\begin{figure}[!h]
\begin{center}
        \includegraphics[width=0.19\textwidth,angle=0]{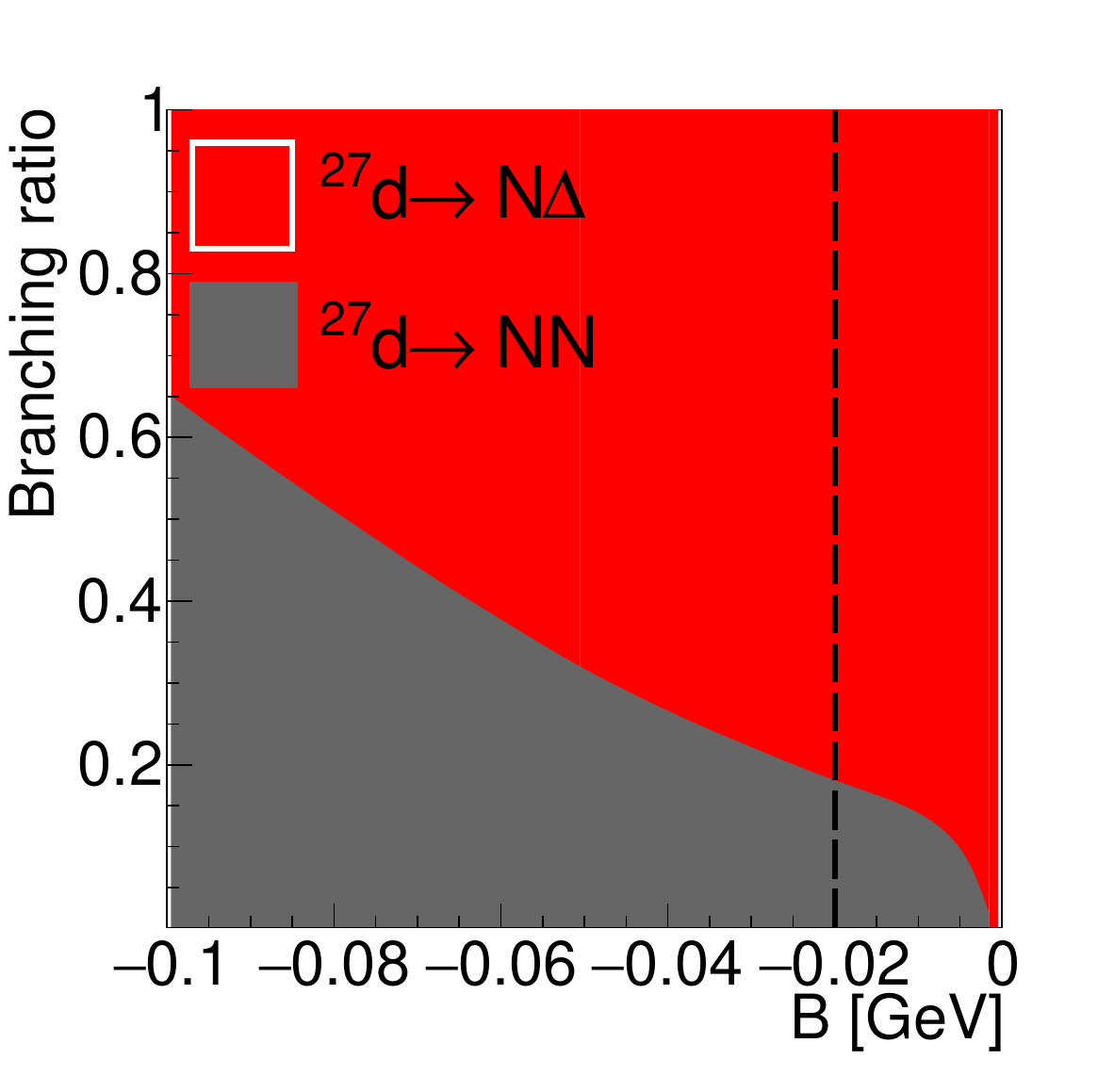}
        \includegraphics[width=0.19\textwidth,angle=0]{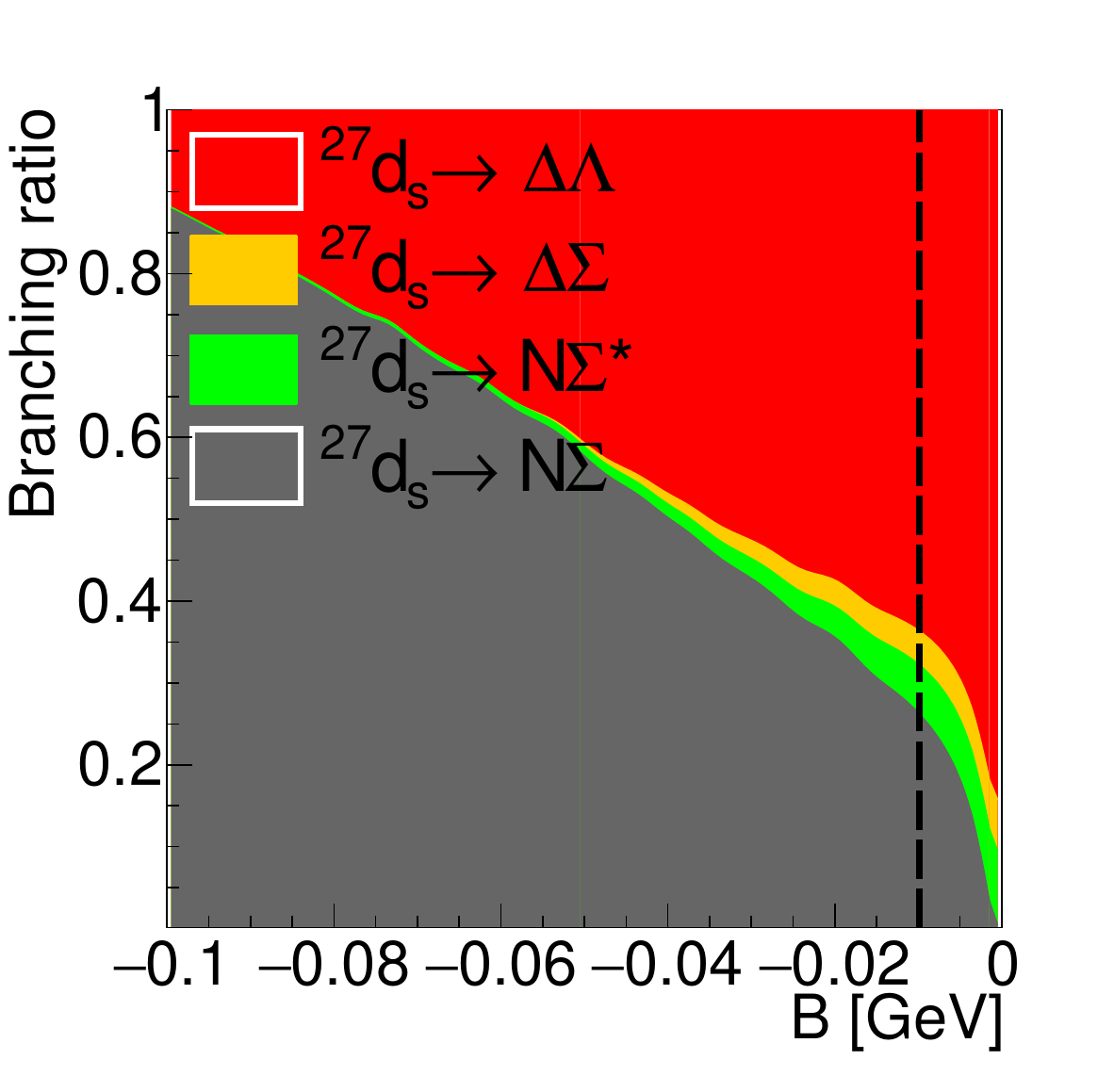}
        \includegraphics[width=0.19\textwidth,angle=0]{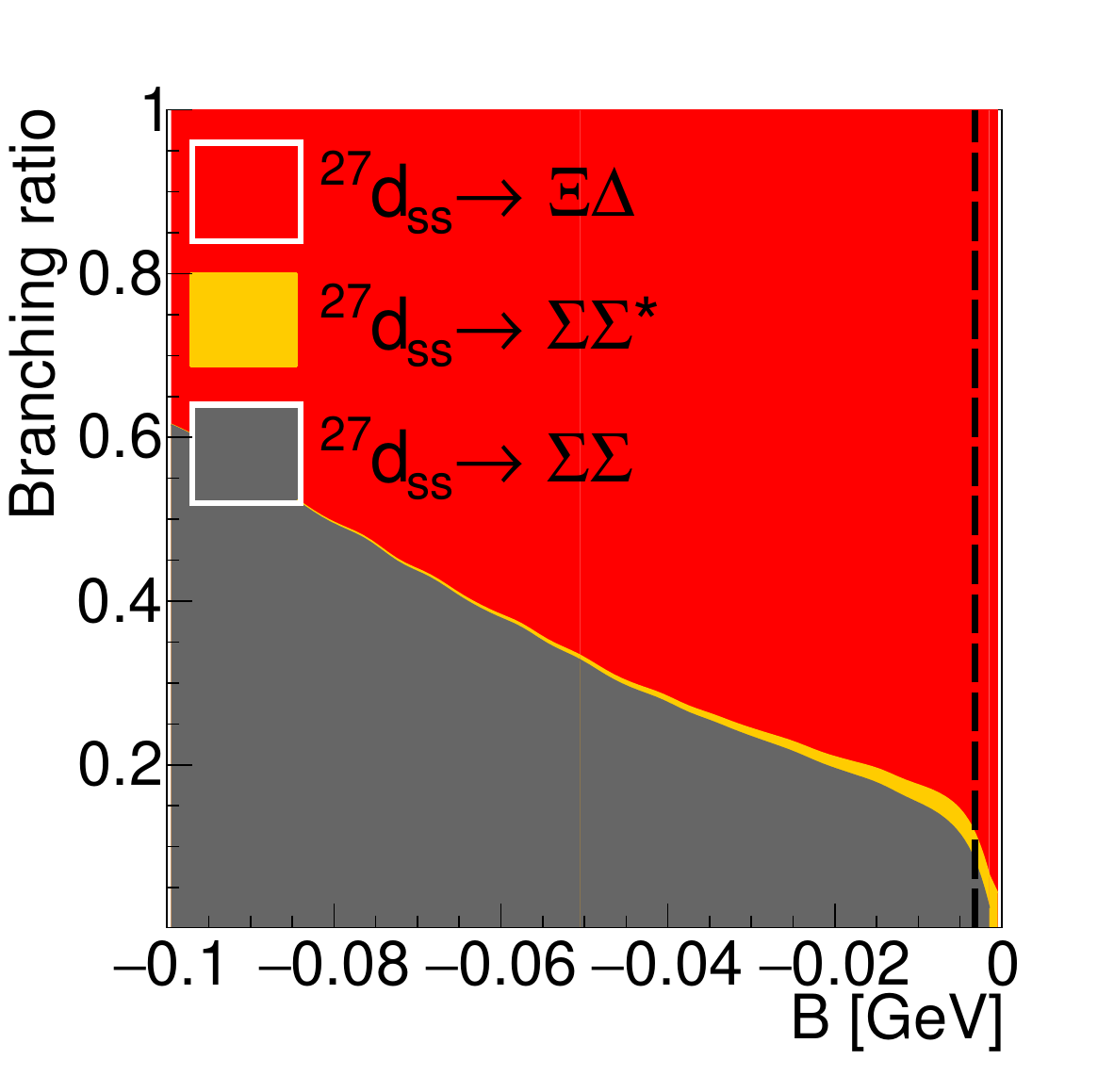}        \includegraphics[width=0.19\textwidth,angle=0]{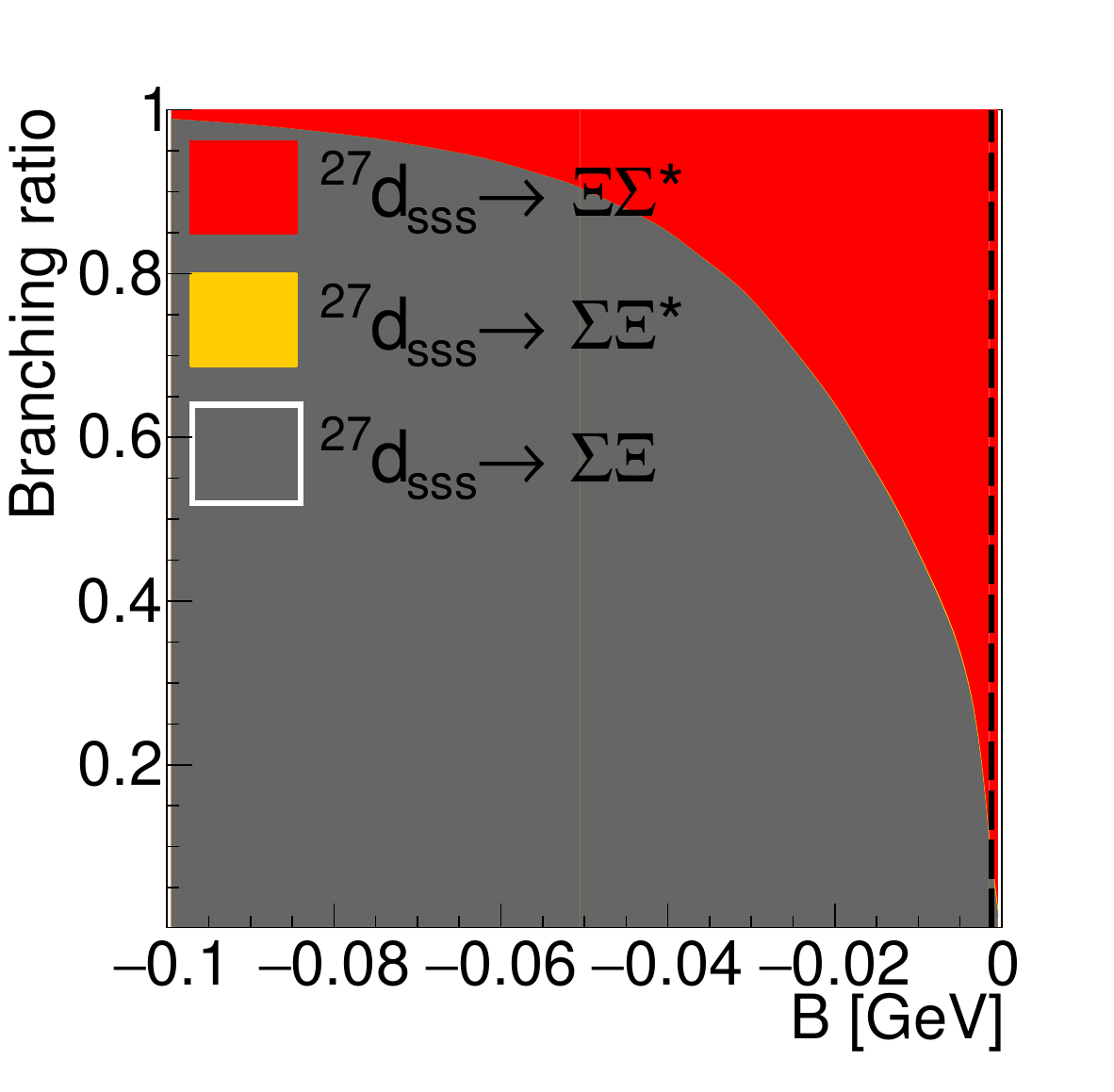}
        \includegraphics[width=0.19\textwidth,angle=0]{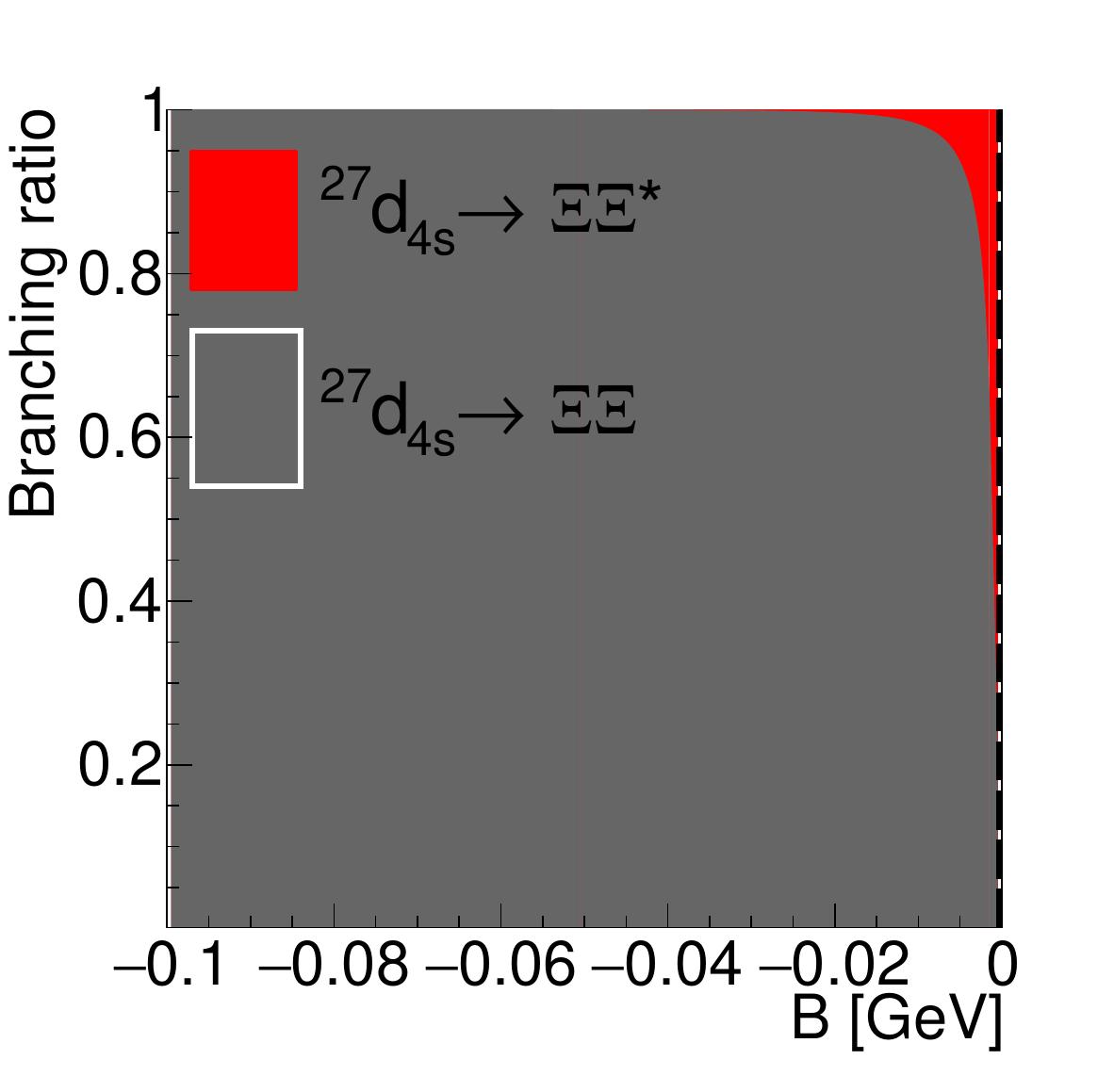}
\end{center}
\caption{same as Fig.~\ref{d27_all_width_M}, but for branching ratios}
\label{d27_all_Br_M}
\end{figure}

Strangeness -2 state with an isospin $I=2$ looks very unfavourable due to the large $\Xi\Delta$ branch as can be seen in Fig.~\ref{d27_all_width_M},~\ref{d27_all_Br_M}. The isospin $I=1$ state also has a $\Xi\Delta$ component, somewhat suppressed by the CG coefficients. Only the isospin $I=0$ state does not have $\Delta$ components. This state also has rather convenient $8\oplus8$ decays, $N\Xi$ and $\Lambda\Lambda$ which can be explored experimentally. As discussed earlier for another 27-plet case one may expect some mixing between states. Due to this uncertainty, we do not show any width/branching calculations for the inner multiplet states, where SU(3) mixing can occur, Table~\ref{Tab27_width_M}.

\begin{figure}[!h]
\begin{center}
        \includegraphics[width=0.4\textwidth,angle=0]{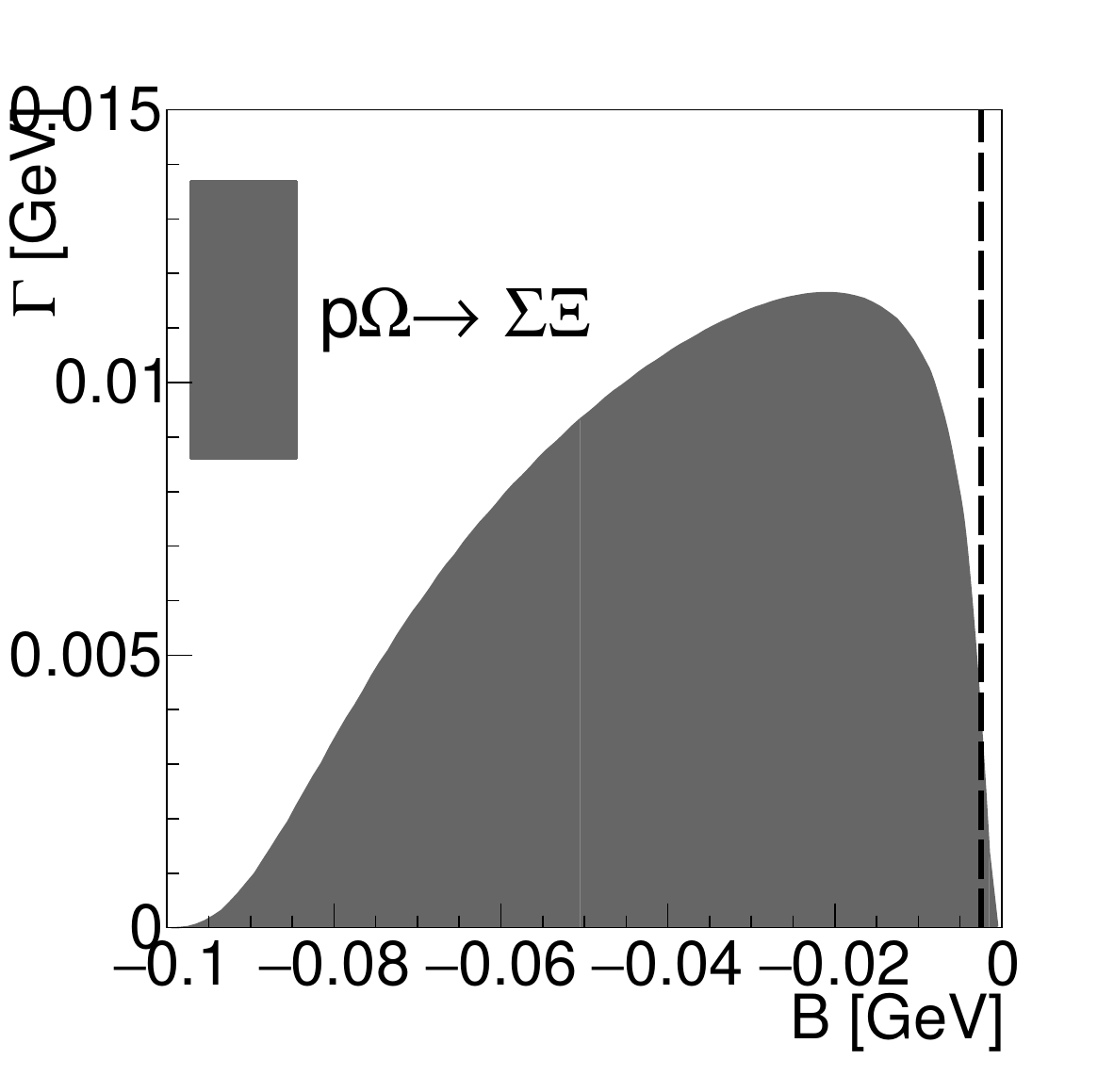}        
\end{center}
\caption{$p\Omega\to\Sigma\Xi$ decay width}
\label{pOmega_width}
\end{figure}

It is very unlikely that strangeness -3 isospin $I=3/2$ state will be observed. A tiny binding of about 1~MeV would be really difficult to see in a 30~MeV wide state. The inner $I=1/2$ state is a lot more interesting. We do not know the composition of that state due to unknown isospin mixing and enormous symmetry breaking due to the mass difference, however, we do know that the $p\Omega$ component is by far the lightest one among all($M(p\Omega)=2611$~MeV, $M(\Lambda\Xi^*)=2648$~MeV, $M(\Xi\Sigma^*)=2705$~MeV,$M(\Sigma\Xi^*)=2721$~MeV). Both Lattice QCD~\cite{NOmegaLQCD} and the early heavy ion correlation function analysis~\cite{NOmega} claimed the existence of this state with a binding energy of about 2.5~MeV, where half of the binding originates from the electromagnetic attraction. So this state, if it exists, is a mixture of an atomic and a hadronic molecule. The latest heavy ion results with increased statistics tend to suggest this state is unbound. If bound, this state is stable against $8\oplus10$ decays, since $\Omega$ is stable against strong decays, however, it is still allowed to decay into $8\oplus8$ via the most prominent $\Lambda\Xi$ channel due to both phase space and the large, $9/10$, associated CG coefficient. For a $\sim 2.5$~MeV binding energy one can expect a measurable $\Gamma(p\Omega\to\Sigma\Xi)\sim 4$~MeV width, see Fig.~\ref{pOmega_width}. This coupled channel effect also needs to be taken into account in the analysis of heavy ion correlation data. 

\begin{table}[ht]

\centering 
\captionsetup{justification=centering,margin=0.5cm}
\protect\caption{$^{27}d$ multiplet width results for the maximum isospin states and expected binding energies from Table\Ref{D27MassTab}.}
\vspace{2mm}

\resizebox{\textwidth}{!}{
\begin{tabular}{lrr|lrr|lrr|lrr|lrr}
\hline

\multicolumn{3}{|c|}{$^{27}d$(2151), B=20~MeV} & \multicolumn{3}{|c|}{$^{27}d_s$(2311), B=10~MeV} & \multicolumn{3}{|c|}{$d_{ss}$(2551), B=3~MeV} & \multicolumn{3}{|c|}{$d_{sss}$(2704), B=1.3~MeV} & \multicolumn{3}{|c|}{$d_{4s}$(2853), B=0.3~MeV} \\

\hline

 decay & $\Gamma$, [MeV]  & BR [\%] & decay & $\Gamma$, [MeV]  & BR [\%] & decay & $\Gamma$, [MeV]  & BR [\%] & decay & $\Gamma$, [MeV]  & BR [\%] & decay & $\Gamma$, [MeV]  & BR [\%]\\
 $\textrm{N}\Delta$ & 91 & 82 & $\Lambda\Delta$ & 34.4 & 64 & $\Xi\Delta$ & 84.5 & 89 & $\Xi\Sigma^*$ & 27.1 & 91 & $\Xi\Xi^*$ & 2.2 & 81\\
  &  &  & $\Sigma\Delta$  & $2.2$ & $4$ & $\Sigma\Sigma^*$  & 3.3 & 3 & $\Sigma\Xi^*$ & 0.2 & 1  & $\Sigma\Omega$ & 0 & 0 \\
  &  &  & $\textrm{N}\Sigma^*$  & 3.2 & 6 &   &  &  &  &  &  & & &  \\
  \hline
 $\textrm{N}\textrm{N}$ & 20 & 18 &  $\textrm{N}\Sigma$ & 14.2 & 26 & $\Sigma\Sigma$ & 7.7 & 8 & $\Sigma\Xi$ & 2.3 & 8  & $\Xi\Xi$ & 0.5 &  19 \\   
\hline
\hline
total & {\bf 111} &  & total  & {\bf 54.0} & & total & {\bf 95.5} & & total &  {\bf 29.6} &  & total & {\bf 2.7} &  \\
\hline
\end{tabular}}\label{Tab27_width_M}

\end{table}


The strangeness -4 state is very interesting. According to CG coefficients, it is mainly made of $\Sigma\Omega$ with $\Xi\Xi^*$ occupying only a quarter of the wavefunction. However, the mass of a $\Sigma\Omega$ configuration is nearly 10~MeV higher than the one of $\Xi\Xi^*$. So, if it exists, this state will be bound relative to the $\Xi\Xi^*$ threshold and with a $\Sigma\Omega$ branch being exactly zero. Also due to the very small binding, and hence the large size of this molecule, the wave function overlap will be tiny. That is why the $8\oplus8$ decay is strongly suppressed. All these factors would lead to an extremely small width of this state, provided it is bound. It should also be mentioned that an atomic $\Sigma^+\Omega^-$ state could be possible. It probably will be bound by about 1~MeV and have a width in the order of ~MeV with two possible decay branches $\Xi\Xi^*$ and $\Xi\Xi$. Both branches require quark rearrangements hence both will be small. The $\Xi\Xi$ has large phase space, but it requires additional spin-flip and angular momentum $L=2$. The $\Xi\Xi^*$ happens in $S-$wave, does not require spin flips, but has tiny available phase space $<10$~MeV. One can search this state in $\Xi\Xi$ pair correlations in heavy ion collisions. The study of this atomic state can give a lot of information about hadronic $\Sigma^+\Omega^-$ interactions which would appear as a small addition to the larger electromagnetic binding.



\section{$\textrm{N}\Delta$ 35-plet}\label{sec:ND35}

\begin{figure}[!h]
\begin{center}
\includegraphics[angle=0,width=0.35\textwidth]{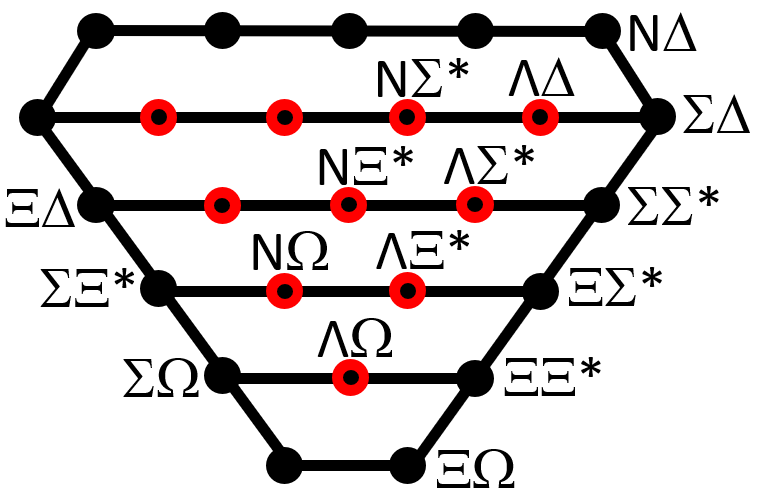}
\end{center}
\caption{$\textrm{N}\Delta$ 35-plet $8\oplus10$}
\label{35_n_delta_multiplet}
\end{figure}

\begin{figure}[!h]
\begin{center}
        \includegraphics[width=0.19\textwidth,angle=0]{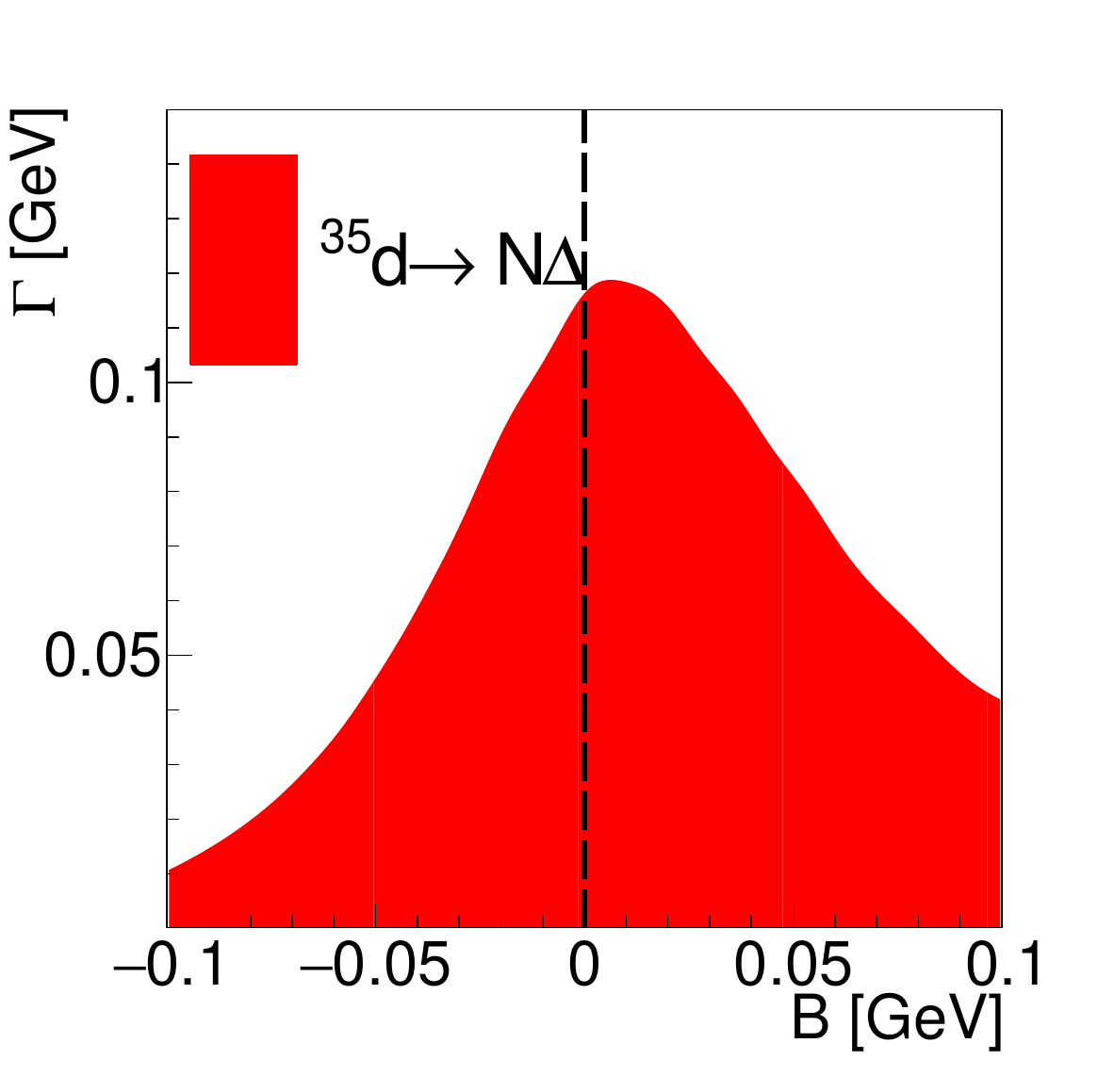}
        \includegraphics[width=0.19\textwidth,angle=0]{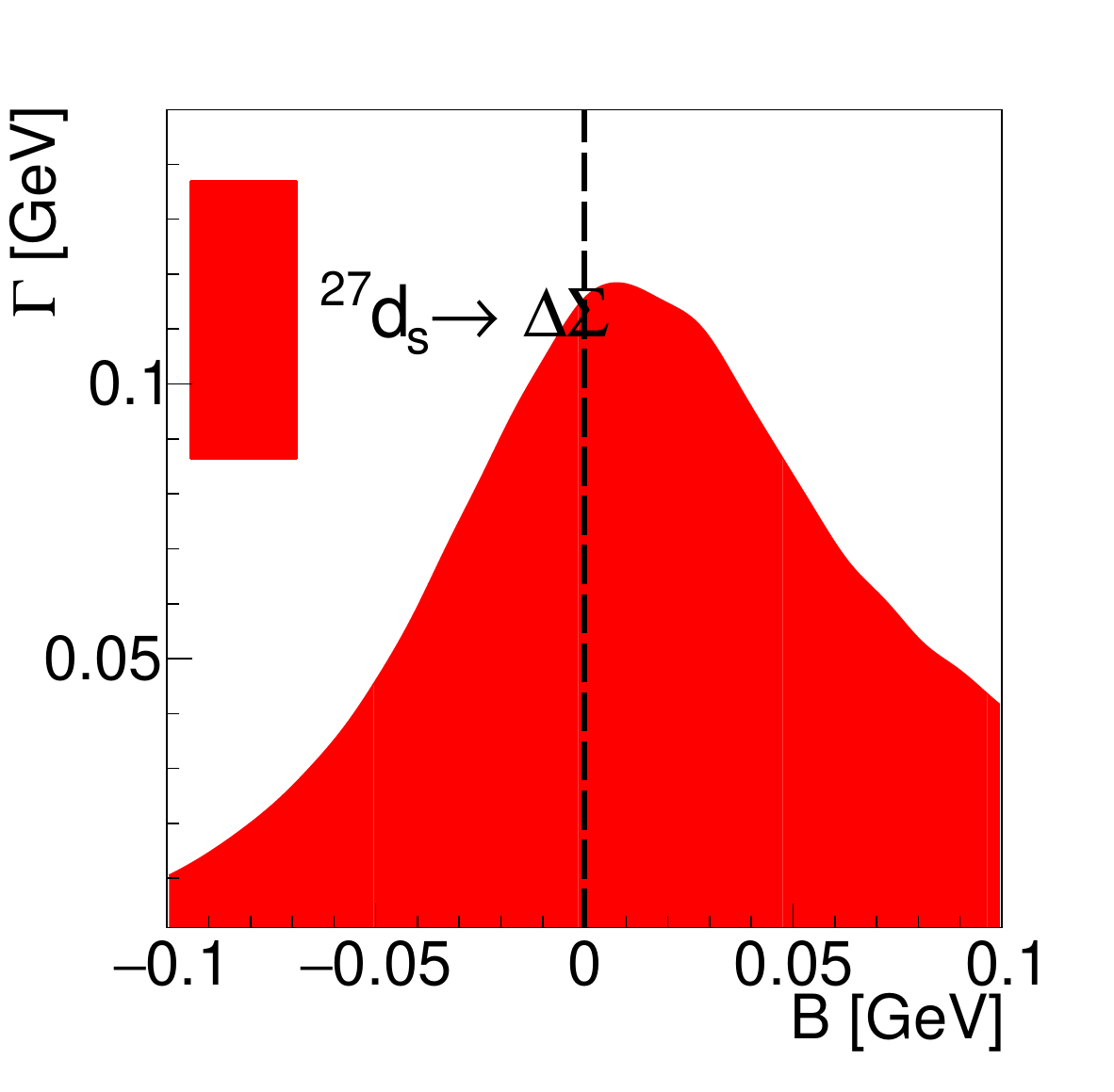}
        \includegraphics[width=0.19\textwidth,angle=0]{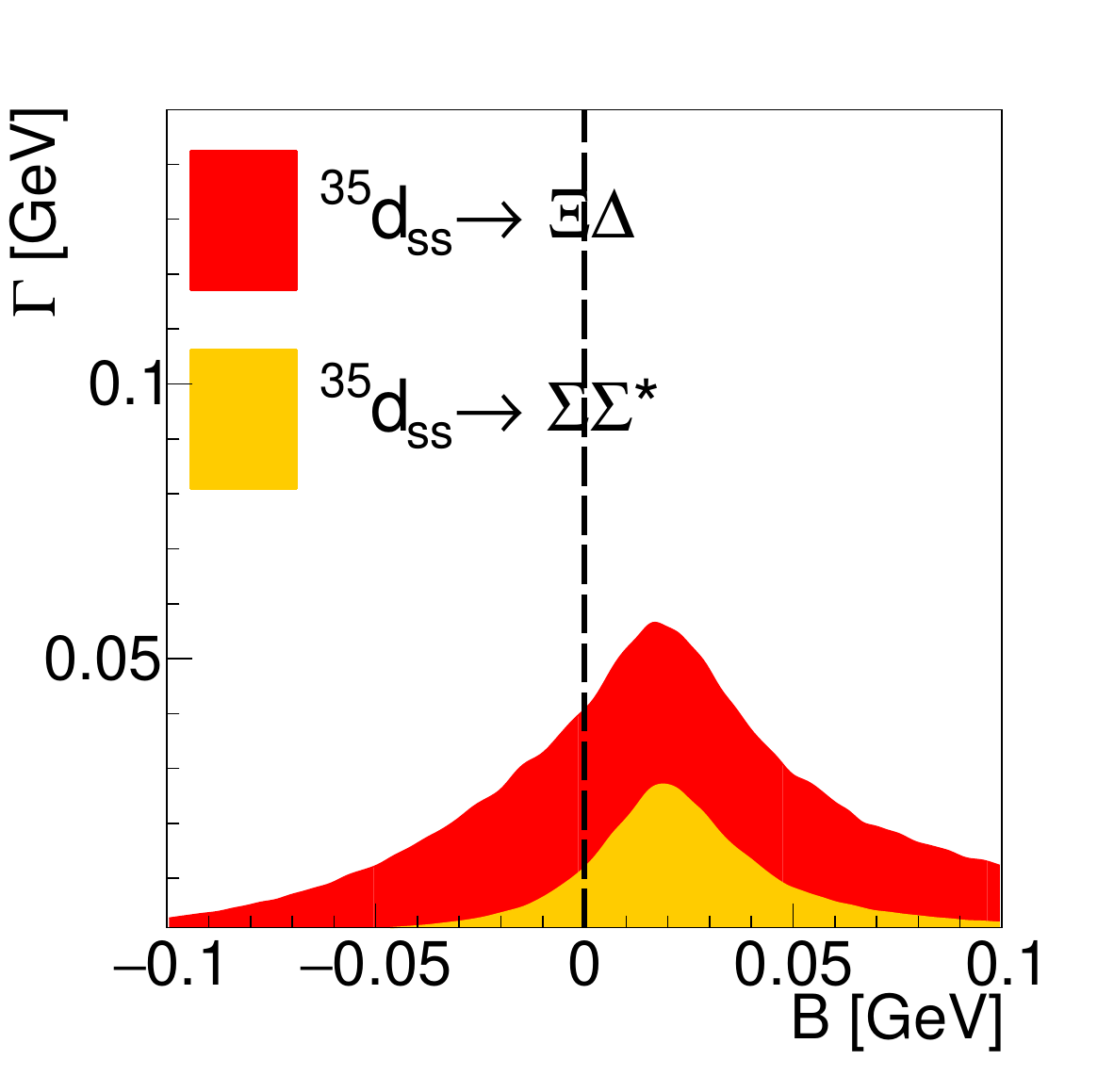}        \includegraphics[width=0.19\textwidth,angle=0]{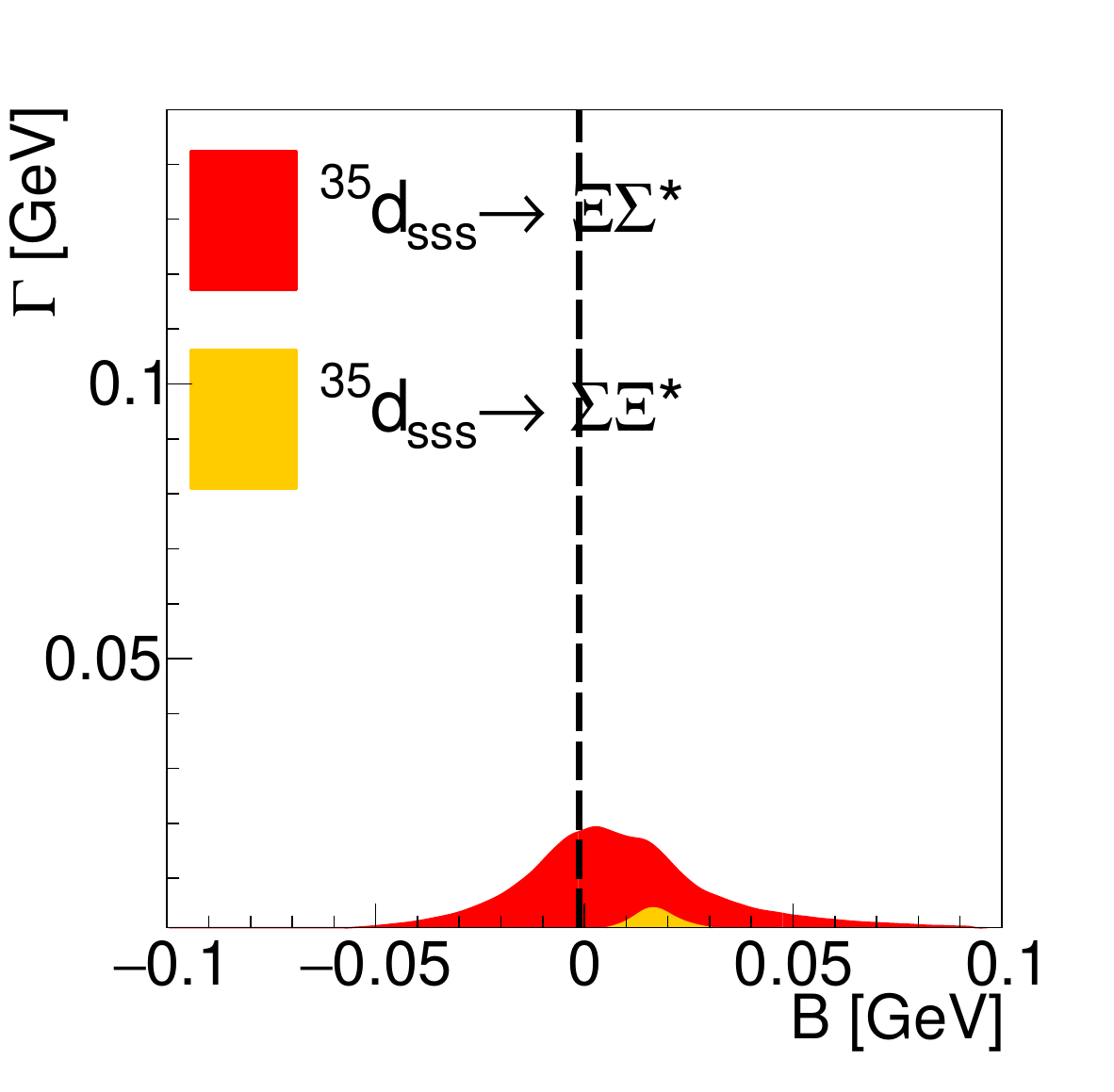}
        \includegraphics[width=0.19\textwidth,angle=0]{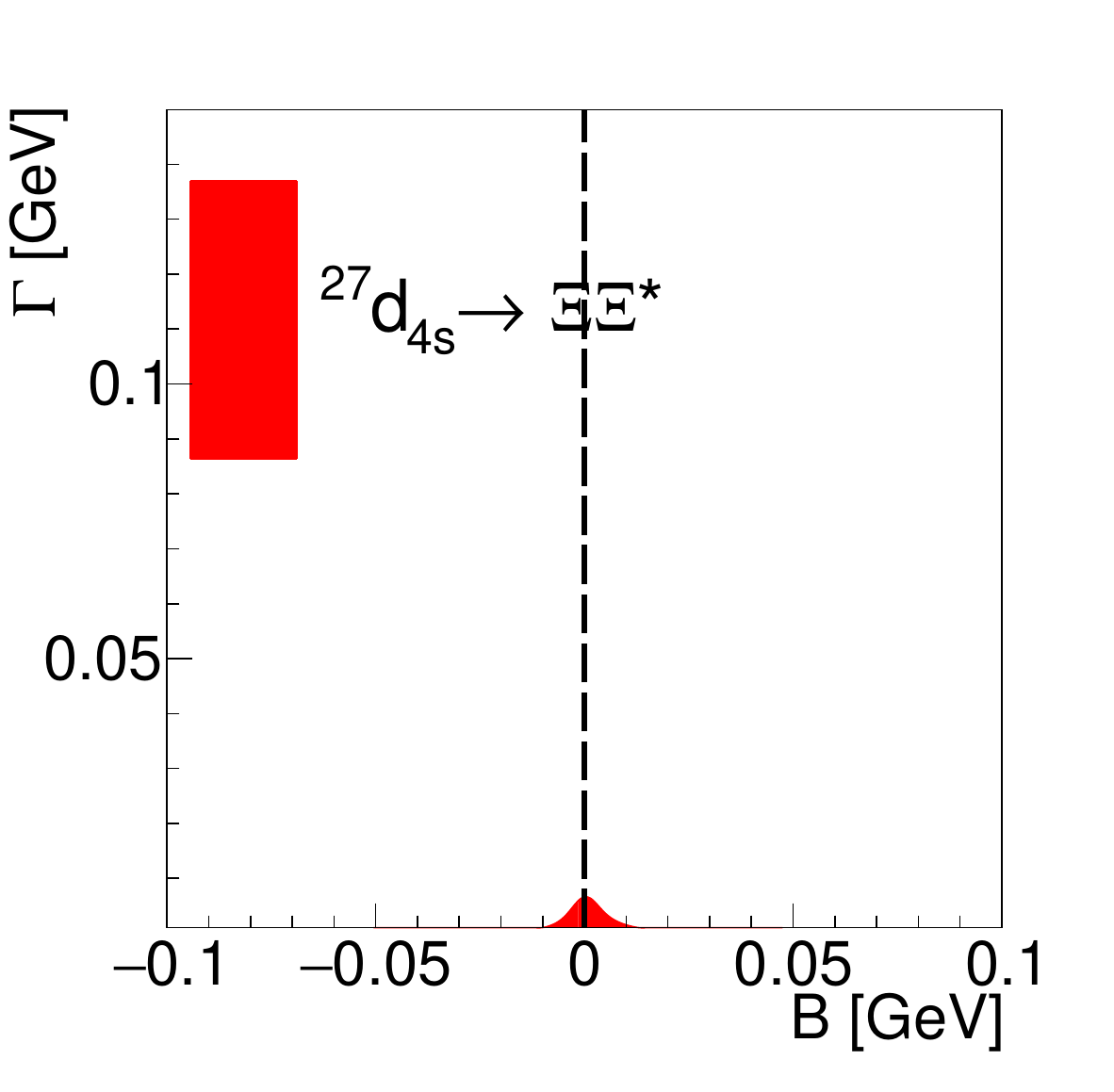}
\end{center}
\captionsetup{justification=centering,margin=0.5cm}
\caption{$d$-35 multiplet total width as a function of binding energy (relative to the lightest member of the Octet-Decuplet pole) for the states with various strangeness increasing from left to right split into major decay branches. The vertical dashed line shows zero}
\label{d35_all_width_M}
\end{figure}

In general, the analogy between $\textrm{N}\Delta$ 27- and 35-plets(Fig.~\ref{35_n_delta_multiplet}) should be similar to that between the $10^*-$ and $27-$ NN multiplets. Since the 27-plet is less bound compared to $10^*-$, the 35-plet should be less bound compare to $\textrm{N}\Delta$ 27-plet. Accessing the members of these multiplet is challenging - they all have large isospin, they do not couple to $8\oplus8$, so production of any of these states require associated particles and all decays are fall-apart many body decays. By analogy to NN-multiplets, one can even expect that all states of this multiplet should be unbound (deuteron is bound, while nn/pp-states of "Demon deuteron" are not). However, recently it was found experimentaly that a spin $S=1$, isospin $I=2$ $\textrm{N}\Delta$ state appears to bound by nearly 20~MeV - only slightly smaller compared to the $\textrm{N}\Delta$ state of a 27-plet~\cite{isoT}. If true, it implies that the other states of this multiplet may also be bound. That is why we decided to calculate possible decay width of these states as well, see Fig.~\ref{d35_all_width_M}, in a similar fashion as a 27-plet. The flavour configurations for these states are presented in Table.~\ref{Tab35}. 
\begin{table}[ht]

\centering \protect\caption{Expected decay branches of the spin $J=1$  SU(3) 35-plet}
\vspace{2mm}
{%
\resizebox{\textwidth}{!}{%
\begin{tabular}{|l|c|c|c|}
\hline
Strangeness  & Max Isospin & Min Isospin & Mass [MeV] \\
\hline
\ 0     & $\textrm{N}\Delta$ & & $\textrm{N}\Delta(2170)$\\
-1    & $\Sigma\Delta$ & $\frac{1}{4}(\sqrt{10}\textrm{N}\Sigma^{*}-\Sigma\Delta+\sqrt{5}\Lambda\Delta)$ & $\Sigma\Delta(2421)$ $\Lambda\Delta(2348)$ $\textrm{N}\Sigma^{*}(2321)$ \\
-2    & $\frac{1}{2}(\sqrt{3}\Sigma\Sigma^{*} + \Xi\Delta)$ &  $\frac{1}{2\sqrt{3}}(2\textrm{N}\Xi^{*}-\Sigma\Sigma^{*}+\sqrt{6}\Lambda\Sigma^{*}-\Xi\Delta)$ & $\Sigma\Sigma^{*}(2572)$ $\Xi\Delta(2547)$ $\Lambda\Sigma^{*}(2499)$ $\textrm{N}\Xi^{*}(2470)$\\
-3    & $\frac{1}{\sqrt{2}}(\Sigma\Xi^{*}+\Xi\Sigma^{*})$ &  $\frac{1}{4}(\sqrt{2}\textrm{N}\Omega-\Sigma\Xi^{*}+3\Lambda\Xi^{*}-2\Xi\Sigma^{*})$ & $\Sigma\Xi^{*}(2721)$ $\Xi\Sigma^{*} (2698)$ $\Lambda\Xi^{*} (2648)$ $\textrm{N}\Omega (2610)$\\
-4    & $\frac{1}{2}(\Sigma\Omega + \sqrt{3}\Xi\Xi^*$) & $\frac{1}{\sqrt{2}}(\Lambda\Omega-\Xi\Xi^{*})$ & $\Sigma\Omega (2861)$ $\Xi\Xi^* (2847)$ $\Lambda\Omega (2788)$\\
-5    & $\Xi\Omega$ & &  $\Xi\Omega (2987)$\\
\hline

\end{tabular}} \label{Tab35}
}
\end{table}

In all cases we expect the widths of these states to be close to the nominal width of their constituents, with not much chance for an experiment to reliably detect any of these states. The two exceptions are $\Lambda\Omega$ and $\Xi\Omega$, which would be stable against strong decays, if bound. These states can be potentially accessed via heavy-ion correlation functions studies. However, accumulating sufficient statistics for the strangeness -4 or -5 states may be extremely challenging.

\section{Summary}

We have developed a theoretical model, employing experimentally constrained parameters, to predict the possible decay branches and partial widths for all members of the $d^*$ hexaquark antidecuplet. For all strange-quark containing members of the antidecuplet, the predicted widths are rather large, with the most promising decay channels including the broad $\Delta$ resonance in the final state. We demonstrated that a $d^*$ Form-Factor, which was first introduced to explain peculiarities in the $d^*\to d\pi\pi$ decay, can also explain the smallness of the $d^*$ width, in agreement  with the qualitative dimensional arguments of A. Gal. The results of the paper will be an important guide for the ongoing search for the $d^*$ anti-decuplet members employing photon-, pion-, and kaon-induced reactions, as well as in high-energy collider experiments.
We have also extended our model to systematically study all other light and strange dibaryons which can appear as a ground state of $8\oplus8$, $8\oplus10$ and $10\oplus10$ baryon multiplets configurations. Several interesting states were identified for experimental searches and strategies in search of these states were proposed.

\section{Acknowledgements}
This work has been supported by the U.K. STFC ST/V002570/1, 	ST/P004008/1 and Strong2020 grants. 

\end{document}